\documentclass{utthesis}

\usepackage{latexsym}
\usepackage[dvips]{graphicx} 
\usepackage{amsfonts}
\usepackage{amscd}
\usepackage{amsmath}    
\usepackage{enumerate}
\usepackage{color}
\usepackage{verbatim}
\usepackage{multirow}

\newcommand{\bra}[1]{\mbox{$\left\langle #1 \right|$}}
\newcommand{\ket}[1]{\mbox{$\left| #1 \right\rangle$}}

\degree{Doctor of Philosophy Thesis}
\department{Department of Physics}
\gradyear{2008}
\author{Xiongfeng Ma}
\title{Quantum cryptography: from theory to practice}

\begin{document}

%


\begin{preliminary}

\maketitle


\begin{abstract}
Quantum cryptography or quantum key distribution (QKD) applies fundamental laws of quantum physics to guarantee secure communication. The security of quantum cryptography was proven in the last decade. Many security analyses are based on the assumption that QKD system components are idealized. In practice, inevitable device imperfections may compromise security unless these imperfections are well investigated.

A highly attenuated laser pulse which gives a weak coherent state is widely used in QKD experiments. A weak coherent state has multi-photon components, which opens up a security loophole to the sophisticated eavesdropper. With a small adjustment of the hardware, we will prove that the decoy state method can close this loophole and substantially improve the QKD performance. We also propose a few practical decoy state protocols, study statistical fluctuations and perform experimental demonstrations. Moreover, we will apply the methods from entanglement distillation protocols based on two-way classical communication to improve the decoy state QKD performance. Furthermore, we study the decoy state methods for other single photon sources, such as triggering parametric down-conversion (PDC) source. Note that our work, decoy state protocol, has attracted a lot of scientific and media interest. The decoy state QKD becomes a standard technique for prepare-and-measure QKD schemes. 

Aside from single-photon-based QKD schemes, there is another type of scheme based on entangled photon sources. A PDC source is commonly used as an entangled photon source. We propose a model and post-processing scheme for the entanglement-based QKD with a PDC source. Although the model is proposed to study the entanglement-based QKD, we emphasize that our generic model may also be useful for other non-QKD experiments involving a PDC source. By simulating a real PDC experiment, we show that the entanglement-based QKD can achieve longer maximal secure distance than the single-photon-based QKD schemes.

We propose a time-shift attack that exploits the efficiency mismatch of two single photon detectors in a QKD system. This eavesdropping strategy can be realized by current technology. We will also discuss counter measures against the attack and study the security of a QKD system with efficiency mismatch detectors.

\end{abstract}


%

\begin{acknowledgements}
The research presented in this Doctor of Philosophy thesis is carried out under the the supervision of Prof.~Hoi-Kwong Lo in the Department of Physics at the University of Toronto. I owe  my most sincere thanks to Hoi-Kwong for sharing his extensive knowledge with me. I can still clearly remember the time when I went to his office every week and struggled to understand the GLLP security analysis, how I was disappointed by my first simulation result, and how happy I was when I finished the simulation work for the decoy state method inspired by his conference paper. I am very grateful for his support of my non-academic life as well.

During my graduate study, I was lucky enough to be surrounded by wonderful colleagues: Jean-Christian Boileau, Ryan Bolen, Kai Chen, Marcos Curty, Fr\'{e}d\'{e}ric Dupuis, Ben Fortescue, Chi-Hang Fred Fung, Leilei Huang, Bing Qi, Li Qian, Kiyoshi Tamaki, Yi Zhao etc. In particular, I would like to thank Bing Qi for enormously helpful and enjoyable discussions about models, experimental setups and security analysis.

I wish to express my warm and sincere thanks to researchers in the field who have helped along the way and influenced the formation of the understanding and approach to quantum cryptography presented in this thesis. I would like to acknowledge that I have benefited very much from thoughtful discussions with Norbert L\"{u}tkenhaus, Jian-Wei Pan, Aephraim M.~Steinberg, Wolfgang Tittel, Gregor Weihs and the members of their research groups.

I would like to thank Ms. Serena Ma for her suggestions and proofreading. Responsibility for any remaining errors and omissions rests entirely with the author.

I gratefully acknowledge the financial support from the Chinese Government Award for Outstanding Self-financed Students Abroad and the Lachlan Gilchrist Fellowship.

Furthermore, my warm thanks are extended to the members of the Department of Physics, the Chinese Students and Scholars Association at the University of Toronto and the Student Diversity Group. With them, I enjoyed a colorful life as a graduate student at the University of Toronto.

Finally, and most importantly, I would like to thank my family for their constant and unending love and support. This thesis is dedicated to my parents, which without them, none of this would have been even possible.
\end{acknowledgements}

\tableofcontents



\end{preliminary}


\chapter{Introduction} \label{Intro}
\begin{quote}
Study the past, if you would divine the future. --- Confucius
\end{quote}

\section{Background}
In this section, we will give a brief overview of quantum information processing and then discuss one of its subfields that this thesis will focus on which is quantum cryptography\footnote{I acknowledge that Subsections \ref{Sub:QIPintro} and \ref{Sub:ClaCry} heavily rely on the Internet to gather information, especially wikipedia.org and quantiki.org.} 


\subsection{Quantum information processing} \label{Sub:QIPintro}
Quantum information processing or quantum information science is an amalgamation of quantum physics and information science. It concerns information science that depends on quantum effects in physics. It includes theoretical issues in communication and computational models as well as  experimental topics in quantum physics, including what can and cannot be done with quantum information. It is an interdisciplinary field, combining ideas in physics, information theory, engineering, computer science, mathematics and chemistry.

A bit; a binary digit, is the base of classical information theory. Regardless of its physical representation, it is always read as either a 0 or 1. For instance, a 1 (true value) is represented by a high voltage, while a 0 (false value) is represented by a low voltage.

A quantum bit, or qubit (sometimes qbit) is a unit of quantum information. That information is described by a state vector in a two-level quantum mechanical system which is formally equivalent to a two-dimensional Hilbert space. A qubit has some similarities to a classical bit, but is fundamentally very different. Like a bit, a qubit can have two possible values, normally a 0 or a 1. The difference is that whereas a bit must be either 0 or 1, a qubit can be 0, 1, or a superposition of both.

Subfields of quantum information processing include:
\begin{itemize}
\item
Quantum computing, which deals on the one hand, with the question how and whether one can build a quantum computer and on the other hand, searching algorithms that harness its power;

\item
Quantum computation, which investigates computational complexity of various quantum algorithms;

\item
Quantum error correction, which is used in quantum computing to protect quantum information from errors due to decoherence and other quantum noise;

\item
Quantum entanglement, which studies entanglement as seen from an information-theoretic point of view;

\item
Quantum cryptography and its generalization, quantum communication, which is the art of transferring a quantum state from one location to another. Note that this is the first quantum information application to reach the level of mature technology and fit for commercialization. This thesis focuses on quantum cryptography.
\end{itemize}

\subsection{Cryptography} \label{Sub:ClaCry}
Nowadays, distant communications play a crucial role in our daily lives. Secure communications become more and more important in many areas, e.g., online purchases, emails and video chats.

Cryptography is the practice and study of encoding and decoding secret messages to ensure secure communications. There are two main branches of cryptography: secret- (symmetric-) key cryptography and public- (asymmetric) key cryptography.

A key is a piece of information (a parameter) that controls the operation of a cryptographic algorithm. In encryption, a key specifies the particular transformation of plaintext into ciphertext, or vice versa during decryption. Keys are also used in other cryptographic algorithms, such as digital signature schemes and message authentication codes.

In practice, due to significant difficulties of distributing keys in secret key cryptography, public-key cryptographic algorithms are widely used in conventional cryptosystems. These encryption schemes can only be proven secure based on the presumed difficulty of a mathematical problem, such as factoring the product of two large primes. We emphasize that no public-key encryption scheme can be secure against eavesdroppers with unlimited computational power. 

One of the most famous quantum computing algorithms is Shor's algorithm \cite{ShorAlgorithm_94}, which can factor a number $N$ in $O((\log N)^3)$ time and $O(\log N)$ space. The algorithm is significant because it implies that public key cryptography might be easily broken, given a sufficiently large quantum computer. RSA \cite{RSA_78}, for example, uses a public key $N$ which is the product of two large prime numbers. One way to crack RSA encryption is by factoring $N$, but with classical algorithms, factoring becomes increasingly time consuming as $N$ grows large; more specifically, no classical algorithm is known that can factor in time $O((\log N)^k)$ for any $k$. By contrast, Shor's algorithm can crack RSA in polynomial time. It has also been extended to attack many other public-key cryptosystems.

In cryptography, the one-time pad is an encryption algorithm where the plaintext is combined with a random key or ``pad" that is as long as the plaintext and used only once. A modular addition is used to combine the plaintext with the pad\footnote{For binary data, the operation XOR amounts to the same thing.}. In 1917, Vernam proposed one-time pad encryption scheme \cite{OneTimePad_Vernam_26}. In 1949, Shannon proved that the one-time pad is information-theoretically secure, no matter how much computing power is available to the eavesdropper \cite{Shannon_OTP_49}. That is, if the key is truly random, never reused and kept secret, the one-time pad provides perfect secrecy. Note that the one-time pad is the only cryptosystem with perfect secrecy.

Despite Shannon's proof of its security, the one-time pad has serious drawbacks in practice:
\begin{enumerate}
\item
it requires a perfectly random key;

\item
secure generation and exchange of the key must be at least as long as the message.
\end{enumerate}

These implementation difficulties have led to one-time pad systems being unpractical and are so serious that they have prevented the one-time pad from being adopted as a widespread tool in information security.

Quantum physics offers a solution to the aforementioned two difficulties for the one-time pad. First, the superposition (uncertainty) nature of quantum mechanics can generate \emph{true} randomness. Secondly, quantum cryptography allows two distant parties to generate secure keys.

\subsection{Quantum cryptography} \label{Sub:Back:QCry}
Quantum cryptography or quantum key distribution (QKD) applies fundamental laws of quantum physics to guarantee secure communication. It enables two legitimate users, commonly named Alice and Bob, to produce a shared secret random bit string, which can be used as a key in cryptographic applications, such as message encryption (for instance, the one-time pad) and authentication. Unlike conventional cryptography, whose security often relies on unproven computational assumptions, QKD promises unconditional security based on the fundamental laws of quantum mechanics.



There are mainly two types of QKD schemes. One is the prepare-and-measure scheme, such as BB84 \cite{BB_84}, in which Alice sends each qubit in one of four states of two complementary bases; B92 \cite{Bennett_92} in which Alice sends each qubit in one of two non-orthogonal states; six-state \cite{sixstate_98} in which Alice sends each qubit in one of six states of three complementary bases. The other is the entanglement based QKD, such as Ekert91 \cite{Ekert_91} in which entangled pairs of qubits are distributed to Alice and Bob, who then extract key bits by measuring their qubits; BBM92 \cite{BBM_92} where each party measures half of the EPR pair in one of two complementary bases. Note that in Ekert91, Alice and Bob estimate the Eve's information based on the Bell's inequality test\footnote{In the original proposal \cite{Ekert_91}, the author claimed that the final key is secure when the Bell's inequality is maximally violated. There are many follow-up works, such as \cite{AGM_Bell_06}.}; whereas in BBM92, similar to BB84, Alice and Bob make use of the privacy amplification to eliminate Eve's information about the final key \cite{LoChauQKD_99}.

QKD needs a quantum channel and a classical channel. The quantum channel can be insecure whereas the classical channel is assumed to be authenticated. Fortunately,
in classical cryptography, unconditionally secure authentication schemes such as the Wegman-Carter authentication scheme \cite{WC_Authen_79,WC_Authen_81} exist. Moreover, those unconditionally secure authentication schemes are efficient: to authenticate an $N$-bit message, only an order $\log{N}$ bits of the shared key are needed. Since a small amount of pre-shared secure bits is needed between Alice and Bob, the goal of QKD is key growing, rather than key distribution. Notice that in the conventional information theory, key growing is an impossible task. Therefore, QKD provides a fundamental solution to a classically impossible problem.

The procedure of the best-known QKD protocol, BB84, is as follows. We assume that Alice uses polarization encoding.
\begin{enumerate}
\item
Alice randomly chooses one of the four states (vertical, horizontal, 45-degree and 135-degree polarizations). Denote the rectangular basis as $Z$ basis and the diagonal basis as $X$ basis. She sends the qubit to Bob through an insecure quantum channel.

\item
Bob randomly chooses $Z$ or $X$ basis to measure the received states. He keeps his measurement result secretly.

\item
Through a public classical channel, Alice and Bob compare the basis and only keep the measurement results that they use the same basis. This step is commonly called \emph{basis reconciliation}. If both of them randomly choose bases, they will discard half of the detection results.

\item
Alice and Bob implement error correction and privacy amplification to extract the final secure key. Later, we will show how to realize this step, which is normally the main focus of a security proof.
\end{enumerate}
Eve may tamper the quantum channel and change/measure the states sent by Alice. The last two steps together is called post-processing. It normally requires an authenticated classical channel. That is, Eve can obtain all information about the classical communication during the post-processing but not modify them.

Proving the security of QKD is a difficult problem in theory. Fortunately, this problem was solved in the last decade, see for example, \cite{Mayers_01, LoChauQKD_99, ShorPreskill_00,Koashi_Uncer_06}. Many security proofs are based on the assumption of idealized QKD system components, such as a perfect single photon source and well-characterized detectors. In practice, inevitable device imperfections may compromise security unless these imperfections are well investigated. Meanwhile, the security of QKD with realistic devices has been studied, see \cite{MayersYao_98,IndividualAttack_00,BLMS_00,FGSZ_01,ILM_07,KoashiPreskill_03,GLLP_04} for examples. For more information about security proofs of QKD, one can refer to Chapter \ref{Chpt:Security}. For a review of quantum cryptography, one may refer to \cite{GRTZ_02}.

Experimental QKD has been successfully demonstrated over 100 km of transmission distance through both commercial telecom fibers and free space \cite{BBBSS_92,Townsend_98,RGGGZ_98,BGKHJTLS_99,GYS_04,Zeilinger_Decoy_07}. Commercial QKD systems are already on the market\footnote{Note that there are three companies, id Quantique, MagiQ and Smartquantum, that have commercial QKD products. However, the security has not been fully addressed yet.}. The main problem in the field is the security and performance of a realistic QKD system.

\subsection{Cryptanalysis and Quantum Cryptanalysis}
Cryptanalysis is the study of methods for obtaining the meaning of encrypted information, without access to the secret information which is normally required to do so. Typically, this involves finding the secret key. In non-technical language, this is the practice of code-breaking or cracking the code, although these phrases also have a specialized technical meaning\footnote{Definition from wikipedia.org.}.

In the quantum analogue, we need to consider loopholes that exist in QKD systems and various attack strategies. The study of attacks has a two-fold meaning. First, it investigates the security in a practical sense. Secondly, it is fundamentally interesting in quantum mechanics. For example, a general physical problem in a practical QKD system with two detectors is the detection efficiency loophole \cite{MSS_EffHole_83,FMS_EffHole_90}. This loophole underlies not only applied technology, such as QKD, but also fundamental physics, such as Bell's inequality testing. Moreover, in practice, it is difficult to build two detectors that have exactly the same characteristics. Our work of time-shift attack (see Section \ref{Sc:Timeshift}) is an illustration of how one can proceed to handle this general problem in the security of QKD.

\section{Preliminary}
In this section, we will provide a general picture of QKD and some terminologies used in the thesis.

\subsection{A QKD scenario} \label{Sub:QKDscenario}
Let us introduce a few generic figures in QKD that we have already used in Section \ref{Sub:Back:QCry}. Alice, the sender, is the one who starts a key transmission. Bob, the receiver, is the one who receives the quantum states and extracts the key sent by Alice. This is just a convention used in the field, but not a strict definition. In some protocols, such as an entanglement based QKD that will be discussed in Chapter \ref{Chpt:Ent}, the roles of Alice and Bob are interchangeable.

The third important character is the eavesdropper, Eve, who play a dark side here. Eve is trying to intrude into the QKD and gain information about the key established between Alice and Bob. One conservative assumption in the QKD is that Eve has full control of both the quantum and classical channels, knows the characteristics of the QKD components very well\footnote{Eve might be the producer of QKD systems.} and has a great computational power. For example, Eve may own a quantum computer. Eve's attack is only limited by quantum mechanics and other physics laws.

Unconditional security is the Holy Grail of QKD, which means the security is proven without any restrictions of Eve's computational ability. As mentioned above, in an unconditional security proof, normally, Eve is assumed to own a powerful quantum computer and have full control of the channels. On the other hand, in most of widely used conventional classical cryptography protocols, security is proven by assuming that Eve has a finite computational power. See for example, RSA \cite{RSA_78}. Thus, with the development of technology and algorithm, the assumption that is made today about computational power does not guarantee security for tomorrow. For instance, Eve may store the encrypted message and decrypt it in the future with better computational power or algorithm. From this point of view, unconditional security is appealed to many real life applications.

\subsection{QKD performance} \label{Sub:QKDperformance}
To compare different QKD protocols or setups, one needs to characterize the performance of QKD. There are two important aspects of QKD performance: key rate and maximal secure distance.

We assume that Alice encodes the quantum information into faint laser pulses. If not (e.g., Alice uses a photon source pumped by a continuous wave laser), then Alice and Bob can manually partition the time domain into pulses. The \emph{key rate} is defined to be the average number of final secure key bits from one pulse. By multiplying the pulse repetition rate (frequency), the key rate gives the speed of key generation.

Due to the loss and noise, all practical QKD systems have a limit of secure distance. That is, beyond a certain distance, a QKD setup with a certain post-processing procedure cannot achieve a positive secure key. The \emph{maximal secure distance} is defined for a certain QKD setup and the post-processing scheme as the maximal QKD transmission distance that can yield a positive key rate.

We emphasize that the mentioned key rate and maximal secure distance here is always based on a guaranteed (proven) security. In many cases, we regard this is the lower bound in the sense that this performance as the least that one can achieve. Considering a performance upper bound\footnote{Beyond a upper bound, one surely cannot obtain a secure key.} of QKD setups and protocols is also an interesting topic. For example, one can refer to Refs.~\cite{FGGNP_97, CLL_Precondition_04}.

For a real life application, certain performance is required. For instance, the state of the art digital speech coding \cite{Wireless_Rappaport_02} typically needs a bit rate around 4-10 kbits/sec. A typical city wide area network must cover an area with a radius of 5-25 km. Later, in the conclusion of Chapter \ref{Chpt:Practical}, we will see that the QKD performance with current technology can achieve these requirements.

\section{Motivation} \label{Sc:ThesisMotivation}
The main objective of this thesis is to bring QKD to real-life applications. To do that, we investigate the security issues of practical QKD systems and propose new techniques to improve QKD performance.

\subsection{QKD security}
As discussed in Section \ref{Sub:Back:QCry}, we need to take into account device imperfections to achieve QKD security.
For example, an imperfect single photon source may open up loopholes for sophisticated attacks, such as photon number splitting attacks \cite{HIGM_95,BLMS_00,LutkenhausJahma_02}.

On the detection side, Eve may launch attacks on the imperfections of detections. For instance, Eve may take advantage of the timing information of signal pulses. We will present a feasible attack with current technology, a time-shift attack, in Section \ref{Sc:Timeshift}.

Thus, in order to guarantee the security of a practical system, QKD components are closely investigated and a realistic model is established. Then, we link our model to the existing security proofs. From there, we can learn about the assumptions that are made to prove security and the requirements for QKD experiments.

\subsection{A gap between theory and experiment}
As mentioned in Section \ref{Sub:QKDperformance}, in real-life applications, high QKD performance is required. Naturally, there are two important aspects of QKD performance: key generation speed (in bits/second) and transmission distance. Correspondingly, we will consider the two criteria, key rate\footnote{Note that developing a QKD system with a high repetition rate is an interesting topic in the field, for example, see Ref.~\cite{TDLFY_10GHz_06}. In this thesis, we will always focus on the key rate unless otherwise stated.} and maximal secure distance, as discussed in Section \ref{Sub:QKDperformance}.

On the theory side, much effort has been spent on the security proof of QKD with imperfect devices \cite{MayersYao_98,IndividualAttack_00,ILM_07,KoashiPreskill_03,GLLP_04}. By directly applying  these security analyses, the QKD performance is very limited. One can refer to the simulation part in Chapter \ref{Chpt:Decoy}.

On the other hand, the transmission distance of QKD experiment has been extended from a few meters in the first QKD experiment to currently more than 150 km. If we apply a standard security analysis, for instance, GLLP, the existing experiment setups can only tolerate a very limited transmission distance (as the simulation results show in Section \ref{Sub:Decoy:Simu}). The key issue here is the security of the experiment. Thus, there is a big gap between the theory and practice of QKD.


This thesis aims to bridge this gap between theory and practice by guaranteeing the security and improving the performance of practical QKD.

Note that in some cases, security is sacrificed to achieve a better QKD performance. In this thesis, we always guarantee the security first and then enhance the performance.

\section{Highlight and Outline}
During my Ph.D. program, I have completed the following projects by collaborating with my colleagues.
\begin{itemize}
\item
In Chapter \ref{Chpt:Security}, there will be reviews of various QKD security proofs and comparison of two standard security proofs of QKD with realistic devices. This work is published in Ref.~\cite{Low_06}.

\item
In Chapter \ref{Chpt:SetupModel}, there will be a discussion on a widely used experiment setup and its model. This work is published in Ref.~\cite{Practical_05}. Here I acknowledge that I benefited very much from discussions about experiment setups with Bing Qi.

\item
In Chapter \ref{Chpt:Decoy}, the decoy state idea and its security proof will be discussed. This work is published in Ref.~\cite{Decoy_05}. In this work, I applied GLLP security analysis to a decoy state QKD and simulated a QKD experiment \cite{GYS_04} to show the improvement given by using decoy states.

\item
In Chapter \ref{Chpt:Practical}, practical decoy state protocols will be discussed. This work is published in Ref.~\cite{Practical_05}. In this work, I applied the idea of the Vaccum+Weak decoy state protocol, which was first proposed by Lo \cite{LoDecoy_03} and considered statistical fluctuations. Furthermore, I designed the experimental parameters and  analyzed data in the decoy state QKD experiment demonstration \cite{ZQMKQ_06,ZQMKQ60km_06}. Hence, it can be concluded that the decoy state idea is highly practical in real life applications.

\item
In Chapter \ref{Chpt:TwoWay}, two post-processing schemes are studied based on two-way classical communication for the decoy state method. This work is published in Ref.~\cite{TwoWay_06}. In this work, I applied the Gottesman-Lo's 2-LOCC\footnote{See Appendix \ref{App:Abb} for the definition of LOCC.} entanglement distillation protocol (EDP) and recurrence scheme to a decoy state QKD and simulated a QKD experiment to show the improvement by using two-way classical communication in a decoy state QKD.

\item
In Chapter \ref{Chpt:Trig}, various decoy state protocols are investigated for triggering parametric down-conversion sources. This work is presented in Ref.~\cite{TriggeringPDC_08}. In this work, I modeled the QKD setup with a triggered PDC source following L\"utkenhaus' work \cite{IndividualAttack_00} and compared various decoy state proposals of triggering PDC QKD.

\item
In Chapter \ref{Chpt:Ent}, QKD with an entangled photon source will be discussed. This work is published in Ref.~\cite{EntanglementPDC_07}. In this work, I built an entangled PDC source model, applied Koashi-Preskill's security analysis and simulated a PDC experiment to show the performance of the entanglement-based QKD in comparison with a triggered single photon source and coherent state QKD.

\item
In Chapter \ref{Chpt:Attack}, quantum attacks and security against these such attacks will be investigated. These works are published in Refs.~\cite{QFLM_TimeShift_07} and \cite{Mismatch_security_08}. Aside from the decoy state method, we also studied other methods for improving the QKD performance, such as the dual detector scheme \cite{QZMLQ_dual_07,QZMLQ_eff_07}. I am not the main contributor of these works. I joined in discussions and helped work out the details.

\item
In Chapter \ref{Chpt:Conclusion}, a summary of my Ph.D.~study is presented and some interesting topics for future research are stated.

\item
In Appendix \ref{AppCh:Abbmath}, the common abbreviations used in the thesis is listed and some detailed mathematical derivations are shown.

\item
In Appendix \ref{Ap:Optimalmu}, the optimization of the source intensity $\mu$ is discussed.
\end{itemize}

\section{Future outlook}
An interesting topic is the natural extension of the current work: further enhancement of the performance of practical QKD systems. Continuous variable QKD is proposed to achieve a higher key rate in short and medium transmission distance. An open question is the security of continuous variable QKD. This is an appealing topic in the field. Modeling and simulations for continuous variable QKD are also interesting.

Another crucial point is that in real life, one needs to consider some extra disturbances (e.g., quantum signals may share the channel with regular classical signals). The final goal is to achieve a customer friendly QKD system that can be easily integrated with the Internet, for instance.

Statistical fluctuations need to be considered in QKD with a finite key length. There is some work on this topic recently, e.g., \cite{Renner_Thesis_05}. An interesting topic is applying Koashi's complementary idea \cite{Koashi_Compl_07} to a finite key QKD and compare it with prior results.

An interesting topic outside quantum cryptography is whether the techniques developed in QKD can be useful in quantum computation. For example, do such models and post-processing schemes also help quantum computation by linear optics realizations?


Finally, quantum information processing is related to the foundation of quantum mechanics. As we know, quantum information (e.g., von Neumann entropy) can help us in understanding quantum entanglement. What about other principles in quantum mechanics?

\chapter{Security analysis} \label{Chpt:Security}
In this chapter, we will review various security proofs. We start with the objective of security proofs and the underlying assumptions in current security proofs. We compare two standard security proofs of the QKD with realistic devices. This work is published in Ref.~\cite{Low_06}.

\section{What are security proofs?} \label{Sc:UncondSecu}
To serve as a secure key in cryptographic uses, there are two criteria:
\begin{enumerate}[(a)]
\item
Alice and Bob share the same key; that is, an \emph{identical key}.

\item
Eve has no information about the key; that is, a \emph{secure key}.
\end{enumerate}
With regards to a careful analysis and the formulation of security, see \cite{Renner_Thesis_05}. For necessary and sufficient conditions for security, see \cite{HHHO_Bound_05}.

The first criterion can be satisfied by performing a classical error correction, for example, by using the Cascade code \cite{BrassardSalvail_93}. After that, Alice and Bob will share an identical key. Next, Alice and Bob will perform privacy amplification, for instance, by random hashing, to eliminate Eve's information about the key.

The goal of current security analyses is to show how much privacy amplification needs to be performed after a certain error correction procedure.

The main task for a security analysis is to figure what the length of the final secure key is and perform hashing to obtain the final key.


\section{Squash model} \label{Sc:Squash}
In this section, we will formalize the widely used squash model in security proofs. Note that the squash model is used in the security proof proposed by Gottesman, Lo, L\"utkenhaus, and Preskill (GLLP) \cite{GLLP_04},  see also \cite{Koashi_NewModel_06,TT_Thres_08,BML_Squash_08}.

\subsection{A calibration problem}
In all the existing QKD security proofs, certain characteristics of sources and detectors are assumed to be known or measurable. However, in reality, such a calibration procedure is a very difficult task. For example, on Alice's side, a good single photon source is not available with current technology although much effort has been made in this field \cite{KBKY_99,LounisMoerner_00,KBBBCDK_03,JPK_04,EAMFZL_05,XAXTJR_06}. On Bob's side, most of security proofs rely on the assumption that Bob measures two conjugate bases (for instance, $X$ and $Z$) of \emph{a qubit}. In real QKD experiments, threshold detectors\footnote{A threshold detector can only tell whether the input signal is vacuum or non-vacuum. For a strict mathematical definition, one can refer to Section \ref{Sub:ThreshDet}.} are widely used. In summary, devices calibration form a gap between the theory and practice of QKD.

In the experiment, to test (calibrate) a source, we need a good (well-characterized) detection system. On the other hand, to characterize a detector, we need a well-tested source. In QKD, we may even want to test these devices in real-time, which makes the task even more difficult.

In most QKD proposals\footnote{One exception approach is the so-called device-independent QKD protocol \cite{AGM_Bell_06} based on Bell's inequality \cite{Bell_Ineq_64}. However, no strict security analysis has been yet provided for this type of QKD protocols. For recent developments of realistic threshold detector models, one can refer to Ref.~\cite{Koashi_NewModel_06}.}, one needs to make sure that Bob's (and sometimes also Alice's) measurement is performed in a two-dimensional Hilbert space. This assumption is another way to state the squash model. We can see that this squash model assumption is \emph{not} easy to avoid. Note that even throwing away the squash model, one needs to have certain assumptions about the side information. Later in Chapter \ref{Chpt:Attack}, we will see that some side information (e.g., timing) may cause fatal security issues in QKD.

\subsection{Squash model}
\begin{figure}[hbt]
\centering \resizebox{12cm}{!}{\includegraphics{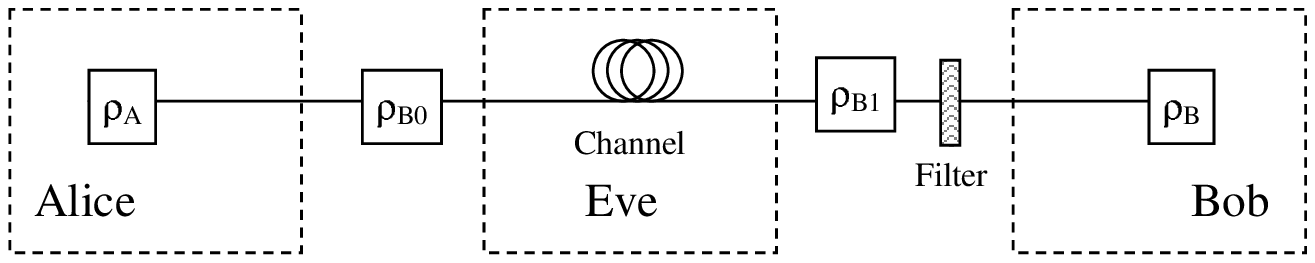}}
\caption{A schematic diagram for the squash model. The filter is the key component of the squash model.} \label{Fig:Squash}
\end{figure}

In theory, the squash model is proposed to avoid the aforementioned calibration problem. As shown in Figure \ref{Fig:Squash}, the scenario that we are talking about here is as follows: Alice prepares her own system $\rho^0_{AB}$. In a prepare-and-measure scheme (e.g., BB84), $\rho_A=Tr_B(\rho^0_{AB})$ determines the basis and key bit value that she will pick up. She then sends the system $\rho_{B0}=Tr_A(\rho^0_{AB})$ to Bob, which is intercepted by Eve. Eve performs some operations and/or measurements on the system and resends a system $\rho_{B1}$ to Bob. After passing through a filter, the state received by Bob is $\rho_B$. That is, Eve prepares a system $\rho_B$ for Bob, generally depending on the system sent by Alice. Finally, Alice and Bob will extract a key from measurements on $\rho_A$ and $\rho_B$. Alice and Bob's detection system follows the squash model.

\textbf{Squash model:} The detection system first performs a filter, projecting the incoming state $\rho$ (with an arbitrary dimension of Hilbert space) into a two-dimensional Hilbert space state $\rho_2$ or output a ``failure" signal. If the projection succeeds, a projection measurement will be performed in a basis\footnote{This basis can be randomly chosen from a conjugated bases set.} in a two-dimensional Hilbert space.

The schematic diagram of the squash model is shown in Figure \ref{Fig:Squash}. As we can see that in the squash model, Bob always receives a qubit or vacuum. In other words, in the squash model, Eve always sends a qubit or vacuum to Bob.

\subsection{Remarks}
\begin{enumerate}
\item
The squash model is reasonable (but not necessarily correct) for threshold detector cases. After treating the double click as a random click event, a threshold detector's response can always be described by a qubit or vacuum measurement outcome.

\item
Even with only one photon, the squash model is still required in the existing security proofs. This is because there are lots of degrees of freedom of a photon, for instance, timing, polarization, phase \cite{LoPreskill_05} and space \cite{QFZMTCL_attack_07}. Thus, by using a perfect photon number resolving detector, one cannot avoid the squash model.

\item
The filter acts as a key component of the squash model. One can model the channel losses and detector efficiency into the failure probability of the filter.

\item
In the squash model, when double clicks\footnote{This is when more than one detector have detection events for one key bit transmission. In general, a double click probability is very small in comparison to dark count probability and detector efficiency.} happen, we assume that Alice and Bob will assign a random bit when they get a double click, due to the strong pulse attack \cite{Lutkenhaus_99DoubleClick}.

\item
In a rigorous security analysis, one needs to experimentally verify whether the squash model gives a good description of a certain detection system. Take a widely used threshold detector for example. One needs to open the detector, examine the components carefully, then write down the quantum operations and compare the operations described by the squash model. Again, we want to emphasize that testing the model is a highly non-trivial task in the experiment.

\item
Another way to avoid the device calibration problem is to propose so called device independent QKD protocols, see for example, Ref.~\cite{AGM_Bell_06}. Up until now, a strict security proof of these device independent QKD protocols is still missing. This is an interesting prospective topic. Recently, security proofs of QKD with a more realistic model, threshold detector model, are presented \cite{Koashi_NewModel_06,TT_Thres_08,BML_Squash_08}. An interesting theoretical question is whether the threshold detector model is equivalent to the squash model.
\end{enumerate}

\section{Entanglement-based QKD} \label{LoChauProof}
In this section, we will review the idea of the Lo-Chau type security proof \cite{LoChauQKD_99} of QKD based on entanglement distillation protocols (EDP) \cite{BDSW_96}.

In the following discussion, we will use $X$ and $Z$ to represent two conjugate bases, which are the Pauli operators:
\begin{equation} \label{EDP:Paulimatrix}
\begin{aligned}
X=\begin{pmatrix}
    0 & 1\\
    1 & 0\\
    \end{pmatrix},~~~~~
Z=\begin{pmatrix}
    1 & 0\\
    0 & -1\\
    \end{pmatrix}
\end{aligned}
\end{equation}
to represent two conjugate bases. The QKD scenario in Lo-Chau's security proof can be described as follows:
\begin{enumerate}
\item
Alice prepares $n$ EPR pairs in one of the four Bell states,
\begin{equation} \label{EDP:BellStates}
\begin{aligned}
|\psi_{00}\rangle&=\frac{1}{\sqrt2}(|00\rangle+|11\rangle)\\
|\psi_{10}\rangle&=\frac{1}{\sqrt2}(|01\rangle+|10\rangle)\\
|\psi_{11}\rangle&=\frac{1}{\sqrt2}(|01\rangle-|10\rangle)\\
|\psi_{01}\rangle&=\frac{1}{\sqrt2}(|00\rangle-|11\rangle),\\
\end{aligned}
\end{equation}
for instance, in $|\psi_{00}\rangle^{\otimes n}$.

\item
Alice sends half of each EPR pair to Bob and keeps the other half in her quantum memory.

\item
After he receives the half EPR pairs, Bob stores all the qubits into his quantum memory.

\item
Alice and Bob perform an EPD protocol \cite{BDSW_96} to distill $m$ ($m\le n$) into nearly perfect EPR pairs.

\item
Alice and Bob measure the EPR pairs in the $Z$ basis to obtain a shared secret key.

\end{enumerate}
The key point of Lo-Chau's security proof is that if in Step 4, Alice and Bob share nearly perfect EPR pairs, the final key is secure. With a quantum computer, the amount of EPR pairs that Alice and Bob can distill is given by:
\begin{equation}\label{EDP:CSSrate}
m=n-r_{err},
\end{equation}
where $r_{err}$ is the amount of information (in bits) cost in the quantum error correction process. Here, $r_{err}$ can be regarded as the number of encrypted bits communicated between Alice and Bob in the post-processing\footnote{In this case, we assume that Alice and Bob encrypt the communication for the error correction.}.

\section{Single-photon-based QKD} \label{Sc:SingleEDP}
In this section, we will review Shor-Preskill's security proof \cite{ShorPreskill_00}. In  Lo-Chau's security, the main drawback is that quantum computers (or at least quantum memories) are required, which are not available with current technology. Based on Lo-Chau's security proof, Shor and Preskill proposed a special EDP scheme, which can be reduced to a prepare-and-measure scheme.



%

The EDP protocol proposed by Shor and Preskill is based on the Calderbank-Shor-Steane (CSS) code \cite{CCS1_96,CCS2_96}.
The basic idea of Shor-Preskill's security proof is to replace Step 4 of Lo-Chau's security proof (see Section \ref{LoChauProof}) by the following procedures:
\begin{enumerate} [(4.a)]
\item
Alice and Bob pick up $k$ testing EDP pairs randomly and both measure in $Z$ basis to estimate bit error rate, $\delta_b$. We call the procedure that corrects this type of error, \emph{bit error correction}.

\item
They pick up another $k$ testing EDP pair randomly and both measuring in $X$ basis to estimate the phase error rate, $\delta_p$. Correspondingly, we call the procedure that corrects this type of error, \emph{phase error correction}.

\item
They abort the protocol if the error rates are too high. Otherwise, they apply a quantum CSS code to correct the bit and phase errors separately. It is here that an important property of the quantum CSS codes is applied: they can decouple the phase correction from the bit correction \cite{ShorPreskill_00}.

\item
They can distill $m$ ($m\le n$) nearly perfect EPR pairs by the quantum error correction procedure.
\end{enumerate}

The key argument in Shor-Preskill's security proof is that since the final $Z$ measurement (see Step 5 in Section \ref{LoChauProof}) commutes with Steps 1-4, Alice and Bob can move this $Z$ measurement ahead of Step 1. Note that this is the reason why CSS codes are applied to decouple bit and phase error corrections\footnote{Note that the CSS code is a linear quantum error correction code. It uses two classical error correction codes (e.g., $C_1$ and $C_2^\perp$ with $C_2\subset C_1$) to protect bit and phase errors separately. For a detailed discussion of the reason why the CSS code can decouple bit and phase error corrections for QKD, one can refer to Ref.~\cite{ShorPreskill_00}.}. After this move, the bit error error correction becomes a regular classical error correction and the phase error correction becomes a privacy amplification. Now the modified procedure will be exactly the same as the BB84 protocol.
\begin{enumerate}
\item
Alice prepares $n$ qubits, each in one of the four eigenstates of $X$ and $Z$. Here, the reason for preparing $X$ eigenstate is to make a symmetry between the bit and phase error rates.

\item
Alice sends the states to Bob.

\item
After he receives the states, Bob measures the states in $X$ or $Z$ bases randomly.

\item
Alice and Bob perform a post-processing scheme to distill $m$ ($m\le n$) into bits of secure key.

\begin{enumerate} [(4.a)]
\item
Alice and Bob pick up $k$ measurement results to estimate the bit error rate, $\delta_b$.

\item
Due to the symmetry of BB84, they can estimate the phase error rate\footnote{Note that $\delta_p=\delta_b$ is true for the case of infinite long key BB84. Later in Section \ref{Sub:Ent:StaFlu}, we will see that this may not be true for a finite key length with statistical fluctuations. Note also that for other protocols, such as the SARG04 protocol \cite{SARG_04}, it is no longer true that $\delta_p=\delta_b$ \cite{TamakiLo_06, FTL_06}.} by $\delta_p=\delta_b$.

\item
If the error rates are too high, they abort the protocol. Otherwise, they apply a classical error correction code to correct all the bit errors.

\item
They apply a privacy amplification (for instance, random hashing) according to the phase error rate, $\delta_p$.
\end{enumerate}


\end{enumerate}

After the error correction and privacy amplification, the key rate is given by \cite{ShorPreskill_00}:
\begin{equation}\label{EDP:CSSrate}
R=qQ_{\mu}\left[1-H_2(\delta_b)-H_2(\delta_p)\right],
\end{equation}
where $q$ is the basis reconciliation factor (1/2 for the BB84 protocol due to the
fact that half of the time, Alice and Bob disagree with the bases, and if one uses
the efficient BB84 protocol \cite{EffBB84_05}, $q\approx1$), $Q_\mu$ is the filter success probability in the squash model\footnote{Basically, $Q_\mu$ is the probability for Bob to obtain a detection (not a vacuum) in a pulse of key transmission. Later, in Section \ref{Sc:Model}, one can see why we use the notation $Q_\mu$ here.} and $H_2(x)$ is the binary entropy function,
\begin{equation} \label{EDP:H2}
\begin{aligned}
H_2(x)=-x\log_2(x)-(1-x)\log_2(1-x).
\end{aligned}
\end{equation}

In summary, there are two main parts of the post-processing, error correction (for bit error correction) and privacy amplification (for phase error correction). These two steps can be understood as follows. First, Alice and Bob apply an error correction, after which they share the same key strings, but Eve may still keep some information about the key. Alice and Bob then perform a privacy amplification to expunge Eve's information from the key.


\section{GLLP security analysis} \label{Sc:GLLP}
In this section, we will review the Gottesman-Lo-L\"utkenhaus-Preskill (GLLP) security analysis idea \cite{GLLP_04}. It gives a security proof of BB84 QKD when realistic devices (such as imperfect single photon sources) are used.

\subsection{Tagged and untagged qubits}
In the original proposal of the BB84 protocol (as well as in Shor-Preskill's security proof), a perfect single photon source is required. Unfortunately, single photon sources are still not available with current technology. For the development of a single photon source, one can refer to Refs.~\cite{KBKY_99,LounisMoerner_00,KBBBCDK_03,JPK_04,EAMFZL_05,XAXTJR_06}. Thus, intuitively, we can think there are two components in an imperfect single photon source, one is good for BB84 and the other is bad. Separating these two components is the main idea of GLLP.


There are two kind of qubits discussed in GLLP, tagged qubits and untagged qubits. Tagged qubits are those that have their basis information revealed to Eve, i.e. tagged qubits are not secure for QKD. On the other hand, untagged qubits are secure for QKD. Note that the idea of the tagged state was (perhaps implicitly) introduced by L\"{u}tkenhaus \cite{IndividualAttack_00}.

The untagged qubits basically come from the idea of a basis-independent source \cite{KoashiPreskill_03}. A basis-independent source means that, to Eve, the quantum states transmitted through the channel are independent of the bases that Alice and Bob are choosing. Whereas the tagged qubits come from basis-dependent sources, whose basis information may be revealed to Eve.

Let us show a concrete example about tagged and untagged qubits. In BB84, qubits coming from single-photon states are untagged, while those from multi-photon states are tagged. This is because Eve, for instance, can perform photon-number splitting attacks \cite{HIGM_95,BLMS_00,LutkenhausJahma_02} to the multi-photon states. This may not true for other protocols. For example, in SARG04 \cite{SARG_04,TamakiLo_06}, two-photon states can be used to extract secure keys.


\subsection{Post-processing} \label{Sub:GLLPpost}
The GLLP post-processing is performed as follows.
%
%
First, Alice and Bob apply error correction to all qubits, sacrificing a fraction $H_2(E_\mu)$ of the raw key, which is represented in the first term of Eq.~\eqref{Post:KeyRate} below. Secondly, in principle, Alice and Bob can distinguish the tagged and untagged qubits (for instance, by measuring the photon numbers on Alice's side), so they can apply the privacy amplification on the tagged state and untagged state separately. One can imagine executing privacy amplification on two different strings, the qubits $s_{tagged}$ and $s_{untagged}$ arising from the tagged qubits and the untagged qubits respectively. Since the privacy amplification is linear (for instance, by linear hashing), the key obtained is the bitwise $XOR$
$$
s_{untagged}\oplus s_{tagged}
$$
of keys that could be obtained from the tagged and untagged qubits separately. If $s_{untagged}$ is private and random, then it does not matter if Eve knows anything about $s_{tagged}$, the sum will be still private and random. Thus, one only needs to apply privacy amplification to the untagged bits.

We define the key generation rate as the ratio of the final key length to the total number of pulses sent by Alice. Applying the GLLP idea to our model, $Q_1$ is the amount of untagged qubits. Thus, the key generation rate is given by \cite{Decoy_05}:
\begin{equation} \label{Post:KeyRate}
R \geq q \{-f(E_{\mu})Q_{\mu}H_2(E_{\mu})+Q_1[1-H_2(e_1)]\},
\end{equation}
where $q$ is the basis reconciliation factor as discussed in Eq.~\eqref{EDP:CSSrate}, $Q_{\mu}$ and $E_{\mu}$ are the overall gain (or filter success probability) and QBER, $Q_1$ and $e_1$ are the gain and error rate of untagged qubits, and $f(x)$ is the error correction inefficiency (see, for example, \cite{BrassardSalvail_93}) as a function of the error rate, normally $f(x)\ge1$ with the Shannon limit $f(x)=1$. For detailed definitions of $Q_{\mu}$, $E_{\mu}$, $Q_1$ and $e_1$, one can refer to Section \ref{Sc:Model}.

Note that one can add $Q_0$ into Eq.~\eqref{Post:KeyRate} by considering other security analysis \cite{Vacuum_05}, see also \cite{Koashi_NewModel_06}.

\subsection{An extension} \label{Sub:GLLPEx}
The original GLLP idea only considers two types of qubits: tagged and untagged. For BB84, it sets a phase error rate, $\delta_p=1/2$ for tagged qubits and $\delta_p=\delta_b$ for the untagged qubits. The idea of applying separate privacy amplification (GLLP) can be naturally extended to the case of more than two classes of qubits \cite{TwoWay_06}, i.e. several kinds of qubits with tag $g$, which generalizes the concept of tagged and untagged qubits. The procedure of data post-processing is similar, an overall error correction step followed by privacy amplification to each case. Therefore, the key generation rate is given by:
\begin{equation} \label{Post:KeyRateEx}
R \geq q \{-f(E_{\mu})Q_{\mu}H_2(E_{\mu})+\sum_g Q_g[1-H_2(e_g)]\}
\end{equation}
where $Q_g$ is the gain of the qubits with tag $g$ and $e_g$ is the corresponding phase error rate. Here, we want to emphasize that $e_g$ is not equal to the bit error rate of the qubits with tag $g$ in general, unless the qubits come from a basis-independent source.

This extension is useful for some post-processing schemes, e.g., SARG04 \cite{SARG_04} and 2-LOCC post-processing schemes \cite{TwoWay_06} (see Chapter \ref{Chpt:TwoWay}).

The above discussion is a review of various security analysis. Next, we will compare two standard security analysis schemes.

\section{GLLP vs.~L\"utkenhaus' security analysis}
In this section, we will compare two data post-processing schemes, L\"utkenhaus \cite{IndividualAttack_00} versus GLLP \cite{GLLP_04}. Here, we use L\"utkenhaus' security analysis, to refer to his work, see Ref.~\cite{IndividualAttack_00}\footnote{We acknowledge that L\"utkenhaus has worked on many security analysis schemes, including ILM \cite{ILM_07} and GLLP \cite{GLLP_04}.}. Note that L\"utkenhaus' security analysis proves the security against individual attacks, while GLLP offers unconditional security. This work is published in Ref.~\cite{Low_06}.

We can rewrite the formula of the key generation rate by L\"{u}tkenhaus' security analysis scheme \cite{IndividualAttack_00}
\begin{equation}\label{TwoSch:Lutkenhaus}
\begin{aligned}
R\geq q\{-Q_\mu H_2(E_\mu)+Q_1[1-\log_2(1+4e_1-4e_1^2)]\},
\end{aligned}
\end{equation}
where the privacy amplification term $\log_2(1+4e_1-4e_1^2)$ comes from collision probability.

Now, we can compare Eqs.~\eqref{Post:KeyRate} and \eqref{TwoSch:Lutkenhaus}. In both key rate formulae, the first term in the bracket is for error correction and the second term is for privacy amplification. The privacy amplification is only performed on the single photon part.  In this manner, L\"utkenhaus \cite{IndividualAttack_00} has already applied the idea of separate privacy amplification.

\begin{figure}[hbt]
\centering \resizebox{12cm}{!}{\includegraphics{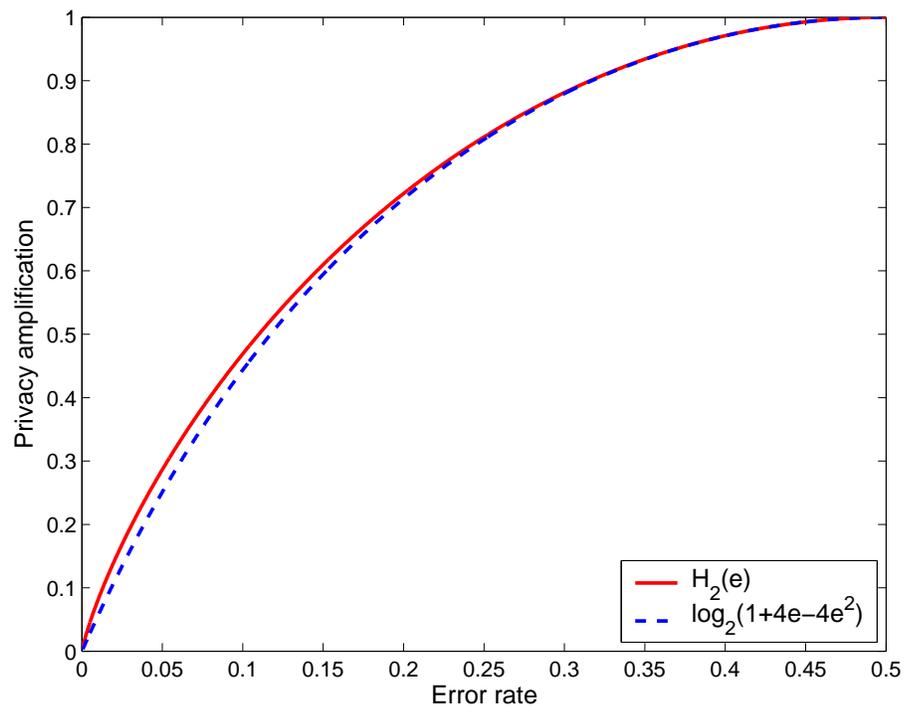}}
\caption{Plot of the privacy amplification parts of GLLP and L\"utkenhaus. The maximal deviation of the two curves is 15.36\% when the error rate is 3.85\%.} \label{TwoSch:fig:ComPri}
\end{figure}

We can see that the only difference between the L\"utkenhaus and GLLP results appears in the privacy amplification part. We compare $H_2(e)$ with $\log_2(1+4e_1-4e_1^2)$ in
Figure~\ref{TwoSch:fig:ComPri}. We can see that the difference of the two functions is quite small. For this reason, in fact, L\"utkenhaus and GLLP give very similar results in the simulations of real experiments \cite{Low_06}.

Based on this observation, we find that there is little to gain by restricting the security analysis to individual attacks, given that the two schemes; L\"{u}tkenhaus vs.~GLLP, provide very close performances. In other words, our view is that one is better off considering unconditional security, rather than restricting to individual attacks.

%
%
%
%
%
%

\chapter{Setup and Model} \label{Chpt:SetupModel}
In this chapter, we will discuss a widely used QKD setup and model. For now, we will focus on the case where a weak coherent state source is used as an imperfect single photon source by Alice. Nevertheless, many concepts from this generic model is useful for other QKD setups. For example, in Chapter \ref{Chpt:Trig}, we will modify this model to fit the case of the QKD with triggered single photon sources.

This work is published in Ref.~\cite{Practical_05}. I acknowledge that I benefited very much from discussions about experiment setups with Bing Qi.

\section{QKD setup} \label{Sc:Setup}
\begin{figure}[hbt]
\centering \resizebox{12cm}{!}{\includegraphics{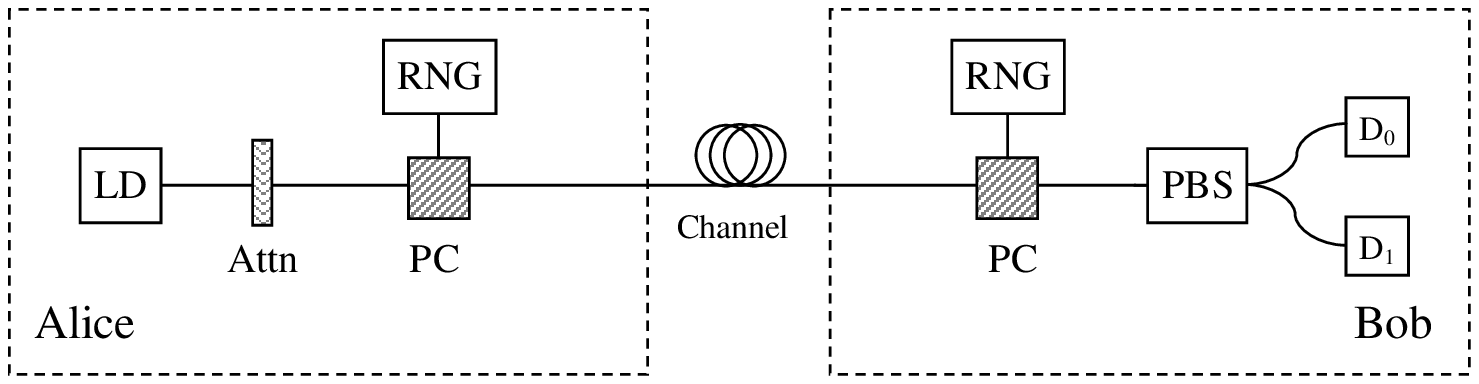}} \caption{A schematic diagram for the coherent state QKD implementation. LD: laser diode; Attn: optical attenuator; RNG: random number generator; PC: polarization controller; PBS: polarization beam splitter; DB$_0$, DB$_1$: single photon detectors.} \label{Fig:Coherent}
\end{figure}

As we pointed out earlier, due to the lack of a perfect single photon source for BB84, a weak coherent state source is widely used. We call this setup a coherent state QKD implementation. Similarly, perfect single photon detectors are commonly replaced by threshold detectors. The setup is shown in Figure \ref{Fig:Coherent}.

As shown in Figure \ref{Fig:Coherent}, the coherent state QKD implementation works as follows.
\begin{enumerate}
\item
Alice uses a weak coherent state photon source. She attenuates the laser beam from a laser diode (LD) with an optical attenuator (Attn). She uses a random number generator (RNG) to generate random bits for her choice of basis and bit values. She encodes one of four polarizations (eigenstates of $X$ and $Z$ bases) by a polarization controller (PC).

\item
Bob receives the quantum states from the channel. He uses a PC as a polarization rotator for choosing his measurement basis, which is also controlled by a RNG. Then he uses a polarization beam splitter (PBS) followed by two single photon detectors (DA$_1$ and DA$_2$) to perform the measurement.
\end{enumerate}


\section{QKD model} \label{Sc:Model}
There are three main parts for a QKD system: source, channel and detection. In this section, we present a widely used QKD system model that follows Ref.~\cite{IndividualAttack_00}. See also Ref.~\cite{Practical_05}. In the model, we assume that Alice sends out quantum signals in pulses. In the case where Alice uses a continuous source, we assume that Alice and Bob manually fit detections into pulses. This model is originally designed for the coherent state QKD, but the channel and detection parts can also be used for other QKD implementations. For example, in Chapter \ref{Chpt:Ent}, we will modify the source part of this model to fit the case of QKD with entangled photon sources.

\subsection{Weak coherent state source}
Highly attenuated lasers are often used as an imperfect single photon source in QKD. This type of source can be well described by a weak coherent state, which is a superposition of number states (aka Fock states) \cite{ScullyZubairy_QOptics_02},
\begin{equation}\label{Model:CoherentState}
\begin{aligned}
\ket{\alpha}=e^{-{|\alpha|^2\over2}}\sum_{n=0}^{\infty}{\alpha^n\over\sqrt{n!}}|n\rangle
\end{aligned}
\end{equation}
Assuming that the phase of the laser is randomized for each pulse, the density matrix of the state emitted by Alice is given by:
\begin{equation}\label{Model:AliceState}
\begin{aligned}
\rho_A &= \frac{1}{2\pi} \int\limits_{0}^{2\pi} d \theta \ket{ |\alpha| e^{i\theta} } \bra{ |\alpha| e^{i\theta} }\\
&= \frac{1}{2\pi} \sum_{n=0}^{\infty}\sum_{m=0}^{\infty}\frac{|\alpha|^{n+m}}{\sqrt{n!m!}} e^{-|\alpha|^2}\ket{n} \bra{m} \int\limits_{0}^{2\pi} d \theta e^{i(n-m)\theta}  \\
&= \sum^{\infty}_{n=0}\frac{\mu^n}{n!}e^{-\mu} \ket{n}\bra{n} \\
\end{aligned}
\end{equation}
where $\theta$ is the phase of the coherent state and $\mu=|\alpha|^2$, defined to be the intensity of the photon source. The photon number follows a Poisson distribution:
\begin{equation}\label{Model:Poisson}
\begin{aligned}
P(n) =  \frac{\mu^{n}}{n!}e^{ - \mu}.
\end{aligned}
\end{equation}
From here, we can see that there are three types of photon states:
\begin{enumerate}
\item
vacuum state: $|0\rangle\langle 0|$

\item
single photon state: $|1\rangle\langle 1|$

\item
multi photon state: $|n\rangle\langle n|$ for $n\ge2$.
\end{enumerate}
Here, we assume the squash model \cite{GLLP_04} as discussed in Section \ref{Sc:Squash}. That is, Eve receives all the pulses sent by Alice. Eve performs some arbitrary operations and sends either a vacuum or a qubit to Bob. Consequently, we denote the qubits coming from these three states as {vacuum}, {single photon} and {multi photon qubits}.

A vacuum qubit is a qubit sent by Eve when Alice sends a vacuum state. In the case without Eve's presence, it is some random qubit stemmed from the dark counts of Bob's detector or other background contributions. Thus, it does not contribute positively to the final secure key. Due to photon-number splitting attacks \cite{HIGM_95,BLMS_00,LutkenhausJahma_02}, multi photon states are not secure for the BB84 protocol. Here is a key observation of this QKD model: \emph{the final secure key can only be extracted from single photon qubits}. Aside from BB84, this is true for most present QKD protocols, such as the B92 \cite{Bennett_92}, six-state \cite{sixstate_98} and  $N$-state \cite{Koashi_Nstate_05} scheme. One exception is the SARG04 protocol \cite{SARG_04}, in which two-photon states can also contribute to the secure key generation rate \cite{TamakiLo_06}.


\subsection{Channel and detection} \label{Sub:Model:ChDet}
We use a beam splitter followed by a perfect single photon detector to model the channel and detection. We define $\eta$ to be the transmittance of the beam splitter. The loss is composed by channel loss, internal loss in Bob's detection system and detector efficiency.  We assume that the channel loss is related to the transmission distance by a loss coefficient $\beta$ measured in dB/km. The transmittance $\eta$ is given by:
\begin{equation}\label{Model:Eta}
\eta=\eta_{B}10^{-\frac{\beta l}{10}}.
\end{equation}
where $\eta_{B}$ denotes the transmittance on Bob's side, including the internal transmission efficiency of optical components and detector efficiency. Here, we assume Bob uses threshold detectors. That is to say, we assume that Bob's detector can tell whether there is a click or not,
but not the actual photon number of the received signal.

In the simulation, we assume independence between the behaviors of the $i$ photons in $i$-photon states. Therefore, the transmittance of the \emph{$i$-photon} state $\eta_i$ with respect to a threshold detector is given by:
\begin{equation}\label{Model:etai}
\eta_i=1-(1-\eta)^i
\end{equation}
for $i=0,1,2,\cdots$.

\textbf{Yield:} Defines $Y_i$ as the yield of an $i$-photon state, i.e., the conditional probability of a detection event at Bob's side, given that Alice sends out an $i$-photon state. Note that $Y_0$ is the background rate which includes detector dark counts and other background contributions.

The yield of the $i$-photon states $Y_i$ mainly comes from two
parts, the background and the true signal. Assuming that the
background counts are independent of the signal photon detection,
then $Y_i$ is given by:
\begin{equation}\label{Model:Yi}
\begin{aligned}
Y_i &= Y_0 + \eta_i - Y_0\eta_i \\
    &\cong Y_0 + \eta_i.
\end{aligned}
\end{equation}
Here, we assume $Y_0$ (typically $10^{-5}$) and $\eta$ (typically
$10^{-3}$) are small.

The {\it gain} of $i$-photon states $Q_i$ is given by:
\begin{equation}\label{Model:Qi}
\begin{aligned}
Q_i &= Y_i\frac{\mu^i}{i!}e^{-\mu}.
\end{aligned}
\end{equation}
The gain $Q_i$ is the probability that Alice sends out an $i$-photon state and Bob obtains a detection.
Then the overall gain, the probability for Bob to obtain a detection event in one pulse, is the sum over all $Q_i$s:
\begin{equation}\label{Model:Gain}
\begin{aligned}
Q_{\mu} &= \sum_{i=0}^{\infty} Y_i\frac{\mu^i}{i!}e^{-\mu}. \\
\end{aligned}
\end{equation}
The overall gain $Q_\mu$ can also be understood as the filter success probability of the squash model that we discussed in Section \ref{Sc:Squash}.

\textbf{Quantum Bit Error Rate (QBER):} The error rate of $i$-photon
states $e_i$ is given by
\begin{equation}\label{Model:ei}
e_i = \frac{e_0 Y_0 + e_{d}\eta_i}{Y_i}
\end{equation}
where $e_{d}$ is the probability that a photon hits the erroneous detector. $e_{d}$ characterizes the alignment and stability of the optical system. Experimentally, even at distances as long as 120 km, $e_{d}$ is relatively independent of the distance \cite{GYS_04}. In the following, we assume that $e_{d}$ is independent of the transmission distance and the background clicks are random. Thus, the error rate of the background is $e_0=1/2$. Then the overall QBER is given by:
\begin{equation}\label{Model:QBER}
\begin{aligned}
E_{\mu} &= \frac{1}{Q_{\mu}} \sum_{i=0}^{\infty} e_iY_i\frac{\mu^i}{i!}e^{-\mu}. \\
\end{aligned}
\end{equation}

In the QKD scenario that we are considering, as discussed in Section \ref{Sub:QKDscenario}, Eve can change $Y_i$ and $e_i$ for her attacks. Without Eve, in a normal QKD, Eqs.~\eqref{Model:etai}, \eqref{Model:Yi}, \eqref{Model:Qi} and \eqref{Model:ei} are satisfied for all $i=0,1,2,\cdots$. Thus, the gain and QBER are given by:
\begin{equation}\label{Model:WithoutEve}
\begin{aligned}
Q_{\mu} &= Y_0 + 1-e^{-\eta\mu} \\
E_{\mu}Q_{\mu} &= e_0 Y_0 + e_{d}(1-e^{-\eta\mu}).
\end{aligned}
\end{equation}
Due the fact that $Q_\mu$ and $E_\mu$ can be measured or tested from the experiment, we will use Eq.~\eqref{Model:WithoutEve} in later simulations.

\subsection{Photon number channel model} \label{Sub:PhNch}
The model described above can be understood in another equivalent model.

\textbf{Photon number channel model:} Alice and Bob have an infinite number of channels. For channel $i$, Alice sends out an $i$-photon state to carry the qubit information, $i=0,1,2\cdots$. In the aforementioned model, Alice chooses which channel to use with a Poisson distribution, shown in Eq.~\eqref{Model:Poisson}, which is determined by her photon source.

Then $Y_i$ and $e_i$ can be regarded as the yield and error rate of channel $i$. Again, in our QKD scenario, Eve has full control of all these channels and she can change the values of $Y_i$ and $e_i$.

Note that one condition for these two models being equivalent is that Alice randomizes the phase of each pulse. It turns out that in some situations, this phase randomization procedure is crucial for security \cite{LoPreskill_05}.


\section{QKD hardware} \label{Sc:Hardware}
Let us examine QKD system elements from a hardware point of view. In the model, we can see that there are a few key components: laser source, channel link and detection system. By having the knowledge of the characteristics of these components, we can fit the model and perform simulations.

\subsection{Laser source}
In QKD experiments, two types of laser pulses are mostly used: telecom wavelength ($\sim$1550nm) and visible light ($\sim$760nm). Note that the $1310nm$ light was also used for QKD experiments. For example, see Ref.~\cite{RGGGZ_98}. Later, we will see that the choice of the wavelength, $\lambda$, determines the channel loss coefficient and detector efficiency.

\subsection{Channel}
There are mainly two types of QKD links: fiber and free space.

For fiber based QKD, the transmission distance is easy to vary. Thus, one can define the channel loss coefficient, $\beta$ in dB/km, which characterizes the loss dependence on transmission distance. For example, the loss coefficient of telecom  fiber is $\beta=0.2$ dB/km. For the visible light, the fiber loss is high, $\beta=2.5$ dB/km \cite{Townsend_98}.

Since commonly used fibers are made of birefringent materials, it is difficult to maintain the polarization. Thus, phase encoding is widely used in fiber based QKD systems. Note that phase encoding\footnote{In a phase encoding scheme, Alice encodes her information into the relative phase between two pulses \cite{Bennett_92}.} is equivalent to the polarization encoding \cite{Bennett_92}.

For free space based QKD, in general, it is difficult to define $\beta$ in dB/km. Instead, the total link loss in dB is commonly used. One main source of loss for the free space QKD implementation is the collection efficiency. Due to atmosphere scattering, the light beam is widened on the receiver arm. For a detailed discussion on how the atmosphere affects the light, one can refer to \cite{Milonni_Atmosp_04}. Note that the atmosphere is almost transparent to the visible light and it is a good medium for polarization maintenance. Later, we will see that the detector efficiency for visible light is normally higher than the one for telecom wavelength. Thus, in general, visible light is commonly chosen for free space based QKD.

\subsection{Detection}
For a detection system, four parameters are important.

\begin{itemize}
\item
$\eta_B$: detection efficiency, including detector efficiency and the internal transmission (coupling) efficiency of optical components inside Bob's box. The typical detection efficiency for a telecom wavelength\footnote{Here, we consider a widely used detection system with single photon detectors based on InGaAs/InP avalanche photodiodes.} is $1\sim5\%$, while for a visible wavelength, it can be as high as 20\%.

\item
$Y_0$: background count rate (probability), including dark counts and other background contributions. Note that if two detectors are used in a QKD system, then $Y_0$ should be the sum of the dark count rates of the two detectors in addition to other background contributions. 

\item
$e_d$: intrinsic detector error probability, which characterizes the alignment and stability of the optical system. In our model,we assume that $e_{d}$ is independent of the transmission distance.

\item
repetition rate: in practice, the repetition rate of detectors limits the key transmission speed. The product of key rate $R$ and repetition rate gives the key generation speed in bits/second. Normally, in an experiment, the laser pulses can be designed to be fast. The repetition rate is mainly limited by the detection system, e.g., the detector dead time and detection time-resolution.
\end{itemize}

In the model, we assume that there are two main sources of QBER, one from $Y_0$, which depends on channel loss\footnote{This part is roughly determined by the ratio $Y_0/\eta$.} and the other from $e_d$, which is independent of channel loss.

Note that there are a few developments in building single photon detectors during recent years, such as superconducting materials based detectors \cite{NIST_super_05} and up-conversion detectors \cite{Yamamoto_up1_05,Gisin_up2_06}

Later in the simulations, we use setup parameters from the QKD experiment completed by Gobby, Yuan and Shields (GYS) \cite{GYS_04}. The key parameters of the experiment setup are listed in Table \ref{Tab:GYSpara}.

\begin{table}[hbt] \centering
\begin{tabular}{|c|c|c|c|c|} \hline
$\lambda$ [nm] & $\beta$ [dB/km] & $\eta_{B}$ & $e_{d}$ & $Y_0$ \\
\hline
1550 & 0.21 & 4.5\% & 3.3\% & $1.7\times 10^{-6}$ \\
\hline
\end{tabular}
\caption{Parameters of the QKD experiment setup from GYS \cite{GYS_04}.} \label{Tab:GYSpara}
\end{table}

\chapter{Decoy state} \label{Chpt:Decoy}

The decoy state method was first proposed by Hwang \cite{Hwang_03} to improve the performance of the coherent state QKD.  We have proven the security of the QKD with decoy states \cite{LoDecoy_03,MasterReport,Decoy_05} and demonstrated its practical advantage. In Hwang's original decoy state method, she suggested the use of a strong coherent state (with $\nu>\mu$) for decoy states. In contrast, we propose using weak coherent states. Subsequently, some practical decoy state protocols with only one or two decoy states are proposed \cite{Practical_05}. We highlight that practical decoy state protocols were also proposed by Wang \cite{Wang_05,Wang2_05}, Harrington, Ettinger, Hughes and Nordholt \cite{HEHN_05}.

The experimental demonstrations for the decoy state method have been completed recently \cite{ZQMKQ_06,ZQMKQ60km_06,LosAlamosDecoy_07,PDC144_07,PanDecoy_07,YSS_Decoy_07,Guo_Decoy_07}. Note that aside from the decoy state method, we also studied other methods to improve the QKD performance, such as the dual detector scheme \cite{QZMLQ_dual_07}.

This work is published in Ref.~\cite{Decoy_05}. By collaborating with Hoi-Kwong Lo and Kai Chen, I apply the GLLP security analysis to a decoy state QKD. With the model described in Section \ref{Sc:Model}, I simulate a QKD experiment \cite{GYS_04} to show the improvement given by using decoy states.

\section{Decoy state} \label{Sc:AsympDecoy}
In this section, we present the QKD with decoy states. By simulating a real experiment setup, we compare two cases: a decoy and non-decoy state QKD.

\subsection{Motivation}
As discussed in Section \ref{Sc:GLLP}, in the GLLP security analysis, Alice and Bob need to determine the portion of tagged and untagged qubits to implement privacy amplification.

From Eq.~\eqref{Post:KeyRate}, we can see that $Q_\mu$ and $E_\mu$ can be measured or tested from the experiment. Alice and Bob need to estimate $Q_1$ and $e_1$ to determine the amount of privacy amplification that is needed.

On the other hand, as we presented in Section \ref{Sc:Model}, Eve has full control of the channel. Thus, she might block out single photon states, which is not good for her attack and make the channel transparent to the multi photon states. Thus, one pessimistic assumption is that all losses and errors come from a single photon state \cite{IndividualAttack_00,GLLP_04}. That is, set $Y_i=1$ and $e_i=0$ for $i\ge2$ in Eqs.~\eqref{Model:Gain} and \eqref{Model:QBER}. Thus, the estimations of $Q_1$ and $e_1$ without decoy states are:
\begin{equation}\label{Decoy:nondecoyQ1e1}
\begin{aligned}
Q_1 &\ge Q_\mu-\sum_{i=2}^{\infty}\frac{\mu^{i}}{i!}e^{ - \mu} \\
e_1 &\le \frac{E_{\mu}Q_{\mu}}{Q_{1}} \\
\end{aligned}
\end{equation}
Here, note that since Alice and Bob cannot distinguish vacuum (background) contribution and single photon state contribution\footnote{Or, they cannot estimate the detection contributions from vacuum qubits, $Q_0$.}, they have to consider these two states together. For a vacuum qubit, since it is a random state, $\delta_b=\delta_p=1/2$. Thus, for the combined state (single photon state and vacuum state), we still have $\delta_b=\delta_p$.


Later in the simulation, we will see that the key rate and maximal secure distance of a coherent state QKD without decoy states are quite limited. In order to lower the amount of necessary privacy amplification, one needs to have a better estimation of $Q_1$ and $e_1$. From Eq.~\eqref{Model:Qi}, we know that in order to estimate $Q_1$, one needs to estimate $Y_1$. Therefore, the question is: \emph{how can Alice and Bob estimate $Y_1$ and $e_1$ accurately?} This is the motivation of the decoy state scheme.

\subsection{Solution}
From the model described in Section \ref{Sc:Model}, there are two observations. First, $Y_i$ and $e_i$ can be changed by Eve, so they are unknowns to Alice and Bob. Secondly, $Q_\mu$ and $E_\mu$  can be determined by Alice and Bob. Thus, Alice and Bob need to estimate $Y_1$ and $e_1$ by using the knowledge of $Q_\mu$ and $E_\mu$. If Eqs.~\eqref{Model:Gain} and \eqref{Model:QBER} are just considered, then Alice and Bob have to assume the worst scenario: all losses and errors come from the single photon state.

We can see that Eqs.~\eqref{Model:Gain} and \eqref{Model:QBER} are linear equations of $Y_i$ and $Y_ie_i$. In addition to the regular signal state, if Alice sends out extra pulses with different intensities, $\mu$, they will obtain more than one linear equation in the form of Eqs.~\eqref{Model:Gain} and \eqref{Model:QBER}. Here comes the key assumption of the decoy state method:
\begin{equation}\label{Decoy:DecoyAss}
\begin{aligned}
Y_i(decoy)=Y_i(signal) \\
e_i(decoy)=e_i(signal).
\end{aligned}
\end{equation}
These extra pulses are called \emph{decoy states}. In the infinite decoy case \cite{Decoy_05}, Alice and Bob perform an infinite number of decoy states, and then they can solve an infinite number of linear equations to obtain $Y_1$ and $e_1$ accurately. We call this case the infinite decoy state protocol. Here, note that with the infinite decoy state, one can strictly show \cite{LoLut_review_07} that the beam-splitting channel model discussed in Section \ref{Sub:Model:ChDet} is a valid assumption.

An intuition on why this can be done: from Eqs.~\eqref{Model:Gain} and \eqref{Model:QBER}, we can see that the contribution from high order terms of $Y_i$ and $e_i$ converge to 0 exponentially\footnote{Actually, $n!$ is quicker than exponential convergence.}. If one only focuses on $Y_1$ and $e_1$, the number of unknowns can be chopped off to a finite number. In the next chapter, we will see that one or two decoy states are sufficient for practical use. In the simulation, we will use Eqs.~\eqref{Model:Yi} and \eqref{Model:ei} for the infinite decoy state case. For a detailed procedure of the decoy state method, one can refer to Section \ref{Sub:Prac:Process}.

In the following discussion, $\mu$ always refers to the intensity (expected photon number) of the signal state used for real key transmission. We will use $\nu$ for the expected photon number of decoy states.

\subsection{Discussion}
In a large parameter regime when the background contribution can be negligible and the error rate is not large, the key rate is roughly in the order of $R=O(\mu\eta)$ from Eq.~\eqref{Post:KeyRate}.

In Appendix \ref{ApSub:nondecoymu}, we show that the optimal $\mu$ for the non-decoy state case is $\mu=O(\eta)$. Thus, the key rate is $R=O(\eta^2)$. That is, the key rate is quadratically dependent on the channel transmission. Note that in general, the channel transmission is quite low, typically less than $1\%$. This is the intrinsic reason why the performance of a QKD without decoy states is very limited.

On the other hand, in Appendix \ref{ApSub:nondecoymu}, we show that the optimal $\mu$ for the infinite decoy state case is $\mu=O(1)$. Thus, the key rate is $R=O(\eta)$. That is, the key rate is linearly dependent on the channel transmission. Note that even with a perfect single photon source, the highest order the key rate can reach is $R=O(\eta)$. Hence, with decoy states, one can treat a weak coherent state as a good single photon source for a QKD.

Note that this conclusion is also true for other photon sources, e.g., triggering PDC sources \cite{TriggeringPDC_08}, see discussions in Chapter \ref{Chpt:Trig}.

\subsection{Simulation} \label{Sub:Decoy:Simu}
We simulate a recent coherent state QKD experiment \cite{GYS_04}. This is to compare the cases with and without decoy states. The parameters of the experiment setup are listed in Table \ref{Tab:GYSpara}.

For both cases, the key rate formula is the same, see Eq.~\eqref{Post:KeyRate}. By using the Cascade protocol \cite{BrassardSalvail_93}, the error correction efficiency is $f(E_\mu)=1.22$. The gain $Q_\mu$ and QBER $E_\mu$ can be measured or tested from the experiment. Therefore, for both cases, we use the same formulae, Eqs.~\eqref{Model:Gain} and \eqref{Model:QBER}. The estimations of $Q_1$ and $e_1$ are different. For the case without decoy states, we use the formulae of Eq.~\eqref{Decoy:nondecoyQ1e1}. For the case with decoy states, we assume that Alice and Bob can estimate $Q_1$ and $e_1$ accurately. In the simulation, we use the formulae of Eqs.~\eqref{Model:Yi} and \eqref{Model:ei}.

\begin{figure}[hbt]
\centering \resizebox{12cm}{!}{\includegraphics{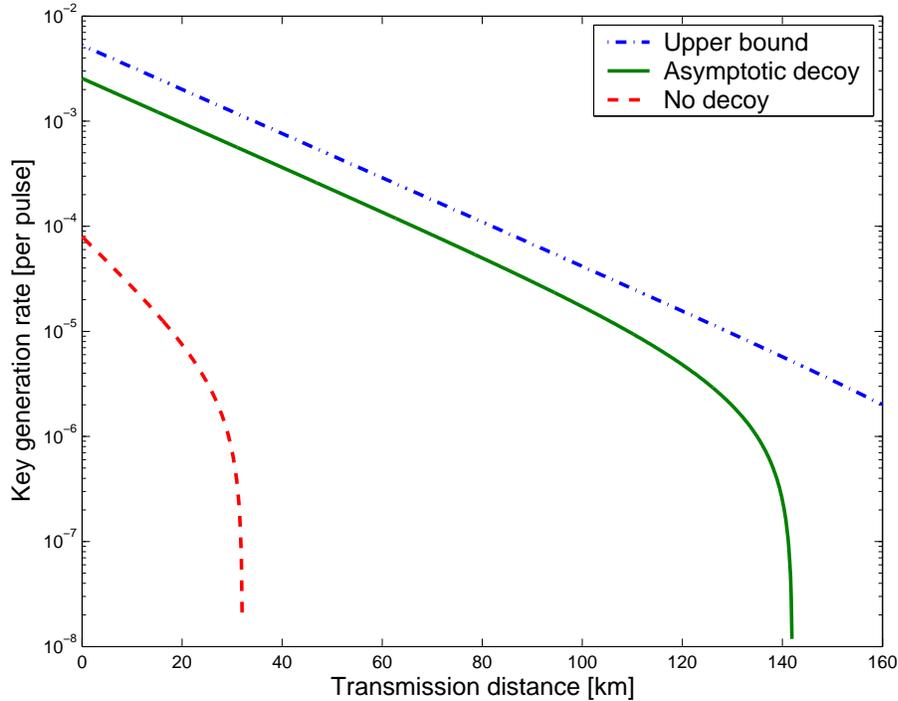}}
\caption{Plot of the key rate as a function of the transmission distance, comparing the coherent state QKD with decoy states and without decoy states. The calculation of the upper bound is shown in Section \ref{Sub:KeyUp}. The experiment setup parameters are listed in Table~\ref{Tab:GYSpara}.}
\label{Fig:AsympDecoy}
\end{figure}

As shown in Appendix \ref{ApSc:muCoherent}, we choose $\mu=0.48$ for the case with decoy states and $\mu=\eta$ for the case without decoy states. The simulation result is shown in Figure \ref{Fig:AsympDecoy}.

From the simulation result, we can see that the decoy state method can improve the QKD performance dramatically.

\begin{enumerate}
\item
With decoy states, the maximal distance can reach 142 km. For comparison, we find that with the prior art method, the maximal secure distance is only about 32 km.

\item
At 0 km distance, the key rates for decoy and no decoy cases are: $2.55\times10^{-3}$ and $7.97\times10^{-5}$. As we can clearly see, the gap between two curves increases when the distance increases.

\item
By comparing the upper bound of the key rate, which is discussed in the next section, one can see that in a large parameter regime (for instance, the distance between 0 km and 120 km), the decoy state protocol can achieve a close performance as the upper bound shown in Section \ref{Sub:KeyUp}.

\item
We checked that our results are stable to small perturbations of the background count rate $Y_0$ and average photon number $\mu$ (both up to a 20\% change).
\end{enumerate}

\section{Upper Bounds} \label{Sc:UpBounds}
As we mentioned in Section \ref{Sub:QKDperformance}, we are interested in maximizing two quantities, key rate and maximal secure distance. In this section, we investigate the upper bounds of these two quantities. By comparing the upper bound performance and the decoy state QKD performance, we want to investigate how much room is left for further improvement.

\subsection{Distance upper bound} \label{Sub:DisUp}
Due to a simple intercept-and-resend attack, an upper bound on the bit error rate of the BB84 protocol with single photon states is 25\%. The maximal secure distance then can be bounded by the distance when the bit error rate of the single photon states $e_1$ reaches 25\%. According to our model, Eq.~\eqref{Model:ei}:
$$
e_1=\frac{e_{d}\eta+\frac12Y_0}{\eta+Y_0}
$$
where $e_{d}$ is the intrinsic error rate of Bob's detectors, $\eta$ is the overall transmittance, and $Y_0$ is the background rate. Thus, $e_1$ exceeds 25\% when
\begin{equation} \label{Model:EtaBound}
\eta\le\frac{0.25Y_0}{0.25-e_{d}}.
\end{equation}
In GYS \cite{GYS_04}'s case, the upper bound of the secure distance is $208$ km by considering the parameters listed in Table \ref{Tab:GYSpara}.

\subsection{Key rate upper bound} \label{Sub:KeyUp}
As for the BB84 protocol, the final secure key can only be derived from single photon qubits. To derive the upper bound of a key generation rate, we assume that Alice and Bob
can distinguish the single photon qubits from other qubits (vacuum and multi photon qubits). Therefore, they can perform the classical data post-processing only onto the single photon qubits. One simple upper bound\footnote{Note that this upper bound is true for any post-processing (based on 1-LOCC or 2-LOCC) Alice and Bob use in BB84.} of key generation rate is given by the \emph{mutual information} between Alice and Bob \cite{MaurerWolf_intrin_99}:
\begin{equation}\label{Model:UpperR}
R^U=Q_1[1-H_2(e_1)],
\end{equation}
where $Q_1$ and $e_1$ are the gain and error rate of single photon states, respectively. The simulation result is shown in Figure \ref{Fig:AsympDecoy}.

Note that the above two upper bounds, Eqs.~\eqref{Model:EtaBound} and \eqref{Model:UpperR}, rely on two assumptions.
\begin{itemize}
\item
Alice and Bob cannot distinguish background counts and true signal counts. That is, they cannot decouple $e_d$ from $e_1$ in Eq.~\eqref{Model:ei}.

\item
A secure key can only be extracted from single photon states. This is true for BB84 and many other protocols. An exception is the SARG04 protocol \cite{SARG_04}.
\end{itemize}

Note that these two bounds are general upper bounds, regardless of the technique used for combating the effect of imperfect devices, such as the decoy state technique.



\section{Discussion}
First, from the simulation, we can see that the decoy state technique can dramatically improve the QKD performance. Later, we will discuss practical protocols for the decoy state QKD and experiment demonstrations. From there, we show that the decoy state method is highly practical.

In comparison to the key rate upper bound, in a large distance regime (for instance, the distance between 0 km and 120 km), the decoy state protocol achieves a close performance to the theoretical limit. Compared to the maximal secure distance upper bound, 208 km, there is a 60 km gap between the theoretical limit and decoy state protocol. Later, by combining two-way classical communication post-processing schemes, we push this maximal secure distance for the infinite decoy state protocol beyond 180 km. From here, we can see that the decoy state protocol pushes the QKD performance close to the theoretical limit.

Therefore, we expect the decoy state protocol to be a standard technique for prepare-and-measure QKD scheme implementations.

Let us recap the key assumptions underlying the security proof for the decoy state QKD: first, there is the squash model and secondly, there is the assumption that Eve cannot distinguish decoy and signal states during key transmission. The second assumption is equivalent to Eq.~\eqref{Decoy:DecoyAss}.
Later in Section \ref{Sc:DecoyExp}, we can see that verifying this assumption is a nontrivial task in real experiments. On the other hand, in Chapter \ref{Chpt:Trig}, we show that this assumption can be loosened by using other single photon sources.

%
%

\chapter{Practical decoy state} \label{Chpt:Practical}
In this chapter, we will discuss practical proposals of the decoy state QKD and experimental demonstrations. Here again, we will focus on the coherent state BB84 QKD.

The work of practical decoy state proposals is published in Ref.~\cite{Practical_05}. In this work, I apply the idea of the Vaccum+Weak decoy state protocol, which was first proposed by Lo \cite{LoDecoy_03}, and consider statistical fluctuations. Here, I would like to highlight the theoretical contributions to the practical decoy state QKD from other groups \cite{HEHN_05,Wang_05,Wang2_05}.

The work for the experimental demonstration is published in Refs.~\cite{ZQMKQ_06,ZQMKQ60km_06}.  In this work, I designed the experimental parameters and analyzed data in the decoy state QKD experiment demonstration. Here, I would like to highlight the experimental demonstrations completed by other groups \cite{ZQMKQ_06,ZQMKQ60km_06,LosAlamosDecoy_07,PDC144_07,PanDecoy_07,YSS_Decoy_07,Guo_Decoy_07}.


Note that aside from the decoy state method, we also studied other methods to improve the QKD performance, such as the dual detector scheme \cite{QZMLQ_dual_07}.

\section{Practical proposals}
The general question in a decoy state scheme with $m$ decoy states can be described by the following mathematical question.
\begin{quote}
\textbf{Question:} Given $2(m+1)$ constrains in the form of Eqs.~\eqref{Model:Gain} and \eqref{Model:QBER}, how do we obtain the lower bound of $R$ given by Eq.~\eqref{Post:KeyRate}?
\end{quote}

%
%

When $m\rightarrow\infty$, Alice and Bob can solve ${Y_1}$ and ${e_1}$ accurately, in principle. This is the infinite case described in Section \ref{Sc:AsympDecoy}.

In the following, we will present three practical decoy methods, the Vacuum+Weak decoy state and one decoy state, and a numerical method. For a general discussion of the two decoy state methods, one can refer to Ref.~\cite{Practical_05}. Note that in Ref.~\cite{Practical_05}, we proved that the Vacuum+Weak decoy state protocol is optimal within the two decoy state methods.

\subsection{Vacuum+Weak decoy} \label{Sub:VW}
In this method, two decoy states are performed to bound $Y_1$ and $e_1$ separately. First, Alice and Bob implement a vacuum decoy state where Alice simply shuts off her photon source. In this case, all detections that Bob obtains are background counts
\begin{equation}\label{Decoy:VaQE}
\begin{aligned}
Q_{vacuum} &= Y_0 \\
E_{vacuum} &= e_0 = \frac12.
\end{aligned}
\end{equation}
The background counts occur randomly, thus its error rate is $e_0=1/2$. The vacuum decoy state allows Alice and Bob to estimate the background rate $Y_0$.

Secondly, they perform a weak  decoy state where Alice uses a weaker intensity $\nu$ ($\nu<\mu$) for the decoy state. In this case, Bob's detections mainly come from two parts: background and single photon contributions. This is because when the intensity is weak, the probability of obtaining a multi photon state is small. With the estimation from the vacuum decoy state, one can estimate $Y_1$ and $e_1$ from the weak decoy state.

Now, let us strictly solve the problem. The gains of the signal state and decoy state are given by Eq.~\eqref{Model:Gain}
\begin{equation} \label{Decoy:QmuQnu}
\begin{aligned}
Q_{\mu}e^{\mu} &= Y_0+\mu Y_1+\frac{\mu^2}{2}Y_2+\frac{\mu^3}{3!}Y_3+\cdots \\
Q_{\nu}e^{\nu} &= Y_0+\nu Y_1+\frac{\nu^2}{2}Y_2+\frac{\nu^3}{3!}Y_3+\cdots 
\end{aligned}
\end{equation}
Considering $\mu^2Q_\nu e^{\nu}-\nu^2Q_\mu e^{\mu}$,  
we find that:
\begin{equation} \label{Decoy:Derive1}
\begin{aligned}
\mu^2Q_{\nu}e^{\nu}-\nu^2Q_{\mu}e^{\mu} &= (\mu^2-\nu^2)Y_0+\mu\nu(\mu-\nu)Y_1+\mu^2\nu^2\frac{\nu-\mu}{3!}Y_3+\cdots,
\end{aligned}
\end{equation}
thus:
\begin{equation}\label{Practical:Y1L}
\begin{aligned}
Y_1 &\ge Y_1^L = \frac{\mu}{\mu\nu-\nu^2}\big(Q_\nu e^{\nu}-Q_\mu e^{\mu}\frac{\nu^2}{\mu^2} -\frac{\mu^2-\nu^2}{\mu^2}Y_0\big) \\
\end{aligned}
\end{equation}
since $\nu<\mu$ and all $Y_i\in[0,1]$.

The upper bound of $e_1$ can be simply derived by Eq.~\eqref{Model:QBER}:
\begin{equation}\label{Practical:e1U}
\begin{aligned}
e_1 &\le e_1^U = \frac{E_\nu Q_\nu e^\nu-e_0Y_0}{Y_1^L\nu}.
\end{aligned}
\end{equation}

Substituting the normal case (without Eve) values, Eqs.~\eqref{Model:WithoutEve}, into these estimations, in the limit of $\nu\ll\mu$, we get:
\begin{equation}\label{Practical:Y1e1Blimit}
\begin{aligned}
Y_1^L &\rightarrow \eta+Y_1 \\
e_1Y_1 &\rightarrow e_0Y_0+e_d\eta \\
\end{aligned}
\end{equation}
which is consistent with the expected value given by Eqs.~\eqref{Model:Yi} and \eqref{Model:ei}. Thus, asymptotically, the Vacuum+Weak decoy method gives a tight lower bound of the key rate. In other words, the infinite decoy state protocol described in Section \ref{Sc:AsympDecoy} is the asymptotic limit of the Vacuum+Weak decoy state protocol.

\begin{figure}[hbt]
\centering \resizebox{12cm}{!}{\includegraphics{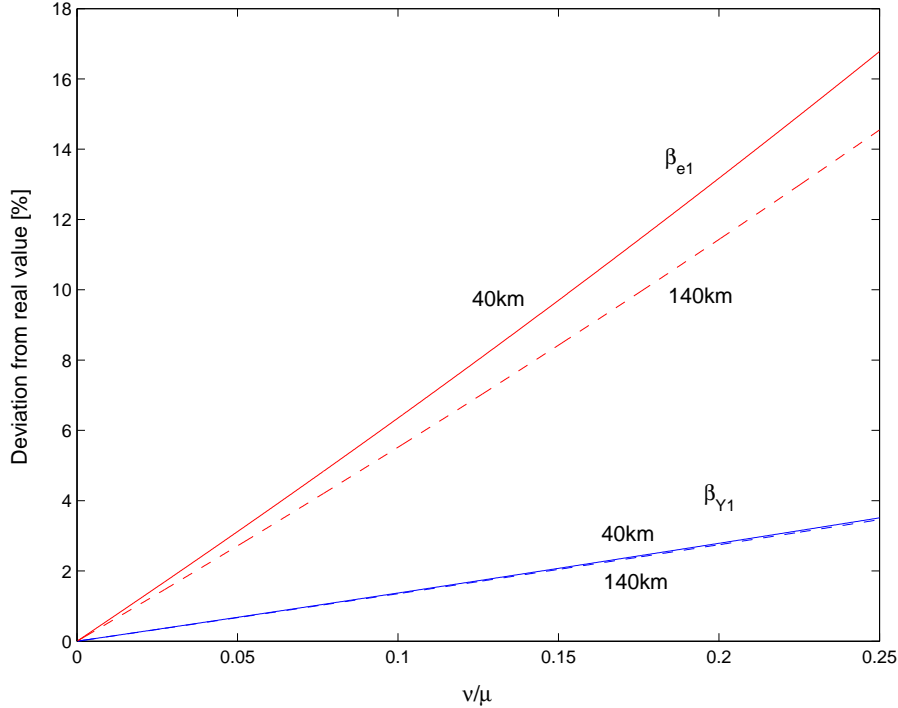}}
\caption{Plot of the relative deviations of $Y_1^L$ and $e_1^U$ from the expected values (i.e., the case $\nu \to 0$) as functions of $\nu/ \mu$ with the fiber length 40 km (solid lines) and 140 km (dashed lines). The bounds $Y_1^L$ and $e_1^U$ are given by Eqs.~\eqref{Practical:Y1L} and \eqref{Practical:e1U}, and the expected values are given by Eqs.~\eqref{Model:Yi} and \eqref{Model:ei}. We consider the Vacuum+Weak protocol here. The expected photon number is $\mu=0.48$ from the optimization calculation of Eq.~\eqref{Optmu:Decoymu} in Appendix \ref{ApSub:decoymu}. The experiment setup parameters are from GYS~\cite{GYS_04}, listed in Table~\ref{Tab:GYSpara}.} \label{Fig:PracDevnu}
\end{figure}

Now let us examine how good these two bounds are by using the parameters listed in Table \ref{Tab:GYSpara}. Here, we define the deviation of the bounds:
\begin{equation}\label{Practical:Y1e1Bdev}
\begin{aligned}
\beta_{Y1} &\equiv \frac{Y_1-Y_1^L}{Y_1} \\
\beta_{e1} &\equiv \frac{e_1^U-e_1}{e_1}. \\
\end{aligned}
\end{equation}
The simulation result is shown in Figure \ref{Fig:PracDevnu}.

From the simulation, we can see that both deviations are relatively independent of the channel transmission distance. The deviation of $e_1^U$ is larger than the one of $Y_1^L$. The choice of a weak decoy state $\nu$ is not very constrained since even with $\nu/\mu\approx1/4$, the deviation is small. In Table \ref{Tab:VWDevR}, we can see that with $\nu/\mu\approx1/4$, the key rate from the Vacuum+Weak decoy state protocol achieves a very close performance of the infinite decoy state case.

\begin{table}[hbt] \centering
\begin{tabular}{|c|c|c|c||c|c|c|c|c|c|}
\hline
Distance & $Y_1$  & $e_1$ & $R_{inf}$ & $Y_1^L$  & $e_1^U$ & $R_{vw}$ \\
\hline
0 km & $4.50\times10^{-2}$ & 3.30\% & $2.55\times10^{-3}$ & $4.34\times10^{-2}$ & 3.88\% & $2.19\times10^{-3}$ \\
\hline
70 km & $1.53\times10^{-3}$ & 3.35\% & $8.28\times10^{-5}$ & $1.47\times10^{-4}$ & 3.95\% & $6.99\times10^{-5}$ \\
\hline
130 km & $8.55\times10^{-5}$ & 4.23\% & $1.96\times10^{-6}$ & $8.23\times10^{-5}$ & 4.91\% & $1.24\times10^{-5}$ \\
\hline
\end{tabular}
\caption{List of the simulation results for three distances: 0 km, 70 km and 130 km, comparing the Vacuum+Weak protocol with the case of the infinite (asymptotic) decoy state. For both protocols, we use $\mu=0.48$. For the Vacuum+Weak decoy state protocol, we use $\nu=0.13$. Parameters of the QKD experiment setup are listed in Table \ref{Tab:GYSpara}.} \label{Tab:VWDevR}
\end{table}

Here, we compare Eqs.~\eqref{Model:Yi}, \eqref{Model:ei}, \eqref{Practical:Y1L} and \eqref{Practical:e1U} by simulating the GYS experiment. We can see that the deviation of the key rate given by the Vacuum+Weak decoy state protocol and infinite decoy state protocol increases when the distance reaches the maximal secure distance. Similar to the conclusion from Figure~\ref{Fig:PracDevnu}, the deviation of $Y_1^L$ from $Y_1$ is small throughout the whole distance regime.

\subsection{One decoy} \label{Sub:Onedecoy}
In some realistic situations, a vacuum decoy state may not be easy to perform, or the background count rate cannot be estimated accurately due to the fact that $Y_0$ is small (typically $10^{-5}$). Consequently, one needs to consider a case without the vacuum decoy state. That is, Alice and Bob only perform a weak decoy state.

We treat the one decoy state method as an imperfect case of the Vacuum+Weak method. Assume that Alice and Bob perform the Vacuum+Weak decoy method, but they prepare very few states as vacuum decoy states. Therefore, they cannot estimate $Y_0$ very well. The one decoy protocol is the same as a Vacuum+Weak decoy state protocol, except that the value of $Y_0$ is unknown. Since Alice and Bob do not know $Y_0$, Eve can pick $Y_0$ as she wishes. We argue that, on physical grounds, it is advantageous for Eve to pick $Y_0$ to be zero. This is because Eve may gather more information on the single-photon signal than the vacuum. Therefore, the bound for the case $Y_0 =0$ should apply to our one decoy protocol.  For this reason, Alice and Bob can derive a bound on the key generation rate, $R$, by substituting $Y_0=0$ in Eqs.~\eqref{Practical:Y1L} and \eqref{Practical:e1U}.

Mathematically, one can treat $Y_0$ as an unknown variable in Eqs.~\eqref{Practical:Y1L} and \eqref{Practical:e1U}, and determine the lower bound of the key generation rate, Eq.~\eqref{Post:KeyRate}, for all possible $Y_0$. By taking the derivative of Eq.~\eqref{Post:KeyRate}, one can find that
\begin{equation}\label{Decoy:OneY1e1}
\begin{aligned}
Y_1^{trial} &= \frac{\mu}{\mu\nu-\nu^2}(Q_\nu e^{\nu}-Q_\mu
e^\mu\frac{\nu^2}{\mu^2}) \\
e_1^{trial} &= \frac{E_\nu Q_\nu e^{\nu}}{Y_1^{trial}\nu}
\end{aligned}
\end{equation}
gives a lower bound of the key rate.

Later, in the next subsection, we will present a numerical method to estimate the key rate $R$. Now we can compare Eq.~\eqref{Decoy:OneY1e1} with the numerical method by simulating the GYS experiment. In this case, we consider three distances: 0 km, 70 km and 130 km.

\begin{table}[hbt] \centering
\begin{tabular}{|c|c|c|c||c|c|c|c|c|c|}
\hline
Distance & $Y_1^{trial}$  & $e_1^{trial}$ & $R_{one}$& $Y_1^{num}$  & $e_1^{num}$ & $R_{num}$ \\
\hline
0 km & $4.34\times10^{-2}$ & 3.89\% & $2.19\times10^{-3}$ & $4.36\times10^{-2}$ & 3.84\% & $2.22\times10^{-3}$ \\
\hline
70 km & $1.48\times10^{-3}$ & 4.40\% & $6.55\times10^{-5}$ & $1.48\times10^{-3}$ & 3.76\% & $7.26\times10^{-5}$ \\
\hline
130 km & $9.93\times10^{-5}$ & 13.0\% & $0$ & $8.33\times10^{-5}$ & 4.34\% & $1.65\times10^{-6}$ \\
\hline
\end{tabular}
\caption{List of simulation results for three distances: 0 km, 70 km and 130 km, comparing the one decoy state protocol with the numerical optimization method shown in the next subsection. For both protocols, we use $\mu=0.48$ and $\nu=0.13$. Parameters of the QKD experiment setup are listed in Table \ref{Tab:GYSpara}.} \label{Tab:OneNumR}
\end{table}

By comparing Tables \ref{Tab:VWDevR} and \ref{Tab:OneNumR}, we can see that the numerical method, shown in the next subsection, can give the highest key rate of the three practical decoy state protocols. However, note that all four methods; infinite decoy, Vacuum+Weak, one-decoy and numerical method, achieve a close QKD performance in a large parameter regime. Here, we have not considered the statistical fluctuations. After considering the statistical fluctuations, the simulation result is shown in Figure \ref{Fig:RVWFluc}.

\subsection{Numerical method}
Both the Vacuum+Weak and one decoy state protocols presented above bound $Y_1$ and $e_1$ separately. With reference to the original question that we were trying to solve in the beginning of this section, what we really want to bound is the key rate of Eq.~\eqref{Post:KeyRate} instead of $Y_1$ and $e_1$ separately.

One natural practical decoy state protocol will be a numerical solution to the question stated in the beginning of this section. To do that, one need to find the lower bound $R$ of Eq.~\eqref{Post:KeyRate} given the constraints of Eqs.~\eqref{Model:Gain} and \eqref{Model:QBER}:
\begin{equation} \label{Practical:QEmuQEnu}
\begin{aligned}
Q_{\mu}e^{\mu} &= Y_0+\mu Y_1+\frac{\mu^2}{2}Y_2+\frac{\mu^3}{3!}Y_3+\cdots \\
Q_{\nu}e^{\nu} &= Y_0+\nu Y_1+\frac{\nu^2}{2}Y_2+\frac{\nu^3}{3!}Y_3+\cdots \\
E_\mu Q_{\mu}e^{\mu} &= Y_0e_0+\mu Y_1e_1+\frac{\mu^2}{2}Y_2e_2+\frac{\mu^3}{3!}Y_3e_3 +\cdots \\
E_\nu Q_{\nu}e^{\nu} &= Y_0e_0+\nu Y_1e_1+\frac{\nu^2}{2}Y_2e_2+\frac{\nu^3}{3!}Y_3e_3 +\cdots. \\
\end{aligned}
\end{equation}
The difference between the Vacuum+Weak and one decoy state protocols is whether $Y_0$ is known or not. 

In order to solve this question numerically, one needs to put a cut-off of $Y_i$ and $e_i$. Later in the simulation, we will consider a cut-off of $i=20$. That is, $Y_i=e_i=0$ for $i>=20$. Note that for $i=20$ and $\mu=1$, the probability is $P(20)=1.51\times10^{-19}$ according to the Poisson distribution of the source photon number given by Eq.~\eqref{Model:Poisson}. For a reasonable finite key transmission, the higher order terms can be neglected.

We present the numerical solutions in Table \ref{Tab:NumDecoy} by using the parameters in Table \ref{Tab:GYSpara}.

\begin{table}[hbt] \centering
\begin{tabular}{|c|c|c|c|c|c|c|c|c|}
\hline
Distance & $Y_1$ & $Y_2$ & $Y_3$ & $e_1$ & $e_2$ & $R$ \\
\hline
0 km & $4.36\times10^{-2}$ & $1.15\times10^{-1}$ & $5.86\times10^{-13}$ & $3.84\%$ & $5.86\times10^{-13}$ & $2.22\times10^{-3}$ \\
\hline
70 km & $1.48\times10^{-3}$ & $4.01\times10^{-3}$ & $5.45\times10^{-4}$ & $3.76\%$ & $4.47\times10^{-3}$ & $7.26\times10^{-5}$ \\
\hline
130 km & $8.33\times10^{-6}$ & $2.15\times10^{-5}$ & $5.86\times10^{-13}$ & $4.34\%$ & $3.17\%$ & $1.65\times10^{-6}$ \\
\hline
\end{tabular}
\caption{Comparison of the numerical result with the infinite decoy state (asymptotic) case and the Vacuum+Weak protocol. For all three protocols, we use $\mu=0.48$. For the two practical decoy state protocol, we use $\nu=0.1$. Parameters of the QKD experiment setup from GYS \cite{GYS_04}.} \label{Tab:NumDecoy}
\end{table}

Here, we have not considered the statistical fluctuations. From Table \ref{Tab:NumDecoy}, we have following remarks:
\begin{enumerate}
\item
If we only consider Eq.~\eqref{Practical:Y1L}, Eve's optimal attack will be setting $Y_i=0$ for $i\ge3$. However, if we consider the numerical decoy state method as shown in Table \ref{Tab:NumDecoy}, Eve might choose $Y_i\neq0$ for $i\ge3$\footnote{In the numerical result, we find that $Y_3$ is always relatively small in comparison to $Y_2$, but the values of $Y_i$ for $i\ge4$ are in the same order of $Y_2$.}.

\item
The result for the numerical decoy state method is relatively stable with a choice of a cut-off $n$. If we choose $n=30$ or $n=40$, the result fluctuates within 3\%. Note that the numerical optimization algorithm that we used here might not be optimal.
\end{enumerate}

\section{Statistical fluctuation analysis} \label{Sc:Stat}
In this section, we will discuss the effect of finite data size on our estimation process for $Y_1$ and $e_1$. We will also discuss how statistical fluctuations might affect our choice of the weak decoy state intensity $\nu$.

All real-life experiments are implemented within a finite period of time. Ideally, we would like to consider a QKD experiment that can be performed within, for instance, a few hours or so. This means that the experiment data size is finite. Shortly, we will see that the statistical fluctuation analysis is a rather complex problem. We do not have a full solution to the problem. Nonetheless, we will provide some rough estimation based on the standard error analysis which suggests that the statistical fluctuation problem of the practical decoy state methods for a QKD experiment appears to be under control, if the experiment is run over only a few hours.

\subsection{What parameters are fluctuating?} \label{Sub:HardStat}
Recall that in Eq.~\eqref{Post:KeyRate}, there are four key parameters: the gain $Q_\mu$ and QBER $E_\mu$ of the signal state and the gain $Q_1$ and error rate $e_1$ of the single photon state.

After key transmission, Bob can count the exact number of clicks and knows the total number of pulses. Hence, the gain of signal state $Q_\mu$, the ratio of the aforementioned two numbers, is measured directly from the experiment. Therefore, they do not need to consider the fluctuation of $Q_\mu$.

In practice, Alice and Bob do not really need to sacrifice testing bits to estimate $E_\mu$. They can directly apply some classical error correction code, for instance, the Cascade \cite{BrassardSalvail_93} code, to correct all bit errors. Then they check whether the error correction is successful or not\footnote{This can be done efficiently by random parity check.}. Afterwards, they can calculate (if necessary) $E_\mu$ by counting the number of errors. Thus, there is no fluctuation for $E_\mu$ as well.

Thus, there is no fluctuation in the error correction part. The difficult part of the statistical fluctuation analysis is in the privacy amplification part. In the following discussion, we will focus on the statistical fluctuation analysis of the Vacuum+Weak decoy state method. To show the complexity of the problem, we will now discuss the following five sources of fluctuations.

\begin{enumerate}
\item
In practice, the intensity of the lasers used by Alice will be fluctuating. In other words, even the parameters $\mu$ and $\nu$ suffer from fluctuations. Without hard experimental data, it is difficult to pinpoint the extent of their fluctuations. Furthermore, the source may not even be a strict coherent state. To simplify our analysis, we will ignore their fluctuations in this thesis.

\item
Up until now, in our analysis, we have assumed that the distribution of the photon number eigenstates (Fock states) in each type of state is fixed, see Eq.~\eqref{Model:Poisson}. For instance, if $N$ signal states of intensity $\mu$ are emitted, we assume that exactly $ N \mu e^{-\mu}$ out of the $N$ signal states are single photons. In real-life, the value of $\mu e^{- \mu}$ is only a probability, the actual number of single photon signals will fluctuate statistically. This fluctuation is dictated by the law of large number. Hence, this problem should be solvable\footnote{It was subsequently pointed out to us by Gottesman and Preskill that the above two sources of fluctuations can be combined into the fluctuations in the photon number frequency distribution of the underlying signal and decoy states. These fluctuations will generally be averaged out to zero in the limit of a large number of signals, provided that there is no systematic error in the experimental set-up.}. For simplicity, we will neglect this source of fluctuations in this thesis.

\item
The yield $Y_i$ may fluctuate in the sense that $Y_i$ for the signal state might be slightly different from $Y'_i$ of the decoy state. Note that if one uses the vacuum state as one of the decoy states, then by observing the yield of the vacuum decoy state, conceptually, one has a very good handle on the yield of the vacuum component of the signal state (in terms of hypergeometric functions). However, note that the background rate is generally rather low (typically $10^{-5}$). Therefore, to obtain a reasonable estimation on the background rate, a rather large number (for instance, $10^7$) of the vacuum decoy states will be needed\footnote{As noted in Ref.~\cite{Decoy_05}, even a $20\%$ fluctuation in the background will have a small effect on the QKD performance.}. Note that, with the exception of the case $i=0$ (the vacuum case), neither $Y_i$ and $Y_i'$ are directly observable in an experiment. In a real experiment, one can measure only some {\it averaged} properties. For instance, the gain $Q_{\mu}$ of the signal state, which can be experimentally measured, has its origin as the weighted averaged yields of the various photon number eigenstates $Y_i$s whereas the $Q_\nu$ for the decoy state is given by the weighted averaged of $Y'_i$s. Relating the observed averaged properties, e.g., $Q_{\mu}$, to the underlying values of $Y_i$s is a challenge. In summary, owing to the fluctuations of $Y_i$ for $i\ge1$, it is not clear to us how to derive a closed form solution to the problem.

\item
The error rates, $e_i$s, for the signal can also be different from the error rates $e_i$s for the decoy state, due to underlying statistical fluctuations. Actually, the fluctuation of $e_1$ appears to be the dominant source of errors in the estimation process. (See, for example, Table \ref{Tab:FlucSim}.) This is because the parameter $e_1$ is rather small (for instance, a few percent) and it appears in combination with another small parameter $Y_1$ in Eq.~\eqref{Model:QBER} for QBER.

\item
In the GLLP analysis \cite{GLLP_04} shown in Eq.~\eqref{Post:KeyRate}, Alice and Bob need to correct phase errors, other than bit-flip errors. From Shor-Preskill's proof \cite{ShorPreskill_00}, we know that the bit-flip error rate and the phase error rate are suppose to be the same only in the asymptotic limit. Therefore, for a finite data set, one has to consider statistical fluctuations. This problem is well studied \cite{ShorPreskill_00}. Since the number of signal states is generally very large, we will ignore this fluctuation from now on. 
\end{enumerate}

Qualitatively, the yields of the signal and decoy states tend to decrease exponentially with distance. Therefore, statistical fluctuations tend to become more and more important as the transmission distance of QKD increases. In general, as the distance of QKD increases, an increasingly larger data size will be needed for the reliable estimation of $Y_1$ and $e_1$ (and hence $R$), thus requiring a longer QKD experiment.

Here, we will neglect the fluctuations due to the first two and the fifth sources listed above. Even though we cannot find any closed form solution for the third and fourth sources of fluctuations, it should be possible to tackle the problem by simulations. Here, we are content with a more elementary analysis. We will simply apply a standard error analysis to perform a rough estimation on the effects of fluctuations due to the third and fourth sources. Note that the
origin of the problem is strictly classical statistical fluctuations. There is nothing quantum in this statistical analysis. While standard error analysis (using essentially normal distributions) may not give a completely correct answer, we expect that it is correct at least in the order of magnitude.

Our estimation, which will be presented below, shows that for a long-distance ($>100$ km) QKD with our Vacuum+Weak decoy state protocol, the statistical fluctuations effect (from the third and fourth sources only) appears to be manageable. This is so, provided that a QKD experiment is run for a reasonable period of time of only a few hours. Our analysis supports the viewpoint that our Vacuum+Weak decoy state protocol is practical for real-life implementations.

We remark on passing, that the actual classical memory space requirement for Alice and Bob is rather modest ($<1 GBytes$) because at long distances, only a small fraction of the signals will give rise to detection events.

We emphasize that we have not fully solved the statistical fluctuation problem for the decoy state QKD. This problem has turned out to be quite complex.
There is other work beinig done to address the statistical fluctuation problem in the decoy state QKD \cite{Wang_05,HHHT_fluc_07}.

\subsection{Standard Error Analysis}
In the following, we will present a general procedure for studying the statistical fluctuations (due to the third and fourth sources noted in the previous subsection) by using the standard error analysis.

Denote the number of pulses (sent by Alice) for signal as $N_{s}$, for the vacuum decoy state as $N_{vac}$ and for the weak decoy state as $N_{w}$. Then, the total number of pulses sent by Alice is given by:
\begin{equation}\label{Stab:NTotal}
\begin{aligned}
N &= N_{s}+N_{vac}+N_{w}.\\
\end{aligned}
\end{equation}
Following that, the parameter $q$ in Eq.~\eqref{Post:KeyRate} is given by:
\begin{equation}\label{Stab:q}
\begin{aligned}
q &= \frac{N_{s}}{2N}. \\
\end{aligned}
\end{equation}
Here, we assume that Alice and Bob perform standard BB84, so there is a factor of 1/2.

In practice, since $N$ is finite, the statistical fluctuations of $Q_1$ and $e_1$ cannot be neglected. All these additional deviations will be related to data sizes $N_{s}$, $N_{vac}$ and $N_{w}$ and in principle, can be obtained from statistic analysis. A natural question prompted by such is as follows. Given the total data size $N=const$, how do we distribute it to $N_{s}$, $N_{vac}$ and $N_{w}$ for maximizing the key generation rate $R$? This question also relates to another one: how do we choose an optimal weak decoy $\nu$ to give a good lower bound of $R$?

In principle, our optimization procedure should look like the following. First, one needs to derive a lower bound of $Q_1$ and an upper bound of $e_1$ (as functions of data size $N_{s}$, $N_{vac}$, $N_{w}$ and $\nu$), taking into account statistical fluctuations. Secondly, one substitutes these bounds into Eq.~\eqref{Post:KeyRate} to calculate the lower bound of the key generation rate, denoted by $R^L$.
Thus, the key rate lower bound $R^L$ is a function of $N_{s}$, $N_{vac}$, $N_{w}$ and $\nu$, and will be maximized when the optimal distribution satisfies
\begin{equation} \label{Stab:RDevN}
\begin{aligned}
\frac{\partial R^L}{\partial N_{s}}
 = \frac{\partial R^L}{\partial N_{vac}} = \frac{\partial
R^L}{\partial N_{w}}
= 0,
\end{aligned}
\end{equation}
given that $N=N_{s}+N_{vac}+N_{w}=const$.

In this statistical fluctuation analysis, our assumptions are as follows:
\begin{enumerate}
\item
Alice knows the exact value of the average photon pair number $\mu$ and $\nu$, which is a fixed number during key transmission.

\item
The distribution of the photon number, Eq.~\eqref{Model:Poisson}, does not fluctuate.

\item
Assume that the QKD transmission is part of an infinite length experiment. Hence, $Q_\mu$ ($E_\mu$) can be regarded as a tested value of the true gain (QBER). Thus, we can use the standard error analysis to address statistical fluctuations.
\end{enumerate}

\subsection{Choice of $N_{s}$, $N_{vac}$, $N_{w}$ and $\nu$}
From the theoretical deviations of $Y_1$ and $e_1$, shown in Figure \ref{Fig:PracDevnu}, reducing $\nu$ may decrease the theoretical deviations.
On the other hand, given a fixed $N_{w}$, reducing $\nu$ will decrease the number of detection events of the decoy states, which in turn, causes a larger statistical fluctuation. Thus, for fixed $N_{s}$, $N_{vac}$ and $N_{w}$, there exists an optimal choice of $\nu$ which maximizes the lower bound of the key generation rate $R^L$:
$$
\frac{\partial R^L}{\partial \nu} = 0
$$
which can be simplified to:
\begin{equation}\label{Stab:Optnu}
\begin{aligned}
\frac{\partial}{\partial\nu}\{\hat{Y}_1^{L}[1-H_2(\hat{e}_1^{U})]\} &=0 \\
\end{aligned}
\end{equation}
where $\hat{Y}_1^{L}$ and $\hat{e}_1^{U}$ are lower bound to $Y_1$ and upper bound to $e_1$ when statistical fluctuations are considered. 

As defined in Eq.~\eqref{Stab:q}, choosing a larger $N_{s}$ leads to a larger factor $q$ in Eq.~\eqref{Post:KeyRate}. On the other hand, choosing large values of $N_{vac}$ and $N_{w}$ can help with better estimations of $Y_1$ and $e_1$. Thus, there is trade-off between $N_{s}$, $N_{vac}$ and $N_{w}$. In order to achieve an optimal $R$, one needs to choose an appropriate set of values $N_{s}$, $N_{vac}$, $N_{w}$ and $\nu$. Given the total data size in Eq.~\eqref{Stab:NTotal}, in principle, one can solve Eqs.~\eqref{Stab:RDevN} and \eqref{Stab:Optnu} to get $N_{s}$, $N_{vac}$, $N_{w}$ and $\nu$. In the later simulation, we will numerically optimize these four parameters.

\section{Simulation}
In practice, solving Eq.~\eqref{Stab:RDevN} is a complicated problem. One problem that we have mentioned in Section \ref{Sub:HardStat} is that the relations between $N_{s}$, $N_{vac}$, $N_{w}$ and estimations of $Q_1$ and $e_1$ are difficult to describe strictly. In the following, we will be content with a rough estimation procedure using the standard error analysis. We will focus the Vacuum+Weak decoy method.

One observation is that Alice and Bob should compare all their detection events of decoy states publicly. In principle, they can also use decoy states to generate the final key. Note that the signal state is chosen to be optimal for key rate generation. In other words, decoy states are not as efficient as signal states to generate the final key. Therefore, it is more efficient for Alice and Bob to use decoy states only for estimations of $Y_1$ and $e_1$.

\textbf{Two assumptions:}
\begin{enumerate}

\item
We assume that the decoy state used in the actual experiment is conceptually only a part of an infinite population of decoy states. There are underlying values for $Q_{\nu}$ and $E_{\nu}$ as defined by the population of decoy states. In each realization, the decoy state allows us to obtain some estimates for those underlying $Q_{\nu}$ and $E_{\nu}$ . Alice and Bob can use the fluctuations of $Q_{\nu}$, $E_{\nu}$ to calculate the fluctuation of the estimates of $Y_1$
and $e_1$.


\item
When the number of events (e.g. the total detection event of the vacuum decoy state) is large
(for instance, $>50$), we assume that the statistical characteristic of a parameter can be described by a {\it normal} distribution.
\end{enumerate}


We will use the experiment parameters in Table \ref{Tab:GYSpara}, and show numerical solutions of Eqs.~\eqref{Stab:NTotal}, \eqref{Stab:RDevN} and \eqref{Stab:Optnu}. We pick the total data size (the number pulses sent by Alice) to be $N=6\times10^9$. 
The GYS experiment \cite{GYS_04} has a repetition rate of $2$ MHz and an uptime of around $50\%$\footnote{Z.~L.~Yuan, private communication.}. Therefore, it should take only a few hours to perform our proposed experiment. The optimal $\mu=0.48$ can be calculated by Eq.~\eqref{Optmu:Decoymu} and we use $f(e)=1.22$.

In a fiber length of 
$103.6$ km ($\eta=3\times10^{-4}$), the optimal weak decoy state intensity $\nu$, pulses distribution of data, and the deviations from the infinite decoy method
are listed in Table \ref{Tab:FlucSim}.
\begin{table}[h]\center
\begin{tabular}{|c|c|c|c|c|c|c|c|c|c|c|}
\hline
$l$ & $\mu$ & $u$ & $N$ & $N_{s}$ & $N_{vac}$ & $N_{w}$ \\
\hline
$103.62$ km & $0.479$ & $10$ & $6\times10^{9}$ & $3.98\times10^{9}$ & $1.76\times10^{9}$ & $2.52\times10^{8}$ \\
\hline
$\eta$ & $\nu$ & $\tilde{B}[bits]$ & $\beta_{Y0}$ & $\beta_{Y1}$ & $\beta_{e1}$ & $\beta_{R}$  \\
\hline
$3\times10^{-4}$ & $0.127$ & $2.17\times10^{4}$ & $48.31\%$ & $7.09\%$ & $97.61\%$ & $74.11\%$  \\
\hline
\end{tabular}
\caption{List of the optimal choice of $\nu$ and pulse number distribution for the Vacuum+Weak decoy state protocol with statistical fluctuation analysis. The pulse number distribution, $N_{s}$, $N_{vac}$ and $N_{w}$, is calculated by Eq.~\eqref{Stab:RDevN}. The optimal weak decoy state intensity is calculated by Eq.~\eqref{Stab:Optnu}. $\tilde{B}$ is the lower bound of the number of the final key bits. All results are obtained by numerical programming using MatLab. The variable $\beta_{Y1}$ denotes the relative deviation in our estimation process of $Y_1$ from its true value by using the data from a finite experiment. This relative deviation originates from finite data with statistical fluctuations. This definition contrasts with the definition of $\beta_{Y1}$ in Eq.~\eqref{Practical:Y1e1Bdev} which refers to the relative difference between the values of $Y_1$ for case i) where $\nu$ is finite and case ii) where $\nu$ approaches zero. Similarly, other $\beta$s denote the relative deviations in our estimates for the corresponding variables in the subscript of $\beta$. We assume that all the statistical fluctuation belongs to the confidence interval of $u=10$ standard deviations (i.e., $1-1.5\times10^{-23}$).  The experiment parameters are listed in Table~\ref{Tab:GYSpara}.}
\label{Tab:FlucSim}
\end{table}

\begin{figure}[hbt]
\centering \resizebox{12cm}{!}{\includegraphics{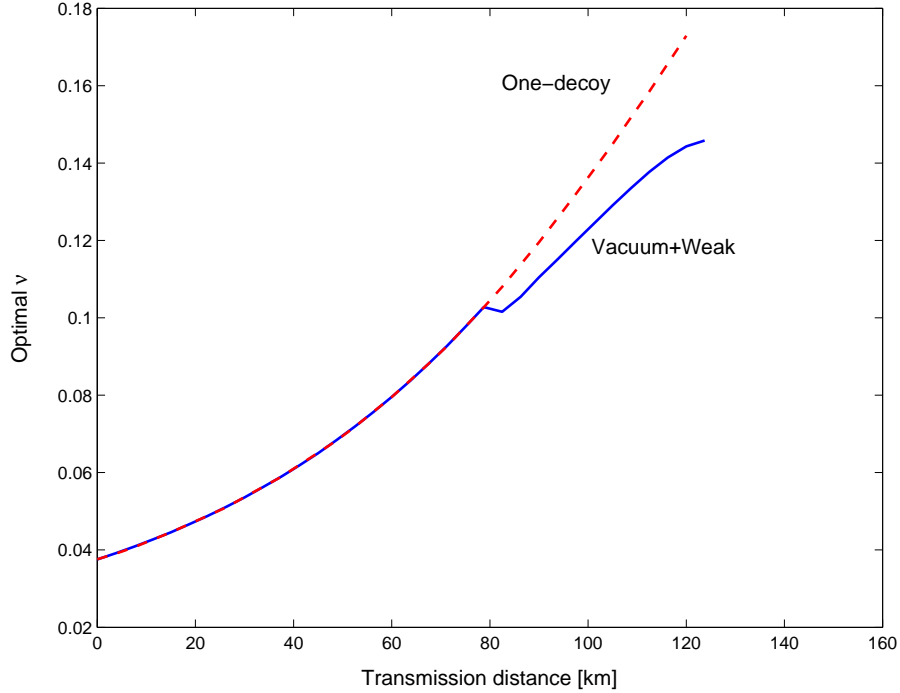}}
\caption{Plot of optimal $\nu$ versus transmission distance. The solid line shows the simulation result of the Vacuum+Weak protocol (Eqs.~\eqref{Practical:Y1L} and \eqref{Practical:e1U}) with statistical fluctuations. The dashed line shows the result for the one decoy state method
(Eq.~\eqref{Decoy:OneY1e1}). Here, we pick the data size (total number of pulses emitted by Alice) to be $N=6\times10^{9}$. We find the optimal $\nu$s for each fiber length by numerically solving
Eqs.~\eqref{Stab:NTotal}, \eqref{Stab:RDevN} and \eqref{Stab:Optnu}. The confidence interval for statistical fluctuation is 10 standard deviations (i.e., $1-1.5\times10^{-23}$). The simulation parameters are listed in Table~\ref{Tab:GYSpara}. The expected photon number of signal state $\mu=0.48$ is calculated by Eq.~\eqref{Optmu:Decoymu}.
} \label{Fig:DevNudis}
\end{figure}

For any fiber length, we can solve Eqs.~\eqref{Stab:RDevN} and \eqref{Stab:Optnu} to get $N_{s}$, $N_E$, $N_{vac}$, $N_{w}$ and $\nu$. Figure~\ref{Fig:DevNudis} shows how the optimal $\nu$ changes with transmission distance.

We have a few remarks on Figure~\ref{Fig:DevNudis}, optimal $\nu$ versus transmission distance.
\begin{enumerate}
\item
The optimal $\nu$ is small ($\sim0.1<\mu$) through the whole distance. In fact, it starts at a value $\nu\approx 0.04$ at zero distance and increases with the transmission distance.

\item
There is small flat step at distance of 82 km. This is due to the fact that the vacuum decoy state becomes useful. From 0 km to 82 km transmission distance regime, the optimal pulse number for the vacuum decoy state $N_{vac}$ is 0. That is, in this regime, one should use the one decoy state protocol instead of the Vacuum+Weak protocol\footnote{Actually, we did this simulation first and found this strange behavior at a distance of 82 km. Then we came up with the one decoy state protocol.} protocol.

\item
As the transmission distance increases, the optimal $\nu$ increases. This is reasonable because in a longer distance, the total transmittance $\eta$ is low, thus Alice and Bob need to put more pulses for decoy states and choose a larger $\nu$ to estimate $Y_1$ and $e_1$ accurately.
\end{enumerate}

\begin{figure}[hbt]
\centering \resizebox{12cm}{!}{\includegraphics{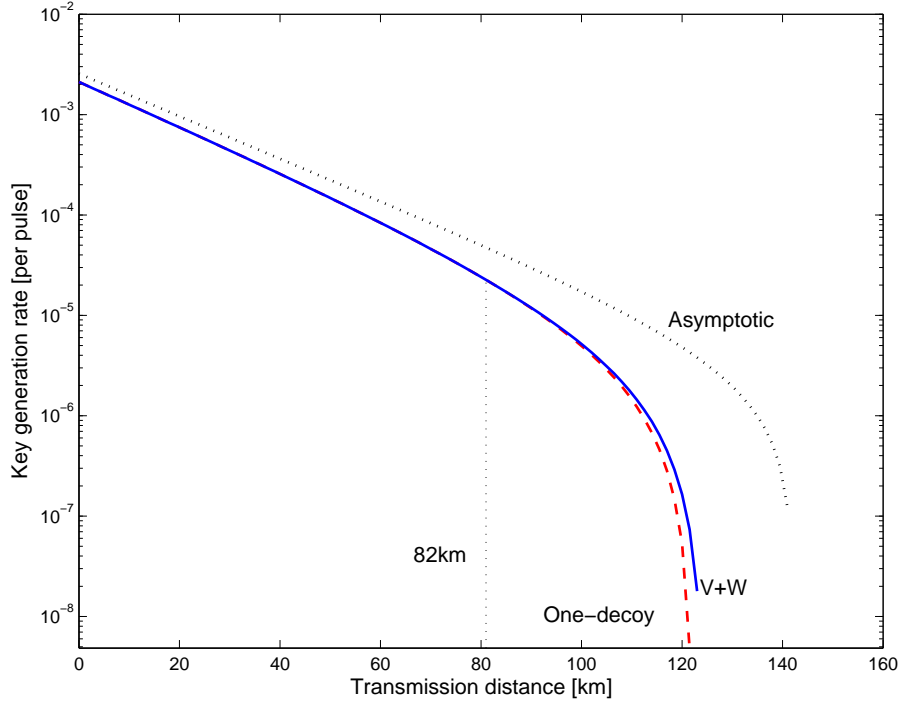}}
\caption{Plot of key generation rate in terms of channel transmission distance. The dotted line shows the key rate of the infinite decoy state method (the asymptotic case of the Vacuum+Weak decoy state protocol). The solid and dashed lines show the key rate of the Vacuum+Weak and one decoy state protocol with statistical fluctuations respectively. The data size is $N= 6 \times 10^9$. The simulation parameters are listed in Table~\ref{Tab:GYSpara}. The expected photon number of signal state $\mu=0.48$ is calculated by Eq.~\eqref{Optmu:Decoymu}.}
\label{Fig:RVWFluc}
\end{figure}

Now, we can put all these elements together to investigate the key generation rate $R$ of Eq.~\eqref{Post:KeyRate}. Figure~\ref{Fig:RVWFluc} shows the key rate of the two practical decoy state protocols with statistical fluctuations in comparison to the infinite decoy state protocol (the asymptotic case). For each distance point, we optimize $\nu$, $N_{s}$, $N_{vac}$ and $N_{w}$ numerically by considering Eqs.~\eqref{Stab:RDevN} and \eqref{Stab:Optnu}.

One can see that even taking into account the statistical fluctuations, both of the Vacuum+Weak and the one decoy state protocols can achieve close performance to the infinite decoy state protocol. Therefore, the following is noted:
\begin{enumerate}
\item
In a large regime of the distance (for instance, the distance between 0 km and 100 km), the two practical decoy state methods with statistical fluctuations achieve a close performance of the asymptotic limit of the infinite decoy state method. This is the case when the channel is not that lossy, the statistical fluctuations are easily controllable. This fact highlights the feasibility of the two practical decoy state protocols.

\item
As shown in Figure~\ref{Fig:DevNudis}, the vacuum decoy state becomes useful at 82 km.

\item
The maximal secure distances of the three curves are 142 km, 125 km and 122 km. Note that with a larger data size, for instance, $N=8.4\times10^{10}$, the maximal secure distance of the Vacuum+Weak decoy state method can achieve 132 km.
\end{enumerate}

We have also simulated other experiment setups and all the results are consistent with the simulation result of the GYS experiment setup shown above. For more details, one can refer to Refs.~\cite{Practical_05,Low_06}.

\section{Experimental demonstrations} \label{Sc:DecoyExp}
The experimental demonstrations for the decoy state methods were first implemented by our group \cite{ZQMKQ_06,ZQMKQ60km_06} and followed by many other groups \cite{LosAlamosDecoy_07,PDC144_07,PanDecoy_07,YSS_Decoy_07,Guo_Decoy_07}.

\subsection{How to generate decoy states}
The only difference of the decoy state QKD setup and the regular setup is that in the decoy state method, Alice needs to prepare decoy states, which have different intensities from the original signal states. Otherwise, the two setups are the same. The regular setup of the QKD without decoy states is discussed in Section \ref{Sc:Hardware}.

There are several ways to generate decoy states. One way to do that is by using an attenuator to change the light intensity. There are two criteria for the attenuator.
\begin{itemize}
\item
The attenuator can change attenuations fast enough\footnote{Or it can switch on and off fast.}. Alice needs to prepare a decoy or signal state randomly in each pulse. Thus, the speed of changing attenuation should not be lower than the QKD repetition rate.

\item
The attenuator will not introduce differences in properties for change of signal and decoy states except for intensities. This is one precondition for the security of QKD with decoy states, as shown in Eq.~\eqref{Decoy:DecoyAss}. In a real experiment, one might need to apply some approximation. For example, an acousto-optic modulator (AOM) may shift the frequency of light. However, if we assume that both signal and decoy states will be shifted with the same amount of frequency, then we can still use AOM to prepare signal and decoy states.
\end{itemize}
For more discussions of using AOM to prepare decoy states, one can refer to Ref.~\cite{ZQMKQ_06}.

Another way to prepare decoy states is by using different laser sources \cite{PanDecoy_07}. In this case, Alice can choose signal and decoy states by switching between different laser sources. Similarly, we require the switch to be fast enough and laser sources having the same properties except for intensities.

\subsection{Experimental data post-processing} \label{Sub:Prac:Process}
The processing of the decoy state QKD is as follows. 

\begin{enumerate}
\item
Alice prepares decoy and signal states and sends them to Bob. Bob measures all pulses in the two conjugate bases.

\item
Bob announces the pulses that he obtains non-vacuum detections. Alice announces the pulses that are used for decoy states. Then they determine all the gains of signal and decoy states.

\item
They perform basis reconciliation. Note that even these detection events that Alice and Bob use different bases, can be used to calculate the gains of signal and decoy states.

\item
They compare all bit values decoy states to determine the QBER(s) of decoy states.

\item
Alice and Bob perform error correction and error testing, after which they can determine the QBER of signal states.

\item
They estimate the necessary amount of privacy amplification. Taking the Vacuum+Weak decoy state protocol for example, they estimate $Y_1$ and $e_1$ by values of $Q_\mu$, $E_\mu$, $Q_\nu$, $E_\nu$ and $Q_{vac}$. In this step, they need to consider statistical fluctuations, for instance, by the procedures described in Section \ref{Sc:Stat}. Then they can plug all the values in Eq.~\eqref{Post:KeyRate} to calculate the amount of key that is needed to sacrifice for privacy amplification. Note that Eq.~\eqref{Post:KeyRate} is for the post-processing with one-way classical communication. In the next chapter, we will show that this result can be improved by introducing two-way classical communication.

\item
They perform privacy amplification to get the final secure key.
\end{enumerate}

Here, we describe the case where the QKD transmission is successful. In practice, Alice and Bob can keep tracking whether the final key is positive or not to determine whether they should continue the post-processing or not. For example, after step 2, they can estimate $Y_1$. If the lower bound $Y_1$ is zero (or even negative), then they abort the post-processing and start QKD again.

\section{Conclusion}
The main conclusion of Chapters \ref{Chpt:Decoy} and \ref{Chpt:Practical} is that the decoy state QKD takes a big step toward practical quantum cryptography. Recall that the motivation of this thesis is to encourage QKD into real-life applications.

Our result shows that we can have the best of both worlds: enjoy both unconditional security and record-breaking experimental performance. The decoy state method can increase key generation rate and extend the distance of QKD dramatically, all within the framework of unconditional security. The general principle of the decoy state QKD developed here can have widespread applications in other set-ups (e.g. open-air QKD or QKD with other photon sources). Later, we will come back to this point.

For practical implementations, we are able to show that with only one or two decoy states, one can achieve most of the benefits of the decoy state method. All the decoy state QKD experiment demonstrations, including our first realization, show that the decoy state idea is easy to implement in real system setups.

Recently, Yuan, Sharpe and Shields implemented an experimental decoy state QKD demonstration that can achieve a 5.51 kbits/s secure key rate through a 25.3 km fiber \cite{YSS_Decoy_07}. Let us compare this result to a couple of typical values in real-life communications. The state of the art digital speech coding \cite{Wireless_Rappaport_02} typically needs a bit rate around 4-10 kbits/sec. A typical city wide area network must cover an area with a radius of 5-25 km. As for other communications, such as video conversation, the bit rate may not be high enough.
We want to point out that the bit rate might not be an essential problem. One can store a long secure key first and then use it for secure communications\footnote{One needs to consider the key management issue in this case.}.

Therefore, we conclude that the practical quantum cryptography is close to real-life applications.

Note that other than the decoy state method, there are other approaches to enhance the performance of the coherent state QKD, such as our dual detector scheme \cite{QZMLQ_dual_07}, QKD with strong reference pulses \cite{Koashi_StrRef_04,TLMB_Strong06} and differential-phase-shift QKD \cite{IWY_DPS02}.

\chapter{Decoy state QKD with 2-LOCC} \label{Chpt:TwoWay}
As shown in the previous two chapters, the decoy state technique is an effective method for improving QKD performance. The data post-processing scheme of the decoy state QKD scheme that we proposed uses one-way classical communication. In this chapter, we develop two data post-processing schemes for the decoy state method using two-way classical communication.
Our numerical simulation results show that the first scheme is able to extend the maximal secure distance from 142 km (by using only one-way classical communication with decoy states) to 181 km.
The second scheme is able to achieve a 10\% greater key generation rate in the whole regime of the distance.
We conclude that the decoy state QKD protocol with two-way classical post-processing is of practical interest.

Here, we only consider a case without statistical fluctuations. For a statistical fluctuation analysis for the decoy state QKD with local operations and two-way classical communication (2-LOCC), one can refer to Ref.~\cite{TwoWay_06}.

This work is published in Ref.~\cite{TwoWay_06}. In this project, I applied the Gottesman-Lo's 2-LOCC EDP and recurrence scheme to the decoy state QKD protocol and simulated a PDC experiment to show the improvement by using two-way classical communication in the decoy state QKD protocol.

\section{2-LOCC EDP}
First, let us review two EDPs based on 2-LOCC (Gottesman-Lo EDP and recurrence EDP) assuming that ideal single-photon (or perfect EPR) sources are used. Later, we will apply these two schemes to the decoy state QKD protocol.

\subsection{Gottesman-Lo EDP} \label{Sub:BP}
\begin{figure}[hbt]
\centering \resizebox{12cm}{!}{\includegraphics{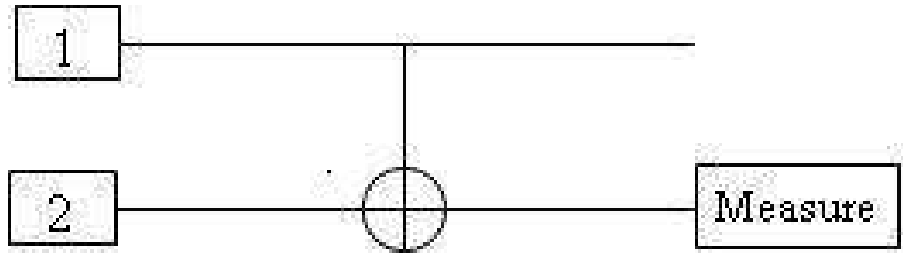}}
\caption{Alice and Bob each choose two qubits of two corresponding EPR pairs and input the quantum circuit as shown above. They discard both control and target qubits if they disagree on the outcomes of measurement on the target qubits. On the other hand, they keep the control qubits as surviving qubits if their measurement outcomes agree.
}%
\label{Fig:BStepCircuit}
\end{figure}

Gottesman and Lo \cite{TwoWay_03} introduced an EDP based on 2-LOCC for use with QKD and showed that it can tolerate a higher bit error rate than 1-LOCC based EDPs. B and P steps are two primitives in the Gottesman-Lo EDP, and the EDP consists of executing a sequence of B and/or P steps, followed by a 1-LOCC EDP. The main objective for extra B and P steps is reducing the bit and/or phase error rates of qubits so that the following 1-LOCC EDP can work to extract secure keys. This is the reason why the Gottesman-Lo EDP is able to tolerate a higher initial bit error rate than 1-LOCC EDPs. The definitions of B and P steps are as follows:

\noindent\textbf{Definition of B step \cite{TwoWay_03}:} (Figure \ref{Fig:BStepCircuit}) After randomly permuting all the EPR pairs, Alice and Bob perform a bilateral XOR (BXOR) between pairs of the shared EPR pairs and measure the target qubits in $Z$ basis. This effectively measures the operator $Z\bigotimes Z$ by Alice and Bob locally, and detects the presence of a single bit flip error. If Alice and Bob's measurement outcomes disagree, they discard the remaining EPR pair. Otherwise, they keep the control qubit.

Since the B step only involves the measurement of $Z\bigotimes Z$, it can be used in the prepare-and-measure protocol, BB84. Classically, the B step simply involves random pairing of the key bits, for instance, $x_1, x_2$ on Alice's side and $y_1, y_2$ on Bob's side and the computation of the parity of each pair of bits, $x_1 \oplus x_2$ and $y_1 \oplus y_2$. Both Alice and Bob announce the parities. If their parities are the same, they keep $x_1$ and $y_1$; otherwise, they discard $x_1$, $x_2$, $y_1$ and $y_2$. We can see that the B step is very simple to implement in data post-processing.

Suppose Alice and Bob input a control qubit $(q_{00}^C, q_{10}^C, q_{11}^C, q_{01}^C)$\footnote{The superscript $C$ and $T$ stand for the control and target qubits, respectively. The subscript $00$, $10$, $11$ and $01$ stand for the case with no error, with a bit error, with both a bit and a phase error, and with a phase error, respectively.}
and a target qubit $(q_{00}^T, q_{10}^T, q_{11}^T, q_{01}^T)$ with bit error rates $\delta_b^C$ and $\delta_p^C$ and phase error rates $\delta_b^T$ and $\delta_p^T$, respectively. After one B step, the survival probability $p_{S}$ is given by:
\begin{equation}\label{oneBpS}
\begin{aligned}
p_{S} &= (q_{00}^C+q_{01}^C)(q_{00}^T+q_{01}^T) + (q_{10}^C+q_{11}^C)(q_{10}^T+q_{11}^T)\\
      &= (1-\delta_b^C)(1-\delta_b^T) + \delta_b^C\delta_b^T,
\end{aligned}
\end{equation}
and the density matrix $(q_{00}', q_{10}', q_{11}', q_{01}')$ of
output control qubit is given by:
\begin{equation}\label{Twoway:oneBmatrix}
\begin{aligned}
q_{00}' &= \frac{q_{00}^Cq_{00}^T + q_{01}^Cq_{01}^T}{p_{S}}\\
q_{10}' &= \frac{q_{10}^Cq_{10}^T + q_{11}^Cq_{11}^T}{p_{S}}\\
q_{11}' &= \frac{q_{10}^Cq_{11}^T + q_{11}^Cq_{10}^T}{p_{S}}\\
q_{01}' &= \frac{q_{00}^Cq_{01}^T + q_{01}^Cq_{00}^T}{p_{S}}.
\end{aligned}
\end{equation}
Eqs.~\eqref{Twoway:oneBmatrix} can be derived from Table II of \cite{BDSW_96}. The corresponding bit error rate $\delta_b$ and phase error rate $\delta_p$ can be obtained from Eq.~\eqref{Twoway:oneBmatrix} by
\begin{equation}\label{Twoway:AfterBstepErr}
\begin{aligned}
\delta_b' &= q_{10}'+q_{11}' = \frac{\delta_b^C\delta_b^T}{p_S}\\
\delta_p' &= q_{11}'+q_{01}'.\\
\end{aligned}
\end{equation}

\begin{figure}[hbt]
\centering \resizebox{12cm}{!}{\includegraphics{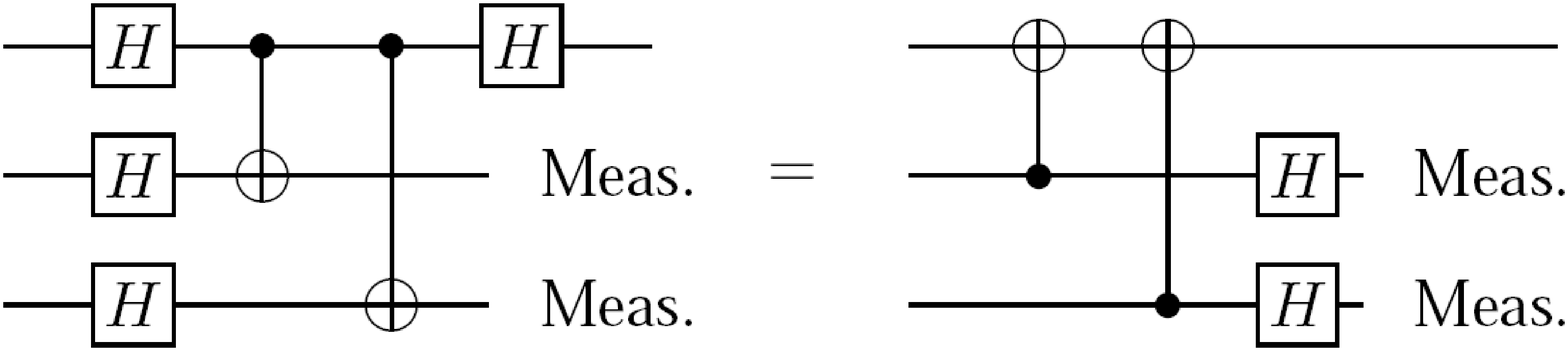}}
\caption{The two circuits are quantum mechanically equivalent. Alice and Bob each choose three qubits of three corresponding EPR pairs and input the quantum circuit as shown above. This figure is originally from Ref.~\cite{TwoWay_03}.
}%
\label{Fig:PStepCircuit}
\end{figure}

\noindent\textbf{Definition of P step \cite{TwoWay_03}:} (Figure \ref{Fig:PStepCircuit}) Alice and Bob randomly permute all the EPR pairs. Afterwards, they group the EPR pairs into sets of three, and measure $X_1X_2$ and $X_1X_3$ on each set (for both Alice and Bob). This can be done (for instance) by performing a Hadamard transform, two bilateral XORs, measurement of the last two EPR pairs, and a final Hadamard transform. If Alice and Bob disagree on one measurement, Bob will conclude the phase error is probably on one of the EPR pairs which was measured, and do nothing; if both measurements disagree for Alice and Bob, Bob assumes the phase error is on the surviving EPR pair and corrects it by performing a $Z$ operation.

Without a quantum computer, Alice and Bob are not able to perform P steps by the quantum circuit shown in the left hand side of Figure \ref{Fig:PStepCircuit}.
In order to implement P steps classically, they can choose a post-processing scheme that does not rely on the measurement result from P steps. That is, they can implement the right hand side quantum circuit of Figure \ref{Fig:PStepCircuit} by simply omitting the measurement step. Thus, when a P step is implemented classically in BB84\footnote{Strictly speaking, this procedure is different from the original P step we described. For simplicity, we use the same name for this simplified version of the P step.}, the phase errors are not detected or corrected (i.e. the phase flip operation $Z$ is not applied). Note that the measurement step in Figure \ref{Fig:PStepCircuit} is not important because the phase errors do not need to be corrected in QKD \cite{ShorPreskill_00}. The phase error correction is used in the security proof. One only needs to show that Alice and Bob could have done the phase error correction but not really need to do it.
From this point of view, P steps are conceptually similar to the privacy amplification.
 
The P step then will be reduced to where Alice and Bob randomly form
trios of the remaining qubits and compute the parity of each trio,
for instance, $x_1 \oplus x_2 \oplus x_3$ by Alice and $y_1 \oplus y_2 \oplus
y_3$ by Bob. They now regard those parities as their new bits for
further processing.

Since before P steps, Alice and Bob will perform random permutation, for
simplicity, we assume the input three qubits have the same density
matrix: $(q_{00}, q_{10}, q_{11}, q_{01})$. After one P step, the
density matrix $(q_{00}', q_{10}', q_{11}', q_{01}')$ of the output
qubit is given by:
\begin{equation}\label{Twoway:onePmatrix}
\begin{aligned}
q_{00}' &= q_{00}^3+3q_{00}^2q_{01}+3q_{10}^2(q_{00}+q_{01})+6q_{00}q_{10}q_{11}\\
q_{10}' &= q_{10}^3+3q_{10}^2q_{11}+3q_{00}^2(q_{10}+q_{11})+6q_{00}q_{10}q_{01}\\
q_{11}' &= q_{11}^3+3q_{10}q_{11}^2+3q_{01}^2(q_{10}+q_{11})+6q_{00}q_{11}q_{01}\\
q_{01}' &= q_{01}^3+3q_{00}q_{01}^2+3q_{11}^2(q_{00}+q_{01})+6q_{10}q_{11}q_{01},\\
\end{aligned}
\end{equation}
which is given in Appendix C of \cite{TwoWay_03}. Therefore, the bit error rate and phase error rate will be given by:
\begin{equation}\label{Twoway:AfterPstepErr}
\begin{aligned}
\delta_b' &= q_{10}'+q_{11}' = 3\delta_b(1-\delta_b)^2+\delta_b^3 \\
\delta_p' &= q_{11}'+q_{01}' = 3\delta_p^2(1-\delta_p)+\delta_p^3. \\
\end{aligned}
\end{equation}

Here, we emphasize that the B and P steps are important elements of
the Gottesman-Lo EDP. After the B and P steps, the Gottesman-Lo EDP will
be the same as the regular 1-LOCC EDP.

\subsection{Recurrence EDP scheme} \label{Rec:Subsection:Rec}
Here, we review another two-way EDP, the recurrence scheme \cite{VV_Recurrence_05}. Similar to the B step in the Gottesman-Lo EDP, the recurrence scheme reduces the bit error rate of the EPR pairs before passing them to the 1-LOCC based EDP for the distillation of maximally-entangled EPR pairs. However, there are two main differences between these two EDP schemes. The first is how the bit error syndrome of a target EPR pair in a bilateral XOR operation is learned. In the Gottesman-Lo EDP, Alice and Bob simply measure the target EPR pair in the $Z$ basis and compare their results to learn about the bit error syndrome
(see Figure \ref{Fig:BStepCircuit}). In the recurrence scheme, Alice and Bob group the bit error syndromes of all target EPR pairs together and learn about all the syndromes using random hashing.
The second difference is that the recurrence scheme requires some extra maximally-entangled EPR pairs to begin with in order to learn about the bit error syndromes, whereas no such extra pairs are required in the Gottesman-Lo EDP. Note that the recurrence methods were studied in various papers, such as \cite{DEJMPS_96,Recurrence_MS_00,Recurrence_ADGJ_01,Reccurence_DNMV_03}.

The procedure of the recurrence protocol is described as follows:
\begin{enumerate}
\item
Alice and Bob perform two BXOR operations on two noisy EPR pairs and one
perfect maximally-entangled EPR pair.
Specifically, the first BXOR is performed on one noisy EPR pair as the source and the perfect EPR pair as the target,
and the second BXOR is performed using the other noisy EPR pair as the source and the same target.

\item
They perform random hashing on the target EPR pairs to learn about the parities
of the noisy EPR pairs. Note that only a portion of the target EPR pairs
have to be measured in order to learn about all the parities. This is
different from the B step approach.

\item
They perform error correction and privacy amplification separately for
even-parity and odd-parity EPR pairs.

\end{enumerate}

In the prepare-and-measure scenario, the first two steps are as follows:
Alice and Bob randomly pair up the key bits,
and for each pair they compute the parity.
They each compress their own sequence of parities by using random hashing,
encrypt the resulting hash values using the one-time pad with some pre-shared secret bits,
and send the encrypted results to each other.
Note that they use the same sequence of secret bits to encrypt their own sequence of hash values.
They learn about the parities of the original noisy EPR pairs by adding the other party's encrypted sequence to their own encrypted sequence of hash values.
Once they know the parities,
they perform
error correction and privacy amplification 
on the even-parity and odd-parity key bits separately.
Note that the secret bits used up in the process should be returned to the secret bits pool by using the newly generated secret bits.


The key generation rate using the recurrence EDP with a single-photon source is
given by:
\begin{equation} \label{Rec:Residue0a}
\begin{aligned}
R = q \left[ -\frac12H_2(p_S)
-\frac12p_SH_2(\frac{\delta_b^C\delta_b^T}{p_S}) + K \right]
\end{aligned}
\end{equation}
where
$q$ is defined similarly as in Eq.\eqref{EDP:CSSrate},
$p_S$ is the probability of obtaining even parity given in Eq.~\eqref{Rec:Survival}, and
$\delta_b^C$($\delta_b^T$) is the bit error rate of the control (target) EPR pair.
Here, the first term in the bracket corresponds to the extra perfect EPR pairs borrowed,
the second term corresponds to error correction, and the third term $K$ corresponds to the privacy amplification given in Eq.~\eqref{Rec:PriRes2}. In Appendix~\ref{App:review-recurrence}, we review the recurrence EDP in detail and develop a key rate formula.

\subsection{Bounds of error rates} \label{Boundary}
Here, we will consider the bounds of error rates (bit error rate
$\delta_b$ and phase error rate $\delta_p$), assuming a laser source that emits a basis-dependent single-photon source. The upper bounds can be
derived by considering some simple attacks (such as intercept-resend
attack) and determining the QBER caused by these attacks.
The lower bounds can be determined by the unconditional security
proof assuming that Eve is performing arbitrary attacks allowed by
the law of quantum mechanics, and Alice and Bob employ a certain post-processing scheme (such as Gottesman-Lo EDP described in
Subsection \ref{Sub:BP}). One lower bound, obtained by considering
Gottesman-Lo EDP, is $18.9\%$ \cite{TwoWay_03}. For BB84, an upper bound,
obtained by considering an intercept-resend attack, is $25\%$.

Here, we consider the lower bound in a general setting where the
error rates are characterized by $(\delta_b,\delta_p)$. In general,
the bit error rate $\delta_b$ can be measured by error testing, but
the phase error rate $\delta_p$ cannot be directly observed from the
QKD experiment. In order to guarantee the security, Alice and Bob
have to bound $\delta_p$ with the knowledge of $\delta_b$. For BB84 with an ideal single-photon source,
due to the symmetry between the $X$ and $Z$ bases, one can show that the
bit error rate and the phase error rate are the same, i.e.
\begin{equation} \label{Twoway:b=p}
\delta_b=\delta_p.
\end{equation}
In general, for other protocols or with non-ideal sources (including
coherent state sources), the bit and phase error rates might be different. For
example, even for BB84, when a basis-dependent source is used,
Eq.~\eqref{Twoway:b=p} may not hold. In this case, according to
Eq.~(9) of \cite{Koashi_05}, due to the concavity of the right hand
side of the equation, it is not difficult to show (see Appendix
\ref{Koashi}) that $\delta_b$ and $\delta_p$ have the relation of
\begin{equation} \label{Bounds:bp}
\sqrt{F}\le\sqrt{(1-\delta_b)(1-\delta_p)}+\sqrt{\delta_b \delta_p},
\end{equation}
where $F$ is the fidelity between the two states with two bases ($X$
and $Z$) sent by Alice, and it is the single parameter that
characterizes the basis dependency of the source. Thus, Alice and
Bob can upper bound $\delta_p$ (denoted as $\delta_p^u$) with this
inequality given $\delta_b$. Clearly, when $\delta_p=\delta_b$, the
inequality will be always satisfied, i.e., $\delta_p=\delta_b$ is a
particular solution of Eq.~\eqref{Bounds:bp}.
Therefore, in general, we have $\delta_p^u\ge\delta_p$. In the
following, we use $\delta_p$ to denote the upper bound $\delta_p^u$
for simplicity.



\begin{figure}[hbt]
\centering \resizebox{12cm}{!}{\includegraphics{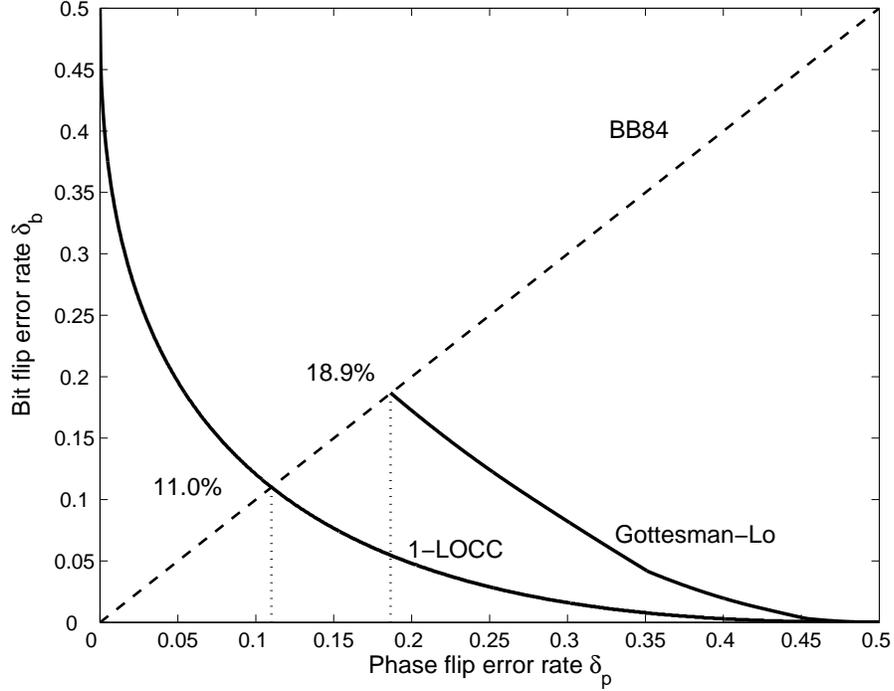}}
\caption{Plot of the secure regions in terms of error rates for the
1-LOCC EDP and Gottesman-Lo EDP. The regions under the solid lines are
proven to be secure due to 1-LOCC EDP, and Gottesman-Lo EDP schemes
(for the region under the solid line and dashed line), respectively.
For 1-LOCC EDP, we use Eq.~\eqref{EDP:CSSrate}. For Gottesman-Lo
EDP, we use Eqs.~\eqref{Twoway:oneBmatrix} and
\eqref{Twoway:onePmatrix}. In the Gottesman-Lo EDP, we numerically optimize the B/P sequence up to 12 steps. } \label{Twoway:Fig:EDP}
\end{figure}

Given a QKD protocol and laser source, Alice and Bob can estimate
the phase error rate $\delta_p$ from the bit error rate $\delta_b$
in accordance to the protocol and source. We investigate the highest
error rates that a data post-processing scheme can tolerate.
Figure~\ref{Twoway:Fig:EDP} shows
the tolerable error rates of the Gottesman-Lo EDP
compared to the 1-LOCC EDP scheme,
illustrating the superior performance of the Gottesman-Lo EDP
over the 1-LOCC EDP.
%
The boundaries of the error rates are found by searching through the regime of:
\begin{equation} \label{Twoway:bpcondition}
\begin{aligned}
\delta_b&\le\delta_p \\
\delta_b+\delta_p&<1/2
\end{aligned}
\end{equation}
such that positive key rates are obtained. The reason that we are
interested in the region specified by the second inequality in
Eq.~\eqref{Twoway:bpcondition} is as follows: We can assume that the
error rates $\delta_b$ and $\delta_p$ are less than $1/2$, otherwise
Alice and Bob can flip the qubits. Furthermore, if $\delta_b+\delta_p \ge
1/2$, the (worst scenario case) state shared by Alice and Bob is a separable state \cite{BDSW_96} and the Gottesman-Lo EDP cannot distill any pure EPR pairs \cite{CLL_Precondition_04}.

The input to the Gottesman-Lo EDP is $(q_{00}, q_{10}, q_{11},
q_{01})$ with $q_{00}+q_{10}+q_{11}+q_{01}=1$, see Subsection
\ref{Sub:BP}. However, Alice and Bob only know $\delta_b=q_{10}+q_{11}$ and
$\delta_p=q_{11}+q_{01}$ from their error testing. There is one free
parameter $q_{11}$. In Appendix C of \cite{TwoWay_03}, the authors
proved that $q_{11}=0$ is the worst case when $\delta_b=\delta_p$.
Following that proof, we can show that $q_{11}=0$ is the worst case
when the condition of Eq.~\eqref{Twoway:bpcondition} is satisfied.
That is, given $(\delta_b, \delta_p)$, if we check the input
$(1-\delta_b-\delta_p, \delta_b, 0, \delta_p)$ for the Gottesman-Lo EDP and obtain a positive key rate, then we can safely claim that the Gottesman-Lo EDP can tolerate the error rates of $(\delta_b,
\delta_p)$.

To determine the tolerable bit error rate of a particular protocol,
one should first obtain the relationship between the bit error rate
and phase error rate, and plot it on Figure~\ref{Twoway:Fig:EDP}.
The intersections between this curve and the boundary curves (the
1-LOCC curve and Gottesman-Lo curve) indicate the tolerable QBER
for the corresponding EDPs. For example, for the BB84 protocol with
a perfect single-photon source, we have $\delta_b=\delta_p$, which
is the dashed line plotted in Figure~\ref{Twoway:Fig:EDP}. We can
immediately read off from the figure that an initial bit error rate
of $18.9\%$ is tolerable using the Gottesman-Lo EDP \cite{TwoWay_03}, while
an error rate of $11.0\%$ is tolerable using the 1-LOCC EDP. In
general, the Gottesman-Lo EDP gives rise to higher tolerable error
rates than the 1-LOCC EDP.

We numerically optimize the B/P sequence up to 12 steps. The result is shown in Figure \ref{Twoway:Fig:EDP}. 

For protocols having constraints on $q_{11}$, such as the six-state
protocol \cite{sixstate_98} and the SARG04 protocol with a
single-photon source \cite{SARG_04,TamakiLo_06,FTL_06}, the tolerable QBER
can go beyond the boundary curves shown in Figure~\ref{Twoway:Fig:EDP}.



\section{Decoy + GLLP + Gottesman-Lo EDP} \label{Sc:BDecoy}
In this section, we propose a 2-LOCC based data post-processing protocol in a form of a sequence of B steps, followed by 1-LOCC error correction and privacy amplification. This new scheme is a generalization of the Gottesman-Lo scheme to a case of imperfect devices. The reasons for skipping P steps here are as follows. First, from the simulation in Section \ref{Boundary}, we found that P steps are not as useful as B steps. Secondly, only considering B steps can simplify the procedure of the data post-processing scheme.

The residual ratio of a post-processing scheme, $r$, is defined by:
\begin{equation}\label{BDecoy:DefResidual}
\begin{aligned}
R &= qQ_{\mu}r
\end{aligned}
\end{equation}
which characterizes the cost of the post-processing scheme.

The procedure of the data post-processing scheme, Decoy + GLLP + B steps, is as follows:
\begin{enumerate}
\item
Alice and Bob perform a sequence of B steps to the sifted key. During this procedure, they will discard a large ratio of the key. The survival key bit ratio is defined to be $\tilde{r}_{B}$.

\item
They calculate the variables (such as QBER, untagged qubits ratio) after the B steps.

\item
They perform an overall error correction, corresponding to the first term in Eq.~\eqref{BDecoy:DGBrate}.

\item
They perform privacy amplification, corresponding to the second term in Eq.~\eqref{BDecoy:DGBrate}.
\end{enumerate}
In the following, we will discuss how the residue of this post-processing scheme is calculated.

In the model described in Section \ref{Sc:Model}, there are three kinds of qubits: vacuum,
single photon and multi photon qubits. We emphasize again here that the final secure key can
only be distilled from untagged qubits (single photon qubits) for the BB84 protocol.

Since either of the two inputs of a B step has three possibilities, the outcomes of a B step then have nine possibilities. Only the case where both inputs are untagged qubits will there be a positive contribution to the final secure key. 
That is, at the end of some B steps, bit error correction and privacy amplification can be
only applied to the remaining qubits that come from the case where both inputs are untagged qubits. In other words, an output qubit after a subsequence of B steps is ``untagged" if a) it passes all B steps and b) it is generated from a case where all initial input qubits are single photon qubits. Therefore, the residue ratio of data post processing can be expressed, according to
Eq.~\eqref{Post:KeyRate}, as:
\begin{equation}\label{BDecoy:DGBrate}
\begin{aligned}
r=\tilde{r}_{B}\{-f(\tilde{\delta})H_2(\tilde{\delta})+\tilde{\Omega}[1-H_2(\tilde{\delta}_p^{untagged})]\}
\end{aligned}
\end{equation}
where $\tilde{\delta}$ is the remaining QBER, $\tilde{r}_{B}$ is overall survival residue, $\tilde{\Omega}$ is the fraction of untagged states in the final survival states\footnote{Without B steps, $\Omega\equiv Q_1/Q_\mu$.} and $\tilde{\delta}_p^{untagged}$ is the phase error rate of the untagged states, after a sequence of B steps. In the following, we will show how these variables change with the performing of B steps.

\textbf{An arbitrary B step:} Let us consider how the various quantities (fraction of untagged states $\Omega$, QBER of overall surviving states $\delta$, bit error rate $\delta_{untagged}$ and
phase error rates $\delta_p$ of the untagged states) are transformed under one step in a B step sequence.

Prior to a B step, the fraction of untagged states is $\Omega$, the overall QBER is $\delta$, the bit error rate of the untagged states is $\delta_{untagged}$, and the phase error rate of the untagged states is $\delta_p$. According to Eq.~\eqref{oneBpS}, the overall survival probability $p_{S}$ and the survival probability of the untagged states $p_{S}^{untagged}$ after one B step are given by:
\begin{equation} \label{BDecoy:1BpS}
\begin{aligned}
p_{S}&=[\delta^2+(1-\delta)^2]\\
p_{S}^{untagged}&=[\delta_{untagged}^2+(1-\delta_{untagged})^2].
\end{aligned}
\end{equation}
Then the residue after one B step is given by:
\begin{equation} \label{B:1Bresidue}
\begin{aligned}
r_{B}=\frac12p_S
\end{aligned}
\end{equation}
where the factor 1/2 stems from the the fact that Alice and Bob only keep one qubit from a survival pair. Subsequently, after a B step, the fraction of untagged states $\Omega'$ is given by:
\begin{equation} \label{BDecoy:1Bfrac}
\begin{aligned}
\Omega'&=\frac{\Omega^2\cdot p_{S}^{untagged}}{p_{S}}.
\end{aligned}
\end{equation}

Overall QBER: the change of the overall QBER $\delta'$ is given by:
\begin{equation} \label{BDecoy:1BQBER}
\begin{aligned}
\delta' &= \frac{\delta^2}{\delta^2+(1-\delta)^2}.
\end{aligned}
\end{equation}

Untagged states: before the first B step, the initial density matrix of the untagged state is
$(1-2e_1+q_{11},e_1-q_{11},q_{11},e_1-q_{11})$, where $e_1$ is the
error rate of single photon states. From Appendix C of \cite{TwoWay_03},
we know that $q_{11}=0$ is the worst case for B steps. Thus we can
conservatively choose $(1-2e_1,e_1,0,e_1)$ as the initial input
density matrix. If only B steps are performed, $q_{11}=0$ will
always be satisfied, according to Eq.~\eqref{Twoway:oneBmatrix}. Therefore, the input untagged qubits for any round of B steps has the form of
\begin{equation} \label{BDecoy:1Binput}
(q_{00}, q_{10}, q_{11}, q_{01}) = (1-\delta_{untagged}-\delta_p,
\delta_{untagged}, 0, \delta_p).
\end{equation}
The bit error rate of untagged state $\delta_{untagged}'$ only
depends on the input $\delta_{untagged}$,
\begin{equation} \label{BDecoy:1Bbun}
\delta_{untagged}'=\frac{\delta_{untagged}^2}{\delta_{untagged}^2+(1-\delta_{untagged})^2}.
\end{equation}
According to Eqs.~\eqref{Twoway:oneBmatrix},
\eqref{Twoway:AfterBstepErr} and \eqref{BDecoy:1Binput}, the phase
error rate of untagged states is
\begin{equation} \label{BDecoy:1Bpun}
\begin{aligned}
\delta_p'&=q_{11}'+q_{01}'\\
&=\frac{2q_{10}q_{11}+2q_{00}q_{01}}{(q_{10}+q_{11})^2+(q_{00}+q_{01})^2}\\
&=\frac{2\delta_p\cdot(1-\delta_{untagged}-\delta_p)}{\delta_{untagged}^2+(1-\delta_{untagged})^2}.
\end{aligned}
\end{equation}

Eqs.~\eqref{BDecoy:1BpS}-\eqref{BDecoy:1Bpun} are valid for a
general B step. Alice and Bob can perform a sequence of B steps as
described above and then perform the error correction and privacy
amplification. Once all of these quantities are obtained, the key generation rate can be calculated from Eq.~\eqref{BDecoy:DGBrate}.

To illustrate the improvement made by introducing B steps, we simulate the GYS experiment \cite{GYS_04}, whose parameters are listed in Table \ref{Tab:GYSpara}. Similar to the simulations in previous chapters, we use $f(e)=1.22$ for the error correction efficiency \cite{BrassardSalvail_93}. 

\begin{figure}[hbt]
\centering \resizebox{12cm}{!}{\includegraphics{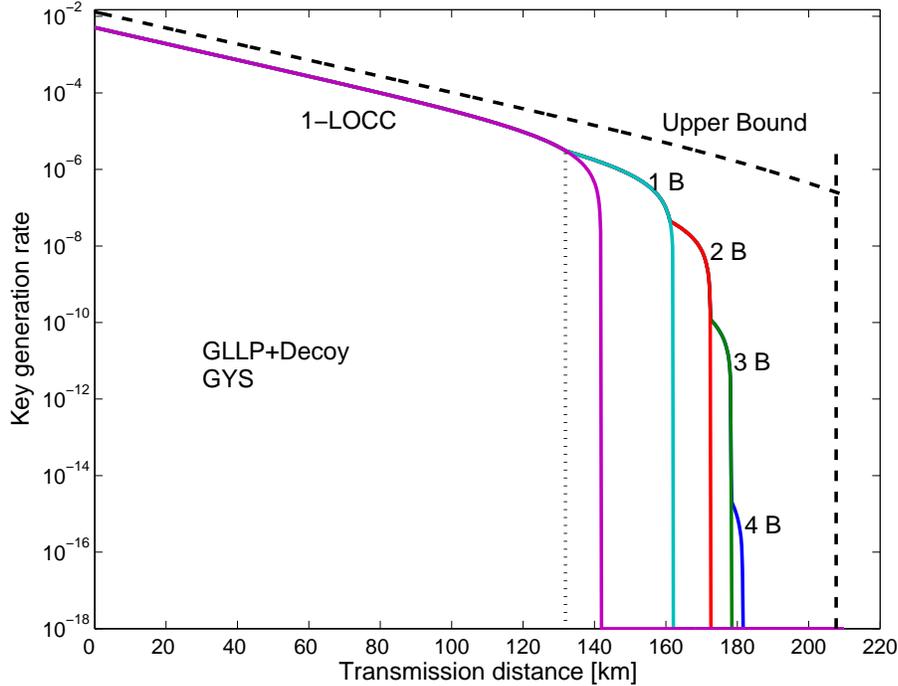}}
\caption{Plot of the key rate as a function of the transmission distance with the data post-processing scheme of GLLP+Decoy+B steps. The simulation parameters are from the GYS experiment \cite{GYS_04} listed in Table \ref{Tab:GYSpara}. The GLLP+Decoy+B steps scheme suppresses the one with 1-LOCC at a distance of 132 km. The maximal secure distance using 4 B steps is $181$ km, which is not far from the upper bound of $208$ km.
Note that B steps are useful only at rather long distances (over $132km$).
}
\label{Fig:5B}
\end{figure}

From Figure~\ref{Fig:5B}, we can see that there is a non-trivial extension of the maximal secure distance after introducing B steps. Note that the key rate of the decoy state protocol with 1 B step is higher than the one with 1-LOCC from a distance of around $132$ km. The maximal secure distance using 4 B steps is $181$ km, which is not far from the upper bound of $208$ km, given in Section \ref{Sub:DisUp}. Even with only 1 B step, the maximal secure distance can be extended from 142 km to 162 km. Thus, B steps are useful in QKD data post-processing.

\section{Decoy + GLLP + Recurrence EDP} \label{Sc:Recurrence}
In this section, we will present another data post-processing scheme based on the recurrence scheme \cite{VV_Recurrence_05}, which is reviewed in Section~\ref{Rec:Subsection:Rec}. Our scheme is a generalization of the recurrence scheme to the case of imperfect sources.

Here, we will use the extended GLLP formula, Eq.~\eqref{Post:KeyRateEx}, in Section \ref{Sub:GLLPEx}. Again, we use the definition of the residual, Eq.~\eqref{BDecoy:DefResidual}:
\begin{equation} \label{Rec:Residue1}
\begin{aligned}
r=-\frac12f(p_S)H_2(p_S)-\frac12p_Sf(\frac{\delta^2}{p_S})H_2(\frac{\delta^2}{p_S})+\sum_{i}\Omega_iK_{i},
\end{aligned}
\end{equation}
where $p_S$ is the even parity possibility given in
Eq.~\eqref{Rec:Survival} with $\delta_b^C=\delta_b^T=\delta$,
$\delta$ is the overall QBER before the recurrence, $f(\cdot)$ is
error correction efficiency, $\Omega_i$ and $K_{i}$ are the
probability and the residue of the qubit groups with label $i$ after
privacy amplification, respectively.

In the post-processing, Alice and Bob first check the parity, corresponding to the first term of
Eq.~\eqref{Rec:Residue1}. Secondly, they apply an overall error
correction to the qubits with even parity, corresponding to the
second term of Eq.~\eqref{Rec:Residue1}. Thirdly, they measure one
of the qubits in the pairs with odd parity to obtain the error
syndrome of another qubit. Afterwards, they can group the surviving
qubits into several groups with labels $i$. Finally, they perform
privacy amplification to each group with label $i$, corresponding to
the last term of Eq.~\eqref{Rec:Residue1}.

In the decoy state protocol, there are three kinds of input qubits: vacuum qubits (V), single-photon qubits (S) and multi-photon qubits (M). The input parameters for recurrence are listed in Table \ref{Rec:Tab:Input}.
\begin{table}[hbt]
\centering
\begin{tabular}{|c|c|c|c|c|}
\hline
Qubit & Fraction & $\delta_b$ & $\delta_p$ & $q_{11}$ \\
\hline
V & $\Omega_V$ & $1/2$ & $1/2$ & $q_{11}^V$ \\
\hline
S & $\Omega$ & $e_1$ & $e_1$ & $a$ \\
\hline
M & $\Omega_M$ & $e_M$ & $1/2$ & $q_{11}^M$ \\
\hline
\end{tabular}
\caption{List of the parameters of three kinds of input qubits for the recurrence scheme. Following Eqs.~\eqref{Model:Qi} and \eqref{Model:Gain}, the fractions of each group are given by $\Omega_V=Q_0/Q_\mu$, $\Omega=Q_1/Q_\mu$ and $\Omega_M=1-\Omega_V-\Omega$. $\Omega_V/2+e_1\Omega+e_M\Omega_M=\delta$ is the overall QBER.}
\label{Rec:Tab:Input}
\end{table}

Thus, the outcome of one round of recurrence will have nine cases.
Clearly, if neither input is a single photon qubits, the outcome
will have no contribution to the final key. Alice and Bob need only
apply Eq.~\eqref{Rec:PriRes2} to calculate the residues, $K_i$, for
the five cases: $V \bigoplus S$, $S \bigoplus V$, $S\bigoplus S$,
$S\bigoplus M$, $M\bigoplus S$. The probabilities of occurrence,
$\Omega_i$, for the five cases are $\Omega_V\Omega$, $\Omega\Omega_V$, $\Omega^2$, $\Omega\Omega_M$, $\Omega_M\Omega$, respectively.
Once we know $K_i$ and $\Omega_i$, we can then determine the overall
residue, $r$, using Eq.~\eqref{Rec:Residue1} (details are shown in
Appendix~\ref{App:residue}):
\begin{equation} \label{Rec:Residue2}
\begin{aligned}
r 
\ge& -B+C-F_a
\end{aligned}
\end{equation}
where
\begin{equation}
\begin{aligned}
B &=
\frac12f(p_S)H_2(p_S)+\frac12p_Sf(\frac{\delta^2}{p_S})H_2(\frac{\delta^2}{p_S})
\\
C &= \frac34\Omega_V\Omega + \Omega^2 (1-e_1+e_1^2) +
\frac12\Omega\Omega_M(2-e_1-e_M+2e_1e_M)
\\
D_1 &=
\frac34\Omega_V\Omega+\frac12\Omega^2(2-e_1)+\frac12\Omega\Omega_M(2-e_M)
\\
D_2 &=
\frac34\Omega_V\Omega+\frac12\Omega^2(1+e_1)+\frac12\Omega\Omega_M(e_M+1)
\\
F_a&=D_1(1-e_1)H_2(\frac{e_1-a}{1-e_1})+D_2e_1H_2(\frac{a}{e_1}) \\
\end{aligned}
\end{equation}
with equality when $q_{11}^V=1/4$ and $q_{11}^M=e_M/2$. In order to
get a lower bound of key generation rate $R$, we maximize $F_a$ over
$a$ by using a bisection method as discussed in Appendix
\ref{App:residue}.


\begin{figure}[hbt]
\centering \resizebox{12cm}{!}{\includegraphics{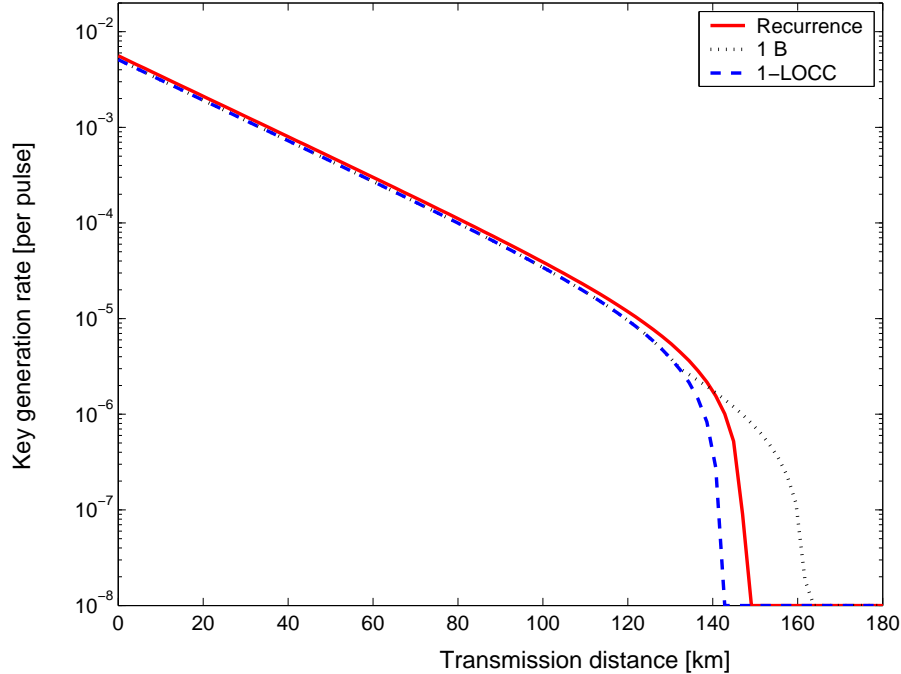}}
\caption{Plot of the key generation rate as a
function of the transmission distance, GLLP+Decoy+Recurrence vs.
GLLP+Decoy+1-LOCC. Recurrence improves the QKD performance over 1-LOCC in the whole regime of the distance.
In particular, the recurrence method increases the key rate by more than $10\%$ in our simulation. The maximal secure
distance for each case is 142.8 km (1-LOCC), 149.1 km (Recurrence),
163.8 km (1 B), respectively. Here, we consider the asymptotic decoy state QKD with an infinitely long experiment.
The parameters used are from the GYS experiment \cite{GYS_04} listed in Table
\ref{Tab:GYSpara}.} \label{Rec:Fig:Comp}
\end{figure}

Figure \ref{Rec:Fig:Comp} shows the key generation rate as a function of the transmission distance for GLLP+Decoy+1-LOCC, GLLP+Decoy+1 B step, and GLLP+Decoy+Recurrence. Recurrence has more than a 10\% improvement of the key rate over 1-LOCC in the whole regime of the distance, and it also increases the maximal secure distance by $6$ km. 

\section{Conclusion}
We have developed two data post-processing schemes for the decoy state QKD using 2-LOCC, one based on B steps and the other based on the recurrence method.
As discussed in Section \ref{Sub:QKDperformance}, the maximal secure distance of QKD is crucial in practical applications, thus
our Decoy+B steps post-processing protocol, which we have shown to
be able to increase the maximal secure distance of QKD from 141 km to 182 km (using parameters from the GYS experiment
\cite{GYS_04}), proves to be useful in
real-life applications. Moreover, our work shows that recurrence
protocols are useful for increasing the key generation rate in a
practical QKD system in the whole regime of the distance. 

In Ref.~\cite{TwoWay_06}, we also show that similar conclusions hold even with statistical fluctuations in the experimental variables for the Decoy+B step scheme. For the Decoy+Recurrence scheme, although we do not have a rigorous argument, physical intuition suggests that similar
conclusions hold in the case of considering statistical fluctuations as well. We conclude that using two-way classical communication is superior to using one-way for our decoy state QKD schemes.

In addition, we provided a region of bit error rates and phase error rates that are tolerable by using the Gottesman-Lo EDP scheme. 

\chapter{Triggering PDC QKD} \label{Chpt:Trig}
Parametric down-conversion (PDC) sources can be used for QKD. One can use a PDC source as a triggered (heralded) single photon source. Recently, there are various practical proposals of the decoy state QKD with triggering PDC sources. In this chapter, we generalize the passive decoy state idea, originally proposed by Mauerer and Silberhorn. The generalized passive decoy state idea can be applied to cases where either threshold detectors or photon number resolving detectors are used. The decoy state protocol proposed by Adachi, Yamamoto, Koashi and Imoto (AYKI) can be treated as a special case of the generalized passive decoy state method. By simulating a recent PDC experiment, we compare various practical decoy state protocols with the infinite decoy protocol and also compare the cases using threshold detectors and photon-number resolving detectors. Our simulation result shows that with the AYKI protocol, one can achieve a key generation rate that is close to the theoretical limit of the infinite decoy protocol. Furthermore, our simulation result shows that a photon-number resolving detector does not appear to be useful for improving the QKD performance in this case. Although our analysis is focused on the QKD with PDC sources, we emphasize that it can also be applied to QKD setups with other triggered single photon sources.

This work is presented in Ref.~\cite{TriggeringPDC_08}. In this work, I modeled the QKD setup with triggered PDC source following the work of L\"utkenhaus \cite{IndividualAttack_00} and compare various decoy state proposals of triggering PDC QKD.

\section{Background}
The coherent state QKD suffers from photon-number splitting (PNS) attacks \cite{HIGM_95,BLMS_00,LutkenhausJahma_02}. As discussed in Section \ref{Sc:AsympDecoy}, a main objective of the decoy state method is to close this loophole of multi photon components in QKD sources. Decoy states can help better estimate the channel properties (e.g., transmittance and error provability). To do that, Alice uses extra states with different light intensities during key transmission. Then Alice and Bob can consider detection statistics from signal and decoy states separately, from which they can better estimate the channel transmittance and error probability. The situation where Alice actively prepares decoy states is called the \textit{active decoy state} method, which is differentiated from the \emph{passive decoy state} method where Alice chooses decoy and signal states by passive measurements. A detailed discussion about the passive decoy state can be found in Section \ref{Sub:Passive}. Note that in the coherent state QKD, one can only use the active decoy state method.

Aside from a coherent state source, a PDC source can be used in a QKD experiment as well. There are two ways to use a PDC source. The first is to use it as a triggered (heralded) single photon source. Alice detects one of the two modes from a PDC source as a trigger \footnote{See Section \ref{Sc:Trig:Impl} for the definition of a trigger.} and actively encodes her qubit information into another mode. We call this implementation \emph{triggering PDC QKD}. The second way is to use it as an entangled photon source for entanglement-based QKD protocols. See Chapter \ref{Chpt:Ent} for more discussion. We call this implementation \emph{entanglement PDC QKD}.

The triggering PDC QKD, similar to the coherent state QKD, suffers from PNS attacks. By applying the GLLP security proof, one can find that the optimal average photon number $\mu$ is in the same order of the overall transmittance $\eta$. Then the key generation rate will be in the order of $\eta^2$. For a rigorous derivation, one can refer to Appendix \ref{ApSc:muTrig}. Thus, the performance of the triggering PDC QKD is very limited.

Since the decoy state idea can substantially enhance the performance of the coherent state
QKD, a natural question will be: ``Can the decoy state idea be applied to the triggering PDC
QKD?" The answer is \emph{yes}. One can apply the infinite decoy state idea
\cite{Decoy_05}, as discussed in Section \ref{Sc:AsympDecoy}, to the triggering PDC QKD. Not surprisingly, with decoy states, the key
generation rate can be $O(\eta)$, which is the same as the order achieved by a single-photon source. Therefore, we expect that the decoy state QKD will become a standard technique not only in the coherent state QKD, but also in QKD with triggering PDC sources. Recently, a few practical decoy proposals for triggering PDC requiring a finite number of decoy states have been proposed \cite{MauererSilberhorn_07,AYKI_07,WWG_07,WWBK_PDC_07}. Note that an experimental demonstration of the decoy state QKD with a triggering PDC source was implemented recently \cite{TrigWang_Exp_08}.

We are interested in comparing various protocols for the triggering PDC QKD. Among the practical decoy protocols for triggering PDC QKD, we find that the one proposed by Adachi, Yamamoto, Koashi and Imoto (AYKI) \cite{AYKI_07} is simple to implement. The AYKI protocol is conceptually similar to the one-decoy state scheme \cite{Practical_05}, as discussed in Section \ref{Sub:Onedecoy}. In the AYKI protocol, Alice and Bob only need to consider the statistics of triggered and non-triggered detection events \footnote{In a non-triggered detection event, Bob gets a detection, but Alice does not get a trigger.} separately, instead of preparing new signals for the decoy states. We emphasize that the AYKI protocol is easy to implement since there is no need for a hardware change.



Other decoy state proposals for the triggering PDC QKD require hardware modifications. For example, the one proposed by Mauerer and Silberhorn \cite{MauererSilberhorn_07} requires photon-number resolving detectors, and the one proposed by Wang, Wang and Guo \cite{WWG_07} requires Alice to pump the laser source at various intensities.

The following is a generalization of the passive decoy state idea proposed by Mauerer and Silberhorn \cite{MauererSilberhorn_07}. The main idea is that Bob can group his detection events in accordance to the public announcement of Alice's detection events. For example, when Alice uses a threshold detector, Bob can group his detection results in accordance to whether Alice gets a detection or not.  The generalized passive decoy state idea can be applied to both cases that use threshold detectors and photon-number resolving detectors. The AYKI protocol can be treated as a special case of the generalized passive decoy state protocol. 



By simulating a recent PDC experiment \cite{PDC144_07}, we compare one case with a perfect photon-number resolving detector and four cases with threshold detectors: no decoy, infinite decoy, weak decoy and AYKI.
Our simulation result shows that in a large regime (for instance, the optical link loss between 0 dB and 25 dB), the performance of AYKI protocol is close to that of the infinite decoy protocol and thus, there is not much room left for improvement after the AYKI protocol has been implemented. Moreover, the QKD performance of the case with the infinite decoy protocol using threshold detectors is close to the case using a perfect photon-number resolving detector. Thus, a photon-number resolving detector does not appear to be useful for triggering PDC QKD.

We emphasize that an advantage of the passive decoy state method is that by passively choosing decoy and signal states, the possibility that Eve can distinguish decoy and signal states is reduced. On the other hand, in active (regular) decoy state experiments, it is more difficult to verify the assumption that Eve cannot distinguish decoy and signal states.

Note that the passive decoy state idea can be combined with the active decoy state idea. In Ref.~\cite{WWBK_PDC_07}, the authors provide a special case where passive and active decoy state ideas are combined. Again, we emphasize that for the coherent state QKD, one can only use active decoy state methods.

Although our analysis is focussed on a QKD with a triggered PDC source, we emphasize that it can also be applied to QKD setups with other triggered single photon sources.



In Section \ref{Sc:Trig:Impl}, we will review the experiment setup of the triggering PDC QKD. In Section \ref{Sc:Trig:Model}, we provide a model for the triggering PDC QKD. In Section \ref{Sc:Trig:Post}, we will study various post-processing schemes for the triggering PDC QKD. In Section \ref{Sc:Trig:Simulation}, we will compare various schemes of the triggering PDC QKD: non-decoy+threshold detectors, infinite decoy+threshold detectors, AYKI and a case with a perfect photon-number resolving detector, by simulating a real PDC experiment.



\section{Experiment setup} \label{Sc:Trig:Impl}
\begin{figure}[hbt]
\centering \resizebox{12cm}{!}{\includegraphics{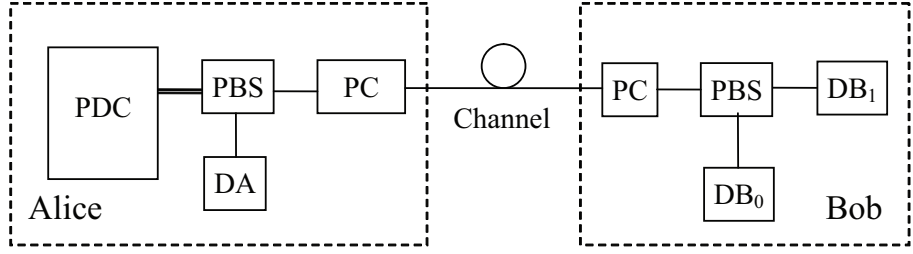}} \caption{A schematic diagram for the
triggering PDC QKD. Alice collects photon pairs emitted from a PDC source and uses a polarization beam splitter (PBS) to separate two polarization modes. She detects one of the two modes with her
detector (DA) as a trigger, modulates the polarization of the other mode by a polarization
controller (PC) and sends it to Bob. On Bob's side, he chooses his basis by a PC and performs a
measurement by his detectors (DB$_0$ and DB$_1$).} \label{Fig:PDCtrig}
\end{figure}

In triggering PDC QKD, a PDC source is used as a triggered single photon source\footnote{Sometimes it is called heralded single photon source.}. 
The schematic diagram is shown in Figure~\ref{Fig:PDCtrig}.

As shown in Figure~\ref{Fig:PDCtrig}, a PDC source generates two modes of photons, which can be separated by a polarization beam splitter (PBS). One mode goes to Alice's own detector (DA in Figure~\ref{Fig:PDCtrig}) as the triggering signal and the other mode is used as a triggered single photon state for the QKD.
When Alice's detector (DA) clicks, we call it a \emph{trigger}. We divide the detection
events on Bob's side into two groups depending on whether Alice gets a trigger or not: triggering
detection events and non-triggering detection events.

Note that Alice can use either a threshold detector or a photon-number resolving detector (DA in Figure~\ref{Fig:PDCtrig}). She only needs to know the number of photons in the trigger mode. Therefore, only one detector is sufficient on Alice's side. Due to the high channel losses, without Eve's interference, Bob is highly likely to receive a vacuum or single photon state. Thus it is sufficient for Bob to use threshold detectors. Threshold single photon detectors can only tell whether there is a click or not, but not the photon numbers. Bob needs to identify polarizations of incoming photons. Here, we assume Alice encodes qubit information in photon polarizations.

In real experiments, there are two types of PDC sources, both of which can be used in a triggering PDC QKD setup. Here, we assume Alice uses a type-II PDC source. The Hamiltonian of the type-II PDC process in the triggering setup shown in Figure~\ref{Fig:PDCtrig} can be
written as \cite{WallsMilburn_94}:
\begin{equation}\label{Impl:HamiltonianTrig}
\begin{aligned}
H &= i\chi a^\dagger b^\dagger+h.c. \\
\end{aligned}
\end{equation}
where $h.c.$ means Hermitian conjugate and $\chi$ is a coupling constant which depends on the
crystal nonlinearity and the amplitude of the pump beam. The operators $a^\dagger$, $b^\dagger$ and $a$, $b$ are the creation and annihilation operators of two modes with different polarizations.

The state coming from a triggering PDC source, with a Hamiltonian of
Eq.~\eqref{Impl:HamiltonianTrig}, can be written as \cite{WallsMilburn_94}:
\begin{equation}\label{Impl:PDCstateL}
\begin{aligned}
|\Psi\rangle=(\cosh\chi)^{-1}\sum_{n=0}^{\infty}(\tanh\chi)^n|n,n\rangle.
\end{aligned}
\end{equation}
Here, we assume that the state is single-mode. The expected photon pair number is given by $\mu=\sinh^2\chi$.  
The probability to get an $n$-photon-pair is:
\begin{equation}\label{Impl:PnL}
\begin{aligned}
P(n)=\frac{\mu^n}{(1+\mu)^{n+1}}.
\end{aligned}
\end{equation}
Here, we assume that the PDC source always sends out photon pairs. That is, the photon number of mode $a$ and $b$ is always the same.

There is a nonzero probability for the PDC source to emit more than one photon pair in a pulse. Thus, Alice may send out multi photon states after she encodes basis and key information by her polarization controller (PC). This is the reason why the triggering PDC QKD suffers from PNS attacks. Later in the next chapter, we will show that when Alice uses the PDC source as an entangled photon source to implement an entanglement based QKD, it will be immune from PNS attacks.

Let us compare triggering PDC QKD and entanglement PDC QKD implementations. For the setup of entanglement PDC QKD, one can refer to Section \ref{Sc:Enta:Impl}. In the triggering PDC QKD, Alice actively encodes the key information, while in the entanglement PDC QKD, Alice measures the polarization of one mode of PDC source directly. The advantage of the triggering PDC QKD here is that it does not rely on the polarization correlations between two modes of the PDC source. It only requires the photon-pair generation of the source, which means entanglement between photon pairs are not important for the triggering PDC QKD. However, in an entanglement PDC QKD implementation, the entanglement between two modes has to be well maintained for QKD transmission. We notice that maintaining entanglement in real experiments is a highly non-trivial task\footnote{A.~M.~Steinberg, private communication.}.

\section{Model} \label{Sc:Trig:Model}
L\"utkenhaus studied the model of triggering PDC QKD \cite{IndividualAttack_00} with threshold detectors. His model is similar to the one of the coherent state QKD, except for a different photon number distribution.

The channel model of triggering PDC QKD is exactly the same as the coherent state QKD. Thus, one can use Eqs.~\eqref{Model:Yi} and \eqref{Model:ei}.

\subsection{On Alice's side} \label{Aliceside}
In the triggering PDC QKD, Alice may use either a threshold detector or a photon-number resolving detector. A \emph{N-photon-resolving} detector is defined to be a detector that can tell 0, 1, $\cdots$, $N$ photons of an incoming signal. For a threshold detector, we have $N=1$, which can only tell the presence of photons, but not the photon numbers. Given an incoming $i$-photon state, the probability for Alice's detector to indicate a $j$-photon state is $\eta_{j\mid i}$, with $\sum_{j=0}^{j=N}\eta_{j\mid i}=1$ for all $i=0,1,\cdots$. In general, $\eta_{j\mid i}$s are real numbers in [0,1]. We define a $j$-photon trigger for a case where Alice's detector indicates a $j$-photon state.

For a triggered PDC photon source, as given in Eq.~\eqref{Impl:PDCstateL}, the probability for Alice's detector to indicate a $j$-photon detection is:
\begin{equation}\label{Model:AliceProbj}
\begin{aligned}
P_{Aj} &= \sum_{i=0}^{\infty}\frac{\mu^i}{(1+\mu)^{i+1}}\eta_{j\mid i}. \\
\end{aligned}
\end{equation}
With the assumption that the PDC source always emits photon pairs, the probability (gain) for Alice getting a $j$-photon detection and Bob getting a detection is:
\begin{equation}\label{Model:Gainj}
\begin{aligned}
Q_{\mu,j} &= \sum_{i=0}^{\infty}Q_{i,j} \\
&= \sum_{i=0}^{\infty}\frac{\mu^i}{(1+\mu)^{i+1}}\eta_{j\mid i}Y_{i}, \\
\end{aligned}
\end{equation}
where the yield $Y_{i}$ is given in Eq.~\eqref{Model:Yi}. 
%
%
%
The quantum bit error rate (QBER) conditioned on Alice's $j$-photon detection, similar to Eq.~\eqref{Model:Gainj}, is given by:
\begin{equation}\label{Model:QBERj}
\begin{aligned}
E_{\mu,j}Q_{\mu,j} &= \sum_{i=0}^{\infty}Q_{i,j}e_i \\
&= \sum_{i=0}^{\infty}\frac{\mu^i}{(1+\mu)^{i+1}}\eta_{j\mid i}Y_{i}e_i. \\
\end{aligned}
\end{equation}
where the error rate $e_{i}$ is given in Eq.~\eqref{Model:ei}.

It is observed that in the triggering PDC QKD setup, shown in Figure~\ref{Fig:PDCtrig}, the quantities $Y_i$ and $e_i$ are independent of Alice's measurement outcome $j$. This is based on the single-mode PDC source assumption described in Eq.~\eqref{Impl:HamiltonianTrig} in Section \ref{Sc:Trig:Impl}. 
Therefore, in Section \ref{Sc:Trig:Post}, we can apply the decoy state idea.

\subsection{Threshold detector} \label{Sub:ThreshDet}
Here, we will discuss a special case where Alice uses a threshold detector. That is,
\begin{equation}\label{Model:thresh}
\begin{aligned}
\eta_{0\mid i} &= (1-Y_{0A})(1-\eta_A)^i \\
& \simeq (1-\eta_A)^i \\
\eta_{1\mid i} &= 1-\eta_{0\mid i} \\
\eta_{j\mid i} &= 0, \text{\space\space\space\space\space\space} \forall j\ge2, \\
\end{aligned}
\end{equation}
where $Y_{0A}$ and $\eta_A$ are the background count rate and the detector efficiency on Alice's side. The approximation is due to the fact that normally, we have $\eta_A\gg Y_{0A}$. That is, we neglect the background contributions on Alice's side.


According to Eqs.~\eqref{Model:Gainj} and \eqref{Model:QBERj}, without Eve's interference, the gains and QBER's of triggered ($j=1$) and non-triggered ($j=0$) detections are given by:
\begin{equation}\label{Model:QEThreshold}
\begin{aligned}
Q_{\mu,0} 
&= \frac{1}{1+\eta_A\mu}-\frac{1-Y_{0B}}{1+(\eta_A+\eta-\eta_A\eta)\mu} \\
Q_{\mu,1} 
&= 1-\frac{1}{1+\eta_A\mu}-\frac{1-Y_{0B}}{1+\eta\mu}+\frac{1-Y_{0B}}{1+(\eta_A+\eta-\eta_A\eta)\mu} \\
E_{\mu,0}Q_{\mu,0} &= e_dQ_{\mu|0} + \frac{(e_0-e_d)Y_{0B}}{1+\eta_A\mu} \\
E_{\mu,1}Q_{\mu,1} &= e_dQ_{\mu|1} + \frac{(e_0-e_d)\eta_A\mu Y_{0B}}{1+\eta_A\mu}. \\
\end{aligned}
\end{equation}
Without Eve's interference, the gains and error rates of the single photon state in two detections are given by:
\begin{equation}\label{Model:qe1Threshold}
\begin{aligned}
Q_{1,0} &= \frac{\mu(1-\eta_A)}{(1+\mu)^2}Y_{1} \\
Q_{1,1} &= \frac{\mu\eta_A}{(1+\mu)^2}Y_{1} \\
e_1Y_1 
&= e_dY_1+(e_0-e_d)Y_{0B} \\
\end{aligned}
\end{equation}
where $Y_1$ and $e_1$ are given in Eqs.~\eqref{Model:Yi} and \eqref{Model:ei}, respectively.

\subsection{Perfect photon-number resolving detector} \label{PrfctPNRD}
Here, we will discuss the case where Alice uses a perfect photon-number resolving detector, which can faithfully tell the number of photons in the incoming signal. That is, $\eta_{j\mid i}=\delta_{ij}$. Thus, from Eqs.~\eqref{Model:Gainj} and \eqref{Model:QBERj}, the gains and QBERs are given by:
\begin{equation}\label{Model:PrfctGQj}
\begin{aligned}
Q_{\mu,i} = Q_{i,i} 
&= \frac{\mu^i}{(1+\mu)^{i+1}}Y_{i} \\
E_{\mu,i}Q_{\mu,i} = e_iQ_{i,i} &= \frac{\mu^i}{(1+\mu)^{i+1}}e_iY_{i}, \\
\end{aligned}
\end{equation}
from where one can directly infer the gains and error rates of the $i$-photon state, $Q_{i,j}=Q_{i,i}\delta_{i,j}$.

\section{Post-processing} \label{Sc:Trig:Post}

Here, we will apply the standard GLLP analysis, as shown in Eq.~\eqref{Post:KeyRate}. All the classical data measured can be grouped according to Alice's detection events, $j=0,1,\cdots,N$. Subsequently, we can apply the GLLP idea \cite{GLLP_04,TwoWay_06} to each group. The final key generation rate will be given by summing over contributions from all groups:
\begin{equation} \label{Post:KeySum}
R = \sum_{j=0}^{N} R_j.
\end{equation}
In each case $j$, one can apply Eq.\eqref{Post:KeyTrig}:
\begin{equation} \label{Post:Keyj}
R_j \geq q \{-f(E_{\mu,j})Q_{\mu,j}H_2(E_{\mu,j})+Q_{1,j}[1-H_2(e_{1})]\},
\end{equation}
where $Q_{0,j}$ and $Q_{1,j}$ are the first and second terms on the right hand side of Eq.~\eqref{Model:Gainj}. Here, the error rate of the single photon state $e_1$ is independent of $j$, see the observation in the end of Section \ref{Aliceside}. Note that the key generation rate from all $j$-photon trigger detections should be non-negative. If any of them contributes a negative key generation rate, we should assign $0$ to it. In this case, Alice and Bob can just discard that type of detection. Based on this observation, we can further simplify Eq.~\eqref{Post:KeySum}. Given that Alice detects more than one photon, the probability of emitting a single photon state in Bob's arm is small\footnote{In Section \ref{Sc:Trig:Impl}, we assume that Alice's PDC source always sends out photon pairs. Given that Alice detects more than one photon on the triggering arm, a single photon state is present on the other arm only when there is a dark count in Alice's detector. Normally, we can assume that the detector efficiency is much higher than the dark count probability on Alice's side. Thus, we neglect the probability of a single photon state with a multi photon trigger.}. As we mentioned in the beginning of this section, only a single photon state can contribute positively to the final key rate.
Thus we can focus on the case $j=0,1$.
\begin{equation} \label{Post:Key01}
R = R_0+R_1,
\end{equation}
where $R_0$ and $R_1$ are given in Eq.~\eqref{Post:Keyj}. Again, both $R_0$ and
$R_1$ should be non-negative, otherwise they should be assigned 0.

In Eq.~\eqref{Post:Keyj}, the gain $Q_{\mu,j}$ and the QBER $E_{\mu,j}$, given in Eqs.~\eqref{Model:Gainj} and \eqref{Model:QBERj}, can be measured or tested from QKD experiments directly. In this section, we will discuss various ways to estimate $Q_{0,j}$, $Q_{1,j}$, and $e_{1}$. We assume that the PDC photon source  and detector characteristics are fixed and known to Alice. That is, $\mu$, the photon number distribution in Eq.~\eqref{Impl:PnL} and $\eta_A$ are fixed and known.



\subsection{Non-decoy states with threshold detectors}
Here, we assume that Alice uses a threshold detector. Thus, Alice only has two measurement outcomes, $j=0,1$. A simple way to estimate $Q_{0,j}$, $Q_{1,j}$, and $e_{1}$ is by assuming that all losses and errors come from the single photon states. This is because Eve can in principle, perform PNS attacks on the multi-photon states.
The gain and error rate of the single photon states in triggered ($j=1$) and non-triggered ($j=0$) detections can be bounded by:
\begin{equation}\label{Post:Q1e1nondecoy}
\begin{aligned}
Q_{1,0} &\ge Q_{\mu,0}-\sum_{i=2}^{\infty}\frac{\mu^i}{(1+\mu)^{i+1}}\eta_{0|i} \\
      &= Q_{\mu,0}-\frac{(1-\eta_A)^2\mu^2}{(1+\eta_A\mu)(1+\mu)^2} \\
Q_{1,1} &\ge Q_{\mu,1}-\frac{\eta_A(2-\eta_A+\mu)\mu^2}{(1+\eta_A\mu)(1+\mu)^2} \\
e_{1,0} &\ge \frac{E_{\mu,0}Q_{\mu,0}}{Q_{1,0}} \\
e_{1,1} &\ge \frac{E_{\mu,1}Q_{\mu,1}}{Q_{1,1}} \\
%
\end{aligned}
\end{equation}
where $\eta_A$ is the efficiency of Alice's detector. The gain $Q_\mu$ and the QBER $E_{\mu}$, given in Eqs.~\eqref{Model:Gainj} and \eqref{Model:QBERj}, can be measured or tested from QKD experiments directly. In the following simulations, we will use Eq.~\eqref{Model:QEThreshold}. Since we assume all errors come from single photon states, one should take the lower bound of the vacuum contribution to be $Q_{0,j}=0$.


\subsection{Infinite active decoy state with threshold detectors}
To perform a privacy amplification process, Alice and Bob need to bound $Q_{0,j}$, $Q_{1,j}$, and $e_{1}$ for Eq.~\eqref{Post:Keyj}. 
From Eq.~\eqref{Model:Gainj}, we know that to bound $Q_{0,j}$ and $Q_{1,j}$, Alice and Bob need to estimate $Y_1$.

The decoy state method provides a good way to estimate $Y_1$ and $e_1$ \cite{Hwang_03,Decoy_05}. The essential idea is that instead of considering each linear equation of $Y_1$ and $e_1$ in the form of Eqs.~\eqref{Model:Gainj} and \eqref{Model:QBERj} separately, Alice and Bob consider all the linear equations simultaneously.

Let us imagine that Alice and Bob obtain an infinite number of linear equations in the form of Eqs.~\eqref{Model:Gainj} and \eqref{Model:QBERj}, e.g., they use an infinite number of intensities $\mu$. In principle, Alice and Bob can solve the equations to get $Y_1$ and $e_1$ accurately. Mathematically, the problem is solvable. The intuition is that the contributions from higher order terms of $Y_i$ and $e_i$ decrease exponentially in Eqs.~\eqref{Model:Gainj} and \eqref{Model:QBERj}. For the case coherent state QKD, one or two decoy states are proven to be sufficient \cite{Practical_05}. Shortly, we will see that one decoy state is sufficient for triggering PDC QKD.

The key underlying assumption of the decoy state method is shown in Eq.~\eqref{Decoy:DecoyAss}.
In other words, Eve sets the same values of $Y_i$ and $e_i$ for the decoy and signal states. This can be guaranteed by the assumption that Eve cannot distinguish decoy and signal states.

In Appendix \ref{ApSc:muTrig}, we will show that the optimal $\mu$ for the infinite decoy state case is in the order of 1, $\mu=O(1)$, which yields final a key rate $R=O(\eta)$. On the other hand, the optimal $\mu$ for the non-decoy case is $\mu=O(\eta)$, which yields a final key rate $R=O(\eta^2)$. Therefore, we expect the decoy state QKD to become a standard technique not only in the coherent state QKD, but also in QKD with triggering PDC sources.


There are various ways to apply the decoy state idea to the triggering PDC QKD \cite{MauererSilberhorn_07,AYKI_07,WWG_07}. Here, we consider the upper bound (infinite decoy state case) of all possible decoy protocols of triggering PDC QKD with threshold detectors: triggering PDC+infinite decoy method \cite{Decoy_05}. In the infinite decoy state method, Alice and Bob perform an infinite number of decoy states by choosing different intensities of the PDC source, $\mu$. They can then solve the linear equations in the form of Eqs.~\eqref{Model:Gainj} and \eqref{Model:QBERj} to estimate $Y_1$ and $e_1$ accurately. Hence, they can calculate each $Q_{0,j}$, $Q_{1,j}$, and $e_{1}$ accurately. In the simulation, we will use Eqs.~\eqref{Model:QEThreshold} and \eqref{Model:qe1Threshold} directly.


\subsection{Weak active decoy state with threshold detectors}
Here, we assume that Alice and Bob use threshold detectors and focus on triggered detection events. Alice uses another intensity $\nu$, for instance, by attenuating the pumping laser, for the weak decoy state. Wang, Wang and Guo proposed a practical decoy method for triggering PDC QKD \cite{WWG_07}, which is essentially applying the Vacuum+Weak decoy state method \cite{Practical_05} described in Section \ref{Sub:VW}. Note that for triggered detection events, the vacuum contribution can be negligible since $\eta_A\gg Y_{0A}$. Thus there is no need to estimate the vacuum contribution here. Therefore, Alice and Bob only need to perform a weak decoy state instead of the Vacuum+Weak decoy states. In this case, only one weak decoy state is sufficient.

Bounds of $Y_1$ and $e_1$ are given by $\mu^2(1+\nu)^3\times Q_{\nu,1}-\nu^2(1+\mu)^3\times Q_{\mu,1}$ in Eqs.~\eqref{Model:Gainj} and \eqref{Model:QBERj}:
\begin{equation}\label{Post:ActiveY1e1}
\begin{aligned}
Y_1 &\ge \frac{1}{\eta_A(\mu-\nu)}[\frac{\mu}{\nu}(1+\nu)^3Q_{\nu|1}-\frac{\nu}{\mu}(1+\mu)^3Q_{\mu|1}] \\
e_1 &\le \min\{\frac{(1+\mu)^2}{\mu}\frac{E_{\mu,1}Q_{\mu,1}}{\eta_AY_{1}}, \frac{(1+\nu)^2}{\nu}\frac{E_{\nu,1}Q_{\nu,1}}{\eta_AY_{1}}\} \\
\end{aligned}
\end{equation}
where $\nu$ is the expected photon pair number of the weak decoy state and $\eta_A$
is the efficiency of Alice's threshold detector. 

It is not difficult to show that when $\nu\rightarrow0$, Eq.~\eqref{Post:ActiveY1e1} approaches the infinite case, Eqs.~\eqref{Model:QEThreshold} and \eqref{Model:qe1Threshold}, described in the previous subsection.

\subsection{Passive decoy state} \label{Sub:Passive}
Recently, Mauerer and Silberhorn proposed a passive decoy state scheme, in which photon-number resolving detectors are required \cite{MauererSilberhorn_07}. Let us recap the heuristic idea of the original passive decoy state scheme briefly here. As discussed in Section \ref{Sc:Trig:Model}, Alice and Bob eventually get different detection events grouped by triggers on Alice's side. The key idea proposed by Mauerer and Silberhorn is that Alice and Bob manually combine the $\{j\}$-trigger detection events to get the decoy states with different photon number statistics and then follow the regular decoy state scheme.

Here, we want to point out that the ``combination" step is unnecessary. In general, each detection event group with a $j$-trigger has a different photon number statistic on the photon source arm. Thus, Alice and Bob need to treat all $\{j\}$-trigger detection events statistics separately. Furthermore, photon-number resolving detectors are not necessary in passive decoy state schemes. Our new generalized passive decoy state scheme is as follows.
\begin{enumerate}
\item
Alice uses a PDC source as her triggered photon source. She detects one of the modes from her PDC source as the trigger and encodes key information into another mode. Due to the detector Alice uses, she will get different trigger events: $j=0,1,\cdots$. When she uses a threshold detector, she will only get $j=0,1$.

\item
As the usual BB84 protocol, Bob measures signals in two different bases. Alice and Bob perform basis reconciliation.

\item
Alice announces her trigger detection results for each pulse: $j$. Bob groups his detection events by the information $j$. For each $j$, they calculate the gain $Q_{\mu,j}$ and test the QBER $E_{\mu,j}$.

Mathematically, they will obtain a set of linear equations in the form of Eqs.~\eqref{Model:Gainj} and \eqref{Model:QBERj}. Notice that the setup parameters, $\mu$ and $\eta_{j\mid i}$s, are known to Alice and Bob. Thus, they can estimate $Y_1$ and $e_1$ by considering Eqs.~\eqref{Model:Gainj} and \eqref{Model:QBERj}. 

\item
The post-processing is applied accordance to Eq.~\eqref{Post:Key01}.
\end{enumerate}

Note that the scheme is called \emph{passive} because Alice does not actively select decoy states. Instead, she determines the decoy states by measuring the trigger mode. Later, we will show that this is one advantage of using the triggering PDC source for QKD. Actually, in this case,  there are no strict definitions of decoy states and signal states. In the original decoy state method \cite{Practical_05}, decoy states are only used to estimate $Y_1$ and $e_1$ and the key is always generated from signal states\footnote{In the coherent state QKD, there is an optimal $\mu$ for a setup. To maximize the final key rate, Alice and Bob should publicly compare all detection results from decoy states.}.  In a triggering PDC QKD case, both the triggered $j=0$ and non-triggered $j=1$ detection events may have positive contributions to the final key generation.

\subsection{Passive decoy state with threshold detectors} \label{AYKI}
Here, we will review the decoy protocol proposed by Adachi, Yamamoto, Koashi and Imoto \cite{AYKI_07} as a special case of the passive decoy state protocol. The AYKI protocol is interesting in practice since it does not involve any hardware change to implement the decoy state idea.

Both Alice and Bob use threshold detectors, thus they have two types of detection events, triggered ($j=1$) and non-triggered ($j=0$). Secure keys can be generated from both types of detection events. Following the passive decoy state method procedure described in the previous subsection, Alice and Bob can estimate $Y_1$ and $e_1$ by considering the statistics of triggered and non-triggered detection events together. This is conceptually similar to the one decoy state idea \cite{Practical_05} described in Section \ref{Sub:Onedecoy}.

By solving two linear equations of Eq.~\eqref{Model:Gainj} with $j=0,1$,
$[1-(1-\eta_A)^2]\times Q_{\mu,0}-(1-\eta_A)^2\times Q_{\mu,1}$, one can get:
\begin{equation}\label{Post:Y1T}
\begin{aligned}
Y_1 &\ge Y_1^L \equiv \frac{(1+\mu)^2}{\mu}[\frac{2-\eta_A}{1-\eta_A}(Q_{\mu,0} -Q_{0,0})-\frac{1-\eta_A}{\eta_A}Q_{\mu,1}] \\
\end{aligned}
\end{equation}
where $Q_{0,0}$ is the vacuum state contribution in non-triggered detection events. One needs to minimize the key rate of Eq.~\eqref{Post:Key01} for $Q_{0,0}$ with the constraint of Eq.~\eqref{Model:QBERj}. Note that this result is essentially Eq.~(14) given in Ref.~\cite{AYKI_07}. We can see that when $\eta_A$ is close to 1 or $\mu$ is small, after neglecting $Q_{\mu,0}$ (background counts), the lower bound $Y_1^L$ is tight (approaches the real value of $Y_1$, see Eq.~\eqref{Model:Yi}):
\begin{equation}\label{Post:Y1Tlim}
\begin{aligned}
\lim_{\eta_A\rightarrow1}Y_1^L = \lim_{\mu\rightarrow0}Y_1^L &= \eta. \\
\end{aligned}
\end{equation}


By neglecting the vacuum state contribution for triggered detection events, $Q_{0,1}=0$, $e_1$ can be simply estimated by:
\begin{equation}\label{Post:e1U}
\begin{aligned}
e_1 \le 
\frac{E_{\mu,1}Q_{\mu,1}}{Q_{1,1}}.
%
%
\end{aligned}
\end{equation}

To get the lower bound of $Y_1$ in Eq.~\eqref{Post:Y1T}, one has to estimate the background contribution $Q_{0,0}$ as well. A simple bound of $Q_{0,0}$ is $0\le Q_{0,0}e_0\le E_{\mu,0}Q_{\mu,0}$ from Eq.~\eqref{Model:QBERj}, where $e_0=1/2$.

Note that the key rate calculated by substituting Eqs.~\eqref{Post:Y1T} and \eqref{Post:e1U} into Eq.~\eqref{Post:Key01} is not optimal. To get a tighter key rate bound, one can numerically calculate the lower bound of Eq.~\eqref{Post:Key01} directly, given the measurement results, Eq.~\eqref{Model:qe1Threshold}.

\subsection{With a perfect photon-number resolving detector}
Here, we discuss a special case where Alice uses a perfect photon-number resolving detector, discussed in Section \ref{PrfctPNRD}.
Now that Alice knows the exact photon number of the source, Alice and Bob only need to focus the post-processing on single photon state detection events. In this case, the BB84 protocol is implemented by single photon states only. Thus, they can directly apply Shor and Preskill's formula \cite{ShorPreskill_00,EntanglementPDC_07}:
\begin{equation} \label{Post:KeyTrig}
R \geq qQ_1 [1-f(e_1)H_2(e_1)-H_2(e_1)].
\end{equation}
Later from the simulation that is shown in Figure~\ref{Fig:Toytr}, we can see that a perfect photon-number resolving detector does not improve the QKD performance dramatically in comparison to the threshold detector case.

\subsection{A few remarks} \label{Remark}
From the analysis of optimal $\mu$ in Appendix \ref{ApSc:muTrig}, one can see that the key rate for a case without decoy states quadratically depends on the channel loss, $R=O(\eta^2)$, while for the case with decoy states, $R=O(\eta)$. This result is consistent with prior work that compared the cases of a coherent state QKD with and without decoy states \cite{Decoy_05}.

In the decoy state security proof \cite{Decoy_05}, the key assumption is that the decoy state and signal state should satisfy Eq.~\eqref{Decoy:DecoyAss}.
This is guaranteed by the assumption that Eve cannot distinguish decoy and signal states. However, in the active decoy state method, Alice may introduce side information that can distinguish decoy and signal states when she actively prepares decoy and signal states. For example, an attenuator on Alice's side, used to prepare different intensities for signal and decoy states, may introduce different frequency shifts for signal and decoy states \cite{ZQMKQ_06}. In general, it is difficult to verify the assumption that Eve cannot distinguish decoy and signal states in real active decoy state experiments.

In the passive decoy state scheme, decoy and signal states are passively determined by Alice's measurement outcome. Alice does not use an extra component (such as in the active decoy state method) to prepare decoy states. This reduces the possibility of side information leakage.
By passively choosing decoy states, Alice prepares same states on Bob's arm\footnote{Strictly speaking, there is one underlying assumption: the PDC source is single-mode.}. In fact, Alice can measure trigger signals after Bob finishes his measurements. Thus, from Eve's point of view, the states transmitted through the channel is independent of Alice's measurement results ($j$). Therefore, in principle, Eve cannot distinguish the decoy and signal states in the passive decoy state QKD.

This is the main advantage in using the passive decoy state methods. Note that for a coherent state QKD, one can only use the active decoy state idea.

\section{Simulation} \label{Sc:Trig:Simulation}
In this section, we will compare the passive decoy state with a perfect number resolving detector and four QKD implementations with threshold detectors:  non-decoy, infinite decoy, weak active decoy and AYKI (passive decoy state).


We deduce experimental parameters from a recent PDC experiment \cite{PDC144_07},
which are listed in Table \ref{Tab:PDC144}. In the following simulations, we will use $q=1/2$ and $f(E_\mu)=1.22$ in Eq.~\eqref{Post:Keyj}. We notice that with the slightly modified experiment setup, a coherent state QKD with decoy states is implemented \cite{PDC144_07}. Thus, it is reasonable to use this experiment setup to simulate the five QKD implementations.

\begin{table}[hbt]
\centering
\begin{tabular}{|c|c|c|c|c|c|c|c|c|c|c|} \hline
Repetition rate & Wavelength & $\eta_{Alice}$ & $\eta_{Bob}$ & $e_{d}$ & $Y_{0B}$ \\
\hline
249MHz & 710 nm & 14.5\% & 14.5\% & 1.5\% & $6.024\times 10^{-6}$ \\
\hline
\end{tabular}
\caption{List of parameters from the 144 km PDC experiment \cite{PDC144_07}. Here, $\eta_{Alice}$ and $\eta_{Bob}$ are the detection efficiencies in Alice and Bob's detection system, not including the optical channel loss. $e_d$ is the intrinsic detector error rate. $Y_{0B}$ is the background count rate of Bob's detection system (for example, if Bob has two detectors, then $Y_{0B}$ will be the sum of the background count rates of the two detectors). The transmission efficiency $\eta$ in Eq.~\eqref{Model:Yi} is given by $\eta_{Bob}$ plus the channel loss. Since Alice owns the PDC source, $\eta_A=\eta_{Alice}$.} \label{Tab:PDC144}
\end{table}

In the simulation, for fair comparison, we always assume Bob uses the same detection setup (with threshold detectors).

\subsection{Without statistical fluctuations}
\begin{figure}[hbt]
\centering \resizebox{12cm}{!}{\includegraphics{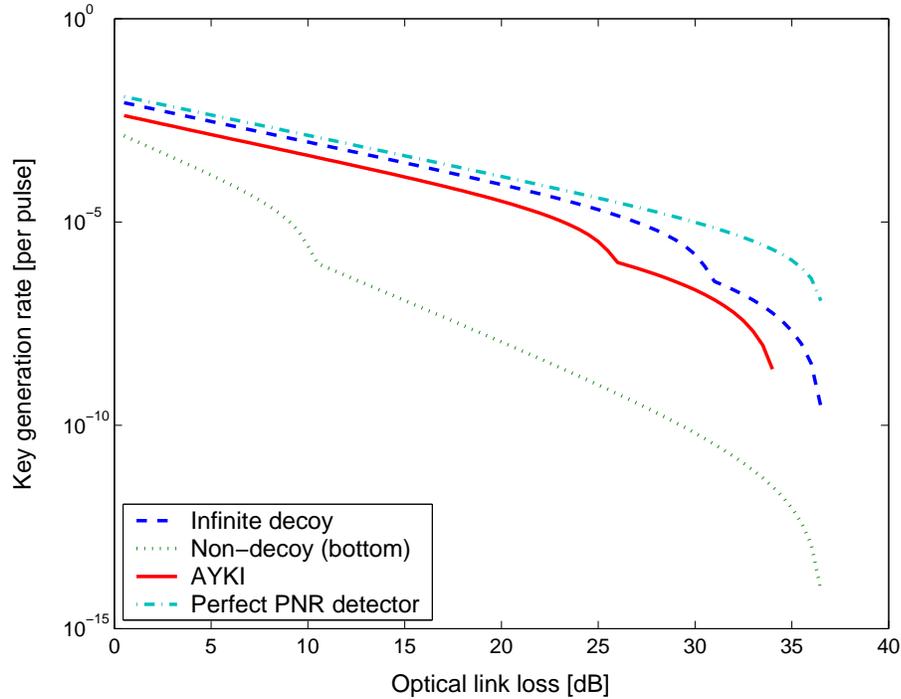}} \caption{Plot of the key
generation rate in terms of the optical loss, comparing four schemes without considering statistical fluctuations: non-decoy, infinite decoy, AYKI and a case with a perfect number resolving detector. Here, we use $q=1/2$ and $f(E_\mu)=1.22$. We numerically optimize $\mu$ for each curve, see Appendix \ref{ApSc:muTrig} for more discussions. Simulation parameters are listed in Table \ref{Tab:PDC144}.} \label{Fig:Toytr}
\end{figure}

In the first simulation, we will consider a case where Alice and Bob perform an infinitely long QKD (no statistical fluctuations). In this case, the weak active decoy state protocol will approach the infinite decoy case, similar to the discussion in Section \ref{Sub:VW}. We assume that Alice is able to adjust $\mu$ (the brightness of the PDC source) in the regime of $[0,1]$ arbitrarily. In the simulation, we numerically optimize $\mu$ for each of the four implementation protocols: non-decoy, infinite decoy, AYKI and a case with a perfect number resolving detector. The simulation result is shown in Figure~\ref{Fig:Toytr}\footnote{Here we simulate a free space QKD setup \cite{PDC144_07}. Since in a free space QKD system, the channel transmittance will depend on not only the distance but also other components, such as the size of the telescope, it is more appropriate to use the optical loss rather than the distance for x-axis of Figure~\ref{Fig:Toytr}.}.

From Figure~\ref{Fig:Toytr}, we have the following remarks.
\begin{enumerate}
\item
In Appendix \ref{ApSc:muTrig}, instead of numerically optimizing $\mu$ as the case was for Figure~\eqref{Fig:Toytr}, we qualitatively investigate the optimal $\mu$ for triggering PDC QKD with and without decoy states. The simulation result is consistent with the qualitative conclusion $R=O(\eta)$ for the case with with decoy states and $R=O(\eta^2)$ for the case without decoy states.

\item
The space between the solid and dashed line in Figure~\ref{Fig:Toytr} indicates
room left for improvement by other decoy protocols with threshold detectors after the
AYKI protocol is implemented. We can see that, in a large regime of the optical link loss (for instance, between 0 dB and 25 dB), the performances of AYKI and the infinite decoy are close. For instance, the AYKI
protocol yields around 50\% of the key rate of the infinite decoy state protocol
when the channel loss is within 20 dB.

\item
By comparing AYKI and a case with a perfect photon-number resolving detector, we can see that even with a perfect photon-number resolving detector on Alice's side, the key rate has not improved dramatically in a large regime of the optical link loss.



\item
The non-decoy protocol is better than the AYKI in the regime close to maximal secure distances. This is because we use the bounds of Eqs.~\eqref{Post:Y1T} and \eqref{Post:e1U} for the AYKI curve. In reality, Alice and Bob can use the bound of Eq.~\eqref{Post:Q1e1nondecoy} directly in this regime.

\item
There is a bump in each curve. This is due to the fact that in the key generation
rate formula Eq.~\eqref{Post:Key01}, the non-triggered detection events have no
contribution to the final secure key after the bump.

\item
At the point of loss=0 dB, the key rates of four cases (from top to bottom) are $1.21\times10^{-2}$, $8.6\times10^{-3}$, $4.2\times10^{-3}$ and $1.3\times10^{-3}$.


\item
At the point of loss=0 dB, the numerical results for optimal $\mu$ for four cases (from top to bottom) are: 1, 0.52, 0.194, 0.0589. The optimal $\mu$ for the case with a perfect threshold detector is always 1, which is reasonable since $\mu=1$ maximizes the single photon state probability. In Appendix \ref{ApSc:muTrig}, we show that the optimal $\mu$s for the infinite decoy and AYKI case are relatively stable in a large regime of the optical link loss (for instance, between 0 dB and 25 dB). The optimal $\mu$ for the no decoy state case decreases with channel loss.

\item
Note that the real $\mu$ used in the experiment \cite{PDC144_07} is $\mu=0.0265$. In general, it is experimentally difficult to increase the brightness ($\mu$) of a PDC source.

\item
All of the four cases can tolerate similar optical losses.
\end{enumerate}


\subsection{With statistical fluctuations}
In a real experiment, the key length is always finite. Alice and Bob should consider statistical fluctuations. As pointed out in Section \ref{Sc:Stat}, the statistical fluctuation analysis is a complicated problem in the decoy state QKD scheme.

Similar to the analysis in Section \ref{Sc:Stat}, we assume a few conditions:
\begin{enumerate}
\item
Alice knows the exact value of the average photon pair number $\mu$, which is a fixed number during key transmission.

\item
The distribution of the photon number, Eq.~\eqref{Impl:PnL}, does not fluctuate.

\item
The QKD transmission is assumed to be part of an infinite length experiment.
\end{enumerate}

Here, we focus on three cases with threshold detectors: infinite decoy, weak decoy and AYKI. We assume that the data size is $6\times10^9$ pulses of Alice's pumping laser. The simulation result is shown in Figure~\ref{Fig:Fluctuation}. From the simulation result, we have the following observations.
\begin{figure}[hbt]
\centering \resizebox{12cm}{!}{\includegraphics{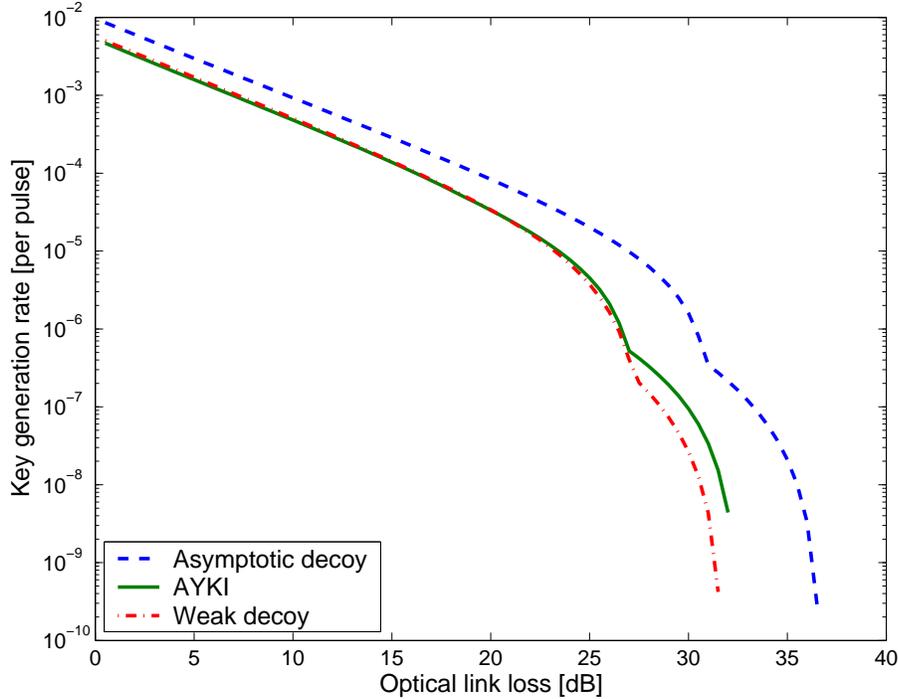}}
\caption{
Plot of the key generation rate in terms of the optical loss, comparing three cases with threshold detectors after considering statistical fluctuations: infinite decoy, weak active decoy and
AYKI. We numerically optimize $\mu$ for each curve. Here, we use $q=1/2$ and $f(E_\mu)=1.22$. In the weak decoy state case, we assume Alice can randomly attenuate her PDC source intensity.  Simulation parameters are listed in Table \ref{Tab:PDC144}. The data size is $6\times10^9$ pumping laser pulses on Alice's side.
} \label{Fig:Fluctuation}
\end{figure}

\begin{enumerate}
\item
Similar to a case without the fluctuation analysis, in a large regime of the optical link loss, the performances of AYKI and the infinite decoy are close.

\item
At the point of loss=0 dB, the key rates of the three cases from top to bottom are $8.6\times10^{-3}$ (infinite), $5.0\times10^{-3}$ (weak) and $4.7\times10^{-3}$ (AYKI).

\item
The maximal tolerable secure optical losses for the three cases are rather similar: 37 dB (infinite), 32.5 dB (AYKI), 32 dB (weak).

\item
The AYKI protocol yields a higher key rate than the weak decoy state protocol when the loss is greater than 16 dB. AYKI is less affected by statistical fluctuations than the weak decoy state because in AYKI, Alice does not need to sacrifice extra pulses for decoy states.


\end{enumerate}

In Section \ref{Remark}, we pointed out that from a practical security point of view, the passive decoy state method has an advantage over active decoy state methods. Moreover, the AYKI method does not require any additional hardware changes to implement the decoy state, while in the weak decoy state case, Alice needs to add an attenuator to create decoy states. Now, from the simulation result, we can see that the AYKI and weak active decoy state method yield a similar QKD performance. Thus, our conclusion is that one should just use the AYKI method instead of the weak decoy state method.

\section{Conclusion}
By investigating the optimal photon source intensity, we find that the triggering PDC QKD setup with decoy states is able to achieve a key rate that linearly depends on the channel transmittance, in comparison to the quadratic dependence for the case without decoy states. Therefore, we expect the decoy state QKD to become a standard technique not only in the coherent state QKD, but also in QKD with triggering PDC sources.

On the practical side, we generalize the passive decoy state idea. The generalized passive decoy state idea can be applied to cases where either threshold detectors or photon number resolving detectors are used. The decoy protocol proposed by Adachi, Yamamoto, Koashi and Imoto (AYKI) can be treated as a special case of the generalized passive decoy state method. In comparison to the active (regular) decoy state methods, the passive one opens less possibility for Eve to distinguish decoy and signal states, which is a key underlying assumption in the security proof of the decoy state QKD scheme. From this sense, the passive decoy state method is more secure than the active decoy state methods in practice.

By simulating a recent PDC experiment, we compared various practical decoy state protocols with the infinite decoy protocol. We also compared cases using threshold detectors and photon-number resolving detectors. Our simulation result shows that with the AYKI protocol, one can achieve a key generation rate that is close to the theoretical limit of infinite decoy protocol. Furthermore, our simulation result suggests that a photon-number resolving detector has little room to improve the QKD performance, in comparison to the case using threshold detectors.

We also considered the statistical fluctuations. We compared infinite decoy protocol, weak active decoy state method and AYKI protocol. The simulation result shows that the weak active decoy state method and AYKI protocol yield a very close QKD performance. In a large regime of the optical link loss, the AYKI protocol can achieve a performance that is close to the infinite decoy case. Since the AYKI protocol requires no hardware changes for triggering PDC QKD, we conclude that AYKI method is a good protocol for triggering PDC QKD experiments.

Although our analysis is focused on QKD with PDC sources, we emphasize that it can also be applied to other QKD setups with triggered single photon sources.


\chapter{Entanglement-based QKD} \label{Chpt:Ent}
A parametric down-conversion (PDC) source can be used as either a triggered single photon source or an entangled photon source in QKD. The triggering PDC QKD was already studied in the previous chapter. However, a model and a post-processing protocol for the entanglement PDC QKD are still missing. Here, we fill in this important gap by proposing such a model and a
post-processing protocol for the entanglement PDC QKD. Although the PDC model is
proposed for studying the entanglement-based QKD, we emphasize that our generic model
may also be useful for other non-QKD experiments involving a PDC source. Since an
entangled PDC source is a basis independent source, we apply Koashi-Preskill's
security analysis to the entanglement PDC QKD. We will also investigate the entanglement
PDC QKD with two-way classical communication. Our results indicate that the recurrence scheme increases the key rate and Gottesman-Lo protocol helps tolerate higher channel losses. By simulating a recent 144 km open-air PDC experiment, we will compare three implementations: entanglement PDC QKD, triggering PDC QKD and coherent state QKD.
The simulation result suggests that the entanglement PDC QKD can tolerate higher
channel losses than the coherent state QKD. The coherent state QKD with decoy states
is able to achieve the highest key rate in the low and medium-loss regions. By applying
Gottesman-Lo two-way post-processing protocol, the entanglement PDC QKD can tolerate
up to 70 dB of combined channel losses (35 dB for each channel) provided that the PDC
source is placed in between Alice and Bob. After considering statistical
fluctuations, the PDC setup can tolerate up to a 53 dB channel loss.

This work is published in Ref.~\cite{EntanglementPDC_07}. In this work, I build an entangled PDC source model, apply Koashi-Preskill's security analysis and simulate a PDC experiment to show the performance of the entanglement-based QKD in comparison with the triggering PDC QKD and coherent state QKD.

\section{Introduction}
As we discussed in Chapter \ref{Chpt:Security}, there are mainly two types of QKD schemes. One is the prepare-and-measure scheme, such as BB84 \cite{BB_84} and the other is the entanglement
based QKD, such as Ekert91 \cite{Ekert_91} and BBM92 \cite{BBM_92}.

With a PDC source, one can realize either prepare-and-measure or entanglement-based QKD protocols \cite{JSWWZ_Ent_00}. To implement a prepare-and-measure QKD protocol, one can use a PDC source as a triggered single photon
source. On the other hand, to implement an entanglement-based QKD protocol, one can use
the polarization entanglement between two modes of light emitted from a PDC source.
We call these two implementations the triggering PDC QKD and entanglement PDC QKD.
With an entangled source, one can also implement QKD protocols based on causality
\cite{MW_Causality_06} and Bell's inequality \cite{AGM_Bell_06}. We notice that the PDC
QKD based on the time-energy entanglement has been exploited \cite{TBZG_EnTime_00}.

Here, we present a model for the entanglement PDC QKD. From the model, we find that an
entangled PDC source is a basis independent source for QKD. Based on this observation, we propose a
post-processing scheme by applying Koashi-Preskill's security analysis \cite{KoashiPreskill_03}.

Recently, a free-space distribution of entangled photons over 144 km was demonstrated
\cite{PDC144_07}. We will simulate this experiment setup and compare three QKD implementations:
entanglement PDC QKD, triggering PDC QKD and coherent state QKD. In the simulation, we will also apply
Gottesman-Lo two-way post-processing protocol \cite{TwoWay_03} and a recurrence scheme
\cite{VV_Recurrence_05}, see also \cite{TwoWay_06}.

The main contributions of this chapter are as follows.
\begin{itemize}
\item
We present a model for the entanglement PDC QKD. Although the model is proposed to
study the entanglement-based QKD, this generic model may also be useful for other
non-QKD experiments involving a PDC source.

\item
From the model, we find that an entangled PDC source is a basis independent source
for QKD. Based on this observation, we propose a post-processing scheme for the
entanglement PDC QKD. Essentially, we apply Koashi-Preskill's security analysis
\cite{KoashiPreskill_03}.

\item
By simulating a PDC experiment \cite{PDC144_07}, we compare three QKD
implementations: entanglement PDC QKD, triggering PDC QKD and coherent state QKD. In
the entanglement PDC QKD, we consider two cases: the source in the middle and source on
Alice's side.

\item
In the case where the PDC source is placed in between Alice and Bob, we find that the
entanglement PDC QKD can tolerate the highest channel losses, up to 70 dB by applying
Gottesman-Lo two-way classical communication post-processing protocol
\cite{TwoWay_03}. Note that a 35 dB channel loss is comparable to the estimated
loss in a satellite to ground transmission in the literature
\cite{Zeilinger_Satellite_03,RGKWK_Space_04,Zeilinger_Space_04,Zeilinger_SpaceGround_04,AMBG_Space_06}.


\item
We consider statistical fluctuations for the entanglement PDC QKD. In this case, the
PDC setup can tolerate up to a 53 dB channel loss.


\item
The coherent state QKD with decoy states is able to achieve the highest key rate in the
low and medium-loss regions.
\end{itemize}

In Section \ref{Sc:Enta:Impl}, we will review two experiment setups of the entanglement PDC
QKD. In Section \ref{Sc:Enta:Model}, the entanglement PDC QKD will be modeled. In Appendix
\ref{QBER}, we calculate the quantum bit error rate in the entanglement PDC QKD. In
Section \ref{Sc:Enta:Post}, a post-processing scheme for the entanglement PDC QKD will be proposed. In Section \ref{Sc:Enta:Simulation}, we will compare the entanglement PDC QKD, the
triggering PDC QKD and the coherent state QKD by simulating a real PDC experiment.
We also apply protocols based on two-way classical communication and consider
statistical fluctuations. In Appendix \ref{ApSc:muEntangle}, the optimal $\mu$ for the entanglement PDC QKD is investigated.

\section{Implementation} \label{Sc:Enta:Impl}
In general, the entangled PDC source does not necessarily belong to one of the two legitimate QKD
users, Alice or Bob. One can even assume that an eavesdropper, Eve, owns the
PDC source. In this section, we will compare two experimental setups of the
entanglement PDC QKD due to the position of the PDC source; in between Alice and Bob
or on Alice's side.

Let us start with a general discussion about an entangled PDC source. With the
rotating-wave approximation, the Hamiltonian of the PDC process can be written as
\cite{KokBraunstein_00}:
\begin{equation}\label{Impl:HamiltonianEn}
\begin{aligned}
H &= i\chi(a^\dagger_Hb^\dagger_V-a^\dagger_Vb^\dagger_H)+h.c. \\
\end{aligned}
\end{equation}
where $h.c.$ means Hermitian conjugate and $\chi$ is a coupling constant depending
on the crystal nonlinearity and the amplitude of the pump beam. The operators $a_i$
and $b_i$ are the annihilation operators for rectilinear polarizations $i\in\{H,V\}$
in modes $a$ and $b$ respectively. Modes $a$ and $b$ are the modes sent to Alice
and Bob, respectively. Notice that the difference between this Hamiltonian and Eq.~\eqref{Impl:HamiltonianTrig} is that in this case, one should consider two freedoms: polarization ($H$ and $V$) and space ($a$ and $b$).


In Section \ref{Sc:Enta:Model}, we will focus on modeling the measurement of the rectilinear
polarization ($Z$) basis. Due to symmetry, all the calculations can be applied to $X$ basis too.

\subsection{Source in the middle}
\begin{figure}[hbt]
\centering \resizebox{12cm}{!}{\includegraphics{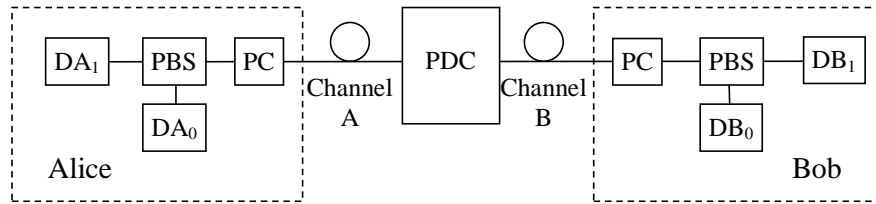}} \caption{A schematic
diagram for the entanglement PDC QKD. Alice and Bob connect to an entangled PDC
source by optical links. They each receive one of two entangled modes coming out
from the PDC source. Both Alice and Bob randomly choose basis (by polarization
controllers) to measure the arrived signals (by single photon detectors). PC:
polarization controller; PBS: polarization beam splitter; DA$_0$, DA$_1$, DB$_0$,
DB$_1$: threshold detectors.} \label{Fig:PDCen}
\end{figure}

First, we consider a case where the PDC source sits in between Alice and Bob. The schematic diagram is shown in Figure~\ref{Fig:PDCen}.

As shown in Figure~\ref{Fig:PDCen}, a PDC source provides two entangled modes, $a$ and $b$, which are
sent to Alice and Bob, respectively. After receiving the signals, Alice and Bob each randomly
choose a basis ($X$ or $Z$) to perform a measurement. A key observation of this setup is that
the state emitted from the PDC source is independent of the bases Alice and Bob that choose for the
measurements.

\subsection{Source on Alice's side} \label{AliceSide}
\begin{figure}[hbt]
\centering \resizebox{12cm}{!}{\includegraphics{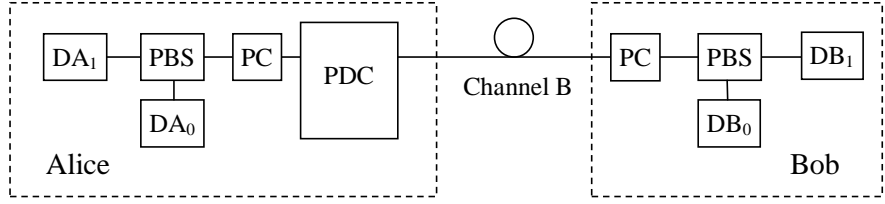}} \caption{A schematic diagram for the
entanglement PDC QKD. Alice measures one of entangled modes coming out from the PDC source and
sends Bob the other mode.} \label{Fig:PDCenA}
\end{figure}

Another case is where Alice owns the PDC source. The schematic diagram is shown in
Figure~\ref{Fig:PDCenA}.

In comparing Figures~\ref{Fig:PDCen} and \ref{Fig:PDCenA}, we can see that the only difference is the position of the PDC source. As we will see Section \ref{Sc:Enta:Post}, the post-processing of these two setups are similar.

Note that in the second setup, Alice's measurement commutes with Bob's measurement.
Thus, we have the same observation as before where the PDC source state is independent of the
measurement bases.

Therefore, for both setups, the entangled PDC source is a basis-independent source. It follows that the entanglement PDC QKD is a basis independent QKD.

\section{Model} \label{Sc:Enta:Model}
In this section, we will model an entangled PDC source, channel and detectors for the
entanglement PDC QKD. We emphasize that this model is applicable for both experiment
setups described in Section \ref{Sc:Enta:Impl}.

\subsection{An entangled PDC source}
From Eq.~\eqref{Impl:HamiltonianEn}, the state emitted from a type-II PDC source can
be written as \cite{KokBraunstein_00}:
\begin{equation}\label{Model:PDCstate}
\begin{aligned}
|\Psi\rangle=(\cosh\chi)^{-2}\sum_{n=0}^{\infty}\sqrt{n+1}\tanh^n\chi|\Phi_n\rangle,
\end{aligned}
\end{equation}
where $|\Phi_n\rangle$ is the state of an $n$-photon-pair, given by:
\begin{equation}\label{Model:PDCn}
\begin{aligned}
|\Phi_n\rangle=\frac{1}{\sqrt{n+1}}\sum_{m=0}^{n}(-1)^m|n-m,m\rangle_a|m,n-m\rangle_b.
\end{aligned}
\end{equation}
For example, when $n=1$, Eq.~\eqref{Model:PDCn} will give a Bell state:
\begin{equation}\label{Model:EPR}
\begin{aligned}
|\Phi_1\rangle &= \frac{1}{\sqrt{2}}(|1,0\rangle_a|0,1\rangle_b-|0,1\rangle_a|1,0\rangle_b) \\
               &= \frac{1}{\sqrt{2}}(|\leftrightarrow\rangle_a|\updownarrow\rangle_b-|\updownarrow\rangle_a|\leftrightarrow\rangle_b), \\
\end{aligned}
\end{equation}
Here, we use the polarizations $|1,0\rangle=|\leftrightarrow\rangle$ and
$|0,1\rangle=|\updownarrow\rangle$ as a qubit basis (Z basis) for QKD. From
Eq.~\eqref{Model:PDCstate}, the probability of getting an $n$-photon-pair is:
\begin{equation}\label{Model:Pn}
\begin{aligned}
P(n)=\frac{(n+1)\lambda^n}{(1+\lambda)^{n+2}}
\end{aligned}
\end{equation}
where we define $\lambda\equiv\sinh^2\chi$. The expected photon pair number is $\mu=2\lambda$,
which is the average number of photon pairs generated by one pump pulse, characterizing the
brightness of a PDC source.


\subsection{Detection} \label{Detection}
Now we need to consider two channels: one for Alice and the other for Bob. We can apply the photon number channel model, described in Section \ref{Sub:PhNch}, to each arm. The yield of an $n$-photon-pair $Y_n$ mainly comes from two parts, the background and the true signal. Assuming that the background counts are independent of the signal photon detection, then $Y_n$ is given by:
\begin{equation}\label{Model:Yn}
\begin{aligned}
Y_n 
    &= [1-(1-Y_{0A})(1-\eta_A)^n][1-(1-Y_{0B})(1-\eta_B)^n] \\
\end{aligned}
\end{equation}
where $Y_{0A}$ and $Y_{0B}$ are the background count rates on the sides of Alice and Bob, respectively. The vacuum state contribution is $Y_0=Y_{0A}Y_{0B}$.
The {\it gain} of the $n$-photon-pair $Q_n$, which is the product of Eqs.~\eqref{Model:Pn} and
\eqref{Model:Yn}, is given by:
\begin{equation}\label{Model:Qn}
\begin{aligned}
Q_n &= Y_nP(n) \\
    &= [1-(1-Y_{0A})(1-\eta_A)^n][1-(1-Y_{0B})(1-\eta_B)^n]\frac{(n+1)\lambda^n}{(1+\lambda)^{n+2}}. \\
\end{aligned}
\end{equation}

The overall gain is given by:
\begin{equation}\label{Model:Enta:Gain}
\begin{aligned}
Q_{\lambda} &= \sum_{n=0}^{\infty} Q_n \\
        &= 1-\frac{1-Y_{0A}}{(1+\eta_A\lambda)^2}-\frac{1-Y_{0B}}{(1+\eta_B\lambda)^2}+\frac{(1-Y_{0A})(1-Y_{0B})}{(1+\eta_A\lambda+\eta_B\lambda-\eta_A\eta_B\lambda)^2}.
\end{aligned}
\end{equation}
Here, the overall gain $Q_\lambda$ is the probability of a coincident detection event given a pump
pulse. Note that the parameter $\lambda$ is one half of the expected photon pair number $\mu$.

The overall quantum bit error rate (QBER, $E_\lambda$) is given by:
\begin{equation}\label{Model:Enta:QBER}
\begin{aligned}
E_{\lambda}Q_{\lambda} 
               =&e_0Q_{\lambda}-\frac{2(e_0-e_{d})\eta_A\eta_B\lambda(1+\lambda)}{(1+\eta_A\lambda)(1+\eta_B\lambda)(1+\eta_A\lambda+\eta_B\lambda-\eta_A\eta_B\lambda)} \\
\end{aligned}
\end{equation}
where $Q_{\lambda}$ is the gain given in Eq.~\eqref{Model:Enta:Gain}. The calculation of the
$E_{\lambda}$ is shown in Appendix \ref{QBER}.

\section{Post-processing} \label{Sc:Enta:Post}
As mentioned in Section \ref{Sc:Enta:Impl}, the entanglement PDC QKD is a basis-independent
QKD. Thus, we can apply Koashi and Preskill's security proof \cite{KoashiPreskill_03}. The key generation rate is given by:
\begin{equation} \label{Post:KeyEn}
R \geq q \{Q_{\lambda}[1-f(\delta_b)H_2(\delta_b)-H_2(\delta_p)]\}.
\end{equation}
where the subscript $\lambda$ denotes for one half of the expected photon number $\mu$, $Q_{\lambda}$ is the overall gain, $\delta_b$ ($\delta_p$) is the bit (phase) error rate, $f(x)$ is the bi-direction error correction efficiency.

Due to the symmetry of $X$ and $Z$ bases measurements, as shown in Section
\ref{Sc:Enta:Impl}, $\delta_b$ and $\delta_p$ are given by:
\begin{equation} \label{Post:deltabp}
\delta_b=\delta_p=E_\lambda,
\end{equation}
where $E_\lambda$ is the overall QBER. This equation is true for the asymptotic
limit of an infinitely long key distribution. Later, in Section \ref{Sub:Ent:StaFlu}, we will see that Eq.~\eqref{Post:deltabp} may not be true when statistical
fluctuations are taken into account.

Note that in Koashi and Preskill's security proof, the squash model
\cite{GLLP_04} is applied. In the squash model, Alice and Bob project the state onto
the qubit Hilbert space before $X$ or $Z$ measurements. For more details of the
squash model, one can refer to \cite{GLLP_04}. In the case where Alice owns the PDC
source, as discussed in Subsection \ref{AliceSide}, the key rate formula of
Eq.~\eqref{Post:KeyEn} has been proven \cite{Koashi_NewModel_06} to be valid for the
QKD with threshold detectors without the squash model, see also
\cite{LoPreskill_NonRan_06}.  We also notice that this post-processing scheme,
Eqs.~\eqref{Post:KeyEn} and \eqref{Post:deltabp}, can be derived from the
security analysis based on the uncertainty principle \cite{Koashi_Uncer_06}.

In Eq.~\eqref{Post:KeyEn}, $Q_\lambda$ can be directly measured from a QKD experiment and
$E_\lambda$ can be estimated by error testing. In the simulation shown in Section \ref{Sc:Enta:Simulation},
we will use Eqs.~\eqref{Model:Enta:Gain} and \eqref{Model:Enta:QBER}.

Note that the post-processing for the entanglement PDC QKD is simpler than the
coherent state QKD and triggering PDC QKD. In the entanglement PDC QKD, all the
parameters needed for the post-processing ($Q_{\lambda}$ and $E_{\lambda}$) can be
directly calculated or tested from the measured classical data. On the other hand, in the coherent PDC QKD and the triggering PDC QKD, Alice and Bob need to know the
value of some experimental parameters ahead of time, such as the expected photon number
$\mu$. They also need to estimate the gain and error rate of the single photon states
$Q_1$ and $e_1$, which make the statistical fluctuation analysis difficult
\cite{Practical_05}, as investigated in Section \ref{Sc:Stat}.

The post-processing can be further improved by introducing two-way classical
communication between Alice and Bob \cite{TwoWay_03,TwoWay_06}. Moreover, the adding
noise technique may enhance the performance \cite{KGR_noise_05}.

\section{Simulation} \label{Sc:Enta:Simulation}
In this section, we will first compare three QKD implementations: entanglement PDC
QKD, triggering PDC QKD and coherent state QKD. Then we will apply post-processing
protocols with two-way classical communication to the entanglement PDC QKD.
Finally, we will consider the statistical fluctuations.

We deduce parameters from a recent PDC experiment \cite{PDC144_07} with respect to the model given
in Section \ref{Sc:Enta:Model}, which are listed in Table \ref{Tab:PDC144}. For the
coherent state QKD, we use $\eta_A=1$ since Alice prepares the states in this case.
In the following simulations, we will use $q=1/2$ and $f(E_\mu)=1.22$
\cite{BrassardSalvail_93}.


The optimal expected photon number $\mu$ of the coherent state QKD is discussed in Ref.~\cite{IndividualAttack_00,Practical_05}. In Appendix \ref{ApSc:muEntangle}, we investigate the optimal $\mu$ ($2\lambda$) for the entanglement PDC QKD. Not surprisingly, we find that the optimal $\mu$ for the entanglement PDC QKD is in the order of 1, $\mu=2\lambda=O(1)$. Thus, the key generation rate given in Eq.~\eqref{Post:KeyEn} depends linearly on the channel transmittance.

\subsection{Comparison of three QKD implementations}
\begin{figure}[hbt]
\centering \resizebox{12cm}{!}{\includegraphics{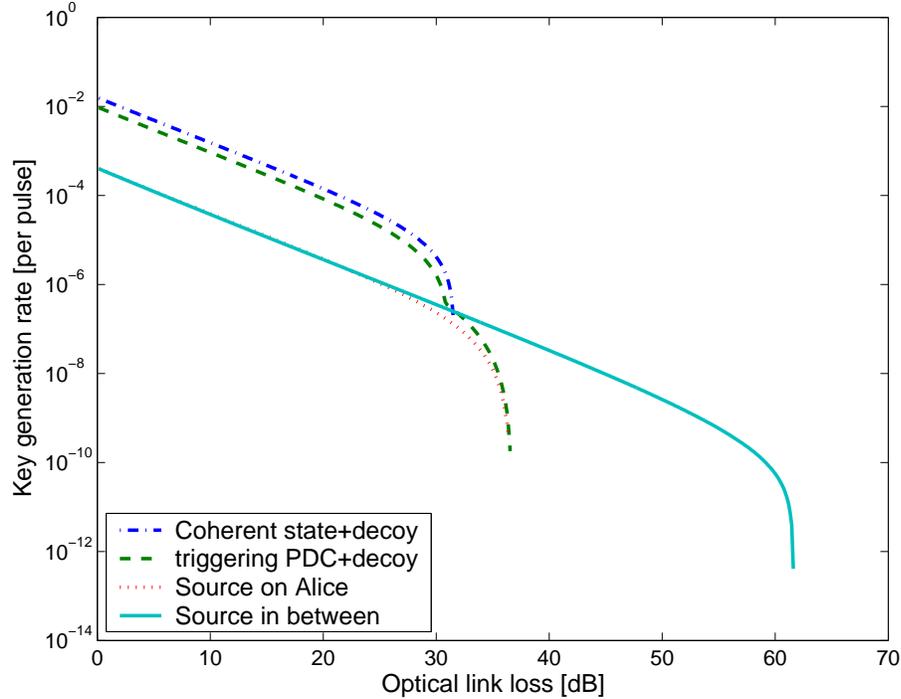}} \caption{Plot of the key generation rate in terms of the optical loss, comparing four cases: coherent state QKD+aysmptotic decoy, triggering PDC+asymptotic decoy, and entanglement PDC QKD (source in the middle and source on Alice's side). For the coherent state QKD+decoy, we use $\eta_A=1$. We numerically optimize $\mu$
($2\lambda$) for each curve. The simulation of triggering PDC QKD with decoy states can be found in Section \ref{Sc:Trig:Simulation}.} \label{Fig:Toyen}
\end{figure}

In the first simulation, we assumed that Alice was able to adjust the expected photon
pair number $\mu$ ($2\lambda$, the brightness of the PDC source) in the region of
$[0,1]$. Thus, we can optimize $\mu$ for the entanglement PDC QKD and the triggering
PDC QKD. The simulation results are shown in Figure~\ref{Fig:Toyen}. For the simulation of triggering PDC QKD with decoy states, one can refer to Section \ref{Sc:Trig:Simulation}.

From Figure~\ref{Fig:Toyen}, we have the following remarks.
\begin{enumerate}
\item
The entanglement PDC QKD can tolerate the highest channel losses in the case where
the source is placed in the middle between Alice and Bob.

\item
The coherent state QKD with decoy states is able to achieve the highest key rate in
the low and medium-loss region. This is because in the coherent state QKD
implementation, Alice does not need to detect any photons, which will effectively
give $\eta_A=1$ in the PDC QKD implementations.

\item
In comparing two cases of the entanglement PDC QKD with a source in the middle and source on
Alice's side, they yield a similar key rate in the low and media- region. However, the source
in the middle case can tolerate higher channel losses.
\end{enumerate}

In the following simulations, we will focus on the case where the entangled PDC
source sits in the middle between Alice and Bob.

\subsection{With two-way classical communication}
\begin{figure}[hbt]
\centering \resizebox{12cm}{!}{\includegraphics{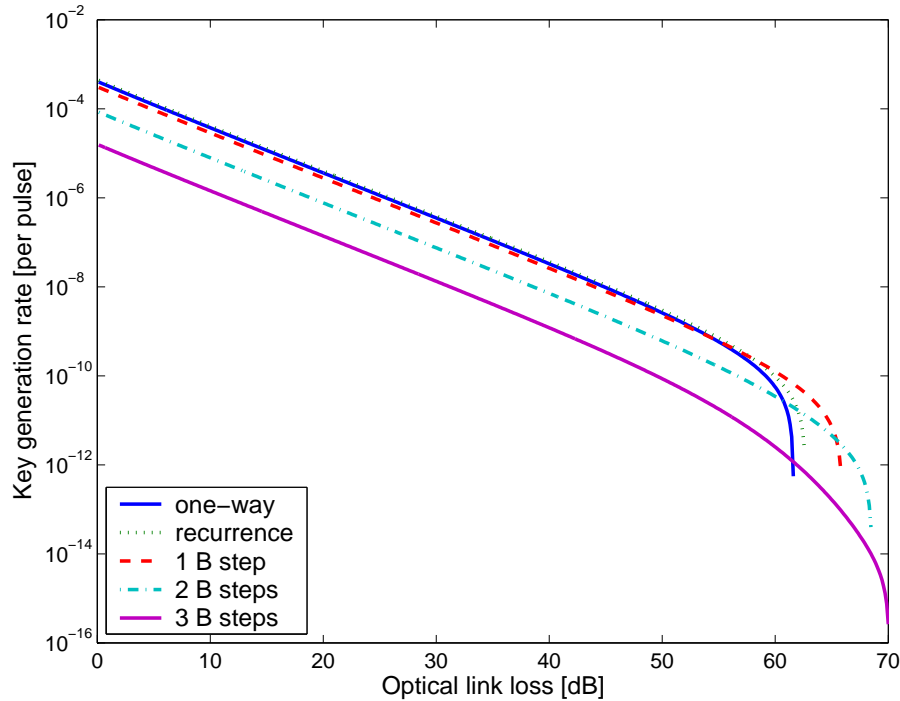}} \caption{Plot of the key generation rate in terms of the optical loss. We apply the
recurrence idea and up to 3 B steps. $\mu$ is numerically optimized for each curve.}
\label{Fig:TwoWay}
\end{figure}

We can also apply the idea of post-processing with two-way classical communication. Similar to the argument in Chapter \ref{Chpt:TwoWay}, we can apply the recurrence idea \cite{VV_Recurrence_05} and the B steps described in Section \ref{Sub:BP}. The simulation results are shown in Figure~\ref{Fig:TwoWay}.

From Figure~\ref{Fig:TwoWay}, we can see that the recurrence scheme can increase the key
rate by around 10\% and extend the maximal tolerable loss by around 1 dB. The PDC
experiment setup can tolerate up to a 70 dB channel loss with 3 B steps. Note that 70 dB
(35 dB in each channel) is comparable to the estimated loss in a satellite to ground
transmission \cite{Zeilinger_SpaceGround_04}.
This result suggests that satellite-ground QKD may be possible.  However, this
simulation assumes an ideal situation where  an infinite number of signals are
transmitted. Moreover, we assume that $\mu$ (the brightness of the PDC source) is a
freely adjustable parameter in the PDC experiment. In a more realistic case where a
finite number of signals are transmitted and $\mu$ is a fixed parameter, the
tolerable channel loss becomes smaller, which will be shown next.

\subsection{Statistical fluctuations} \label{Sub:Ent:StaFlu}
In Eq.~\eqref{Post:deltabp}, we assume that $\delta_b$ and
$\delta_p$ are the same due to the symmetry between $X$ and $Z$
measurements. Alice and Bob randomly choose to measure in $X$ or $Z$
basis. Then asymptotically, $\delta_b$ is good estimate of
$\delta_p$. However, in a realistic QKD experiment, only a finite
number of signals are transmitted. Thus $\delta_p$ may slightly
differ from $\delta_b$.
We assume that Alice and Bob do not perform error testing explicitly. Instead, they
obtain the bit error rate directly from an error correction protocol (e.g., the
Cascade protocol \cite{BrassardSalvail_93}). In such a case, there is no fluctuation
in the bit error rate $\delta_b=E_\lambda$. On the other hand, the phase error rate
may fluctuate to a certain value of $\delta_p=\delta_b+\epsilon$. Following the
fluctuation analysis of Ref.~\cite{ShorPreskill_00}, we know that the probability of
getting an $\epsilon$ bias is
\begin{equation} \label{Simulation:Confi}
P_{\epsilon} \leq \exp[{-\frac{\epsilon^2n}{4\delta_b(1-\delta_b)}}],
\end{equation}
where $n=NQ_\lambda$ the number of detection events, the product of total number of pulses $N$ and
the overall gain $Q_\lambda$.

In the 144 km PDC experiment \cite{PDC144_07}, the repetition rate of the pump pulse is
249MHz as given in Table \ref{Tab:PDC144}. As discussed in
Ref.~\cite{Zeilinger_SpaceGround_04}, the typical time of a ground-satellite QKD
allowed by satellite visibility is 40 minutes. Here, we assume the experiment runs
10 minutes, which means the data size (the number of the pumping pulses) is $N=1.5\times10^{11}$. By taking this data size, we considered the fluctuations for the entanglement PDC QKD.

In a realistic case, the brightness of the PDC source $\mu$ cannot be set freely.
In the 144 km PDC experiment \cite{PDC144_07}, the expected photon pair number is
$\mu=2\lambda=0.053$. After taking $\mu=0.053$ and the data size of
$N=1.5\times10^{11}$ for the fluctuation analysis, the simulation result is shown in
Figure~\ref{Fig:Fluctuation}.

\begin{figure}[hbt]
\centering \resizebox{12cm}{!}{\includegraphics{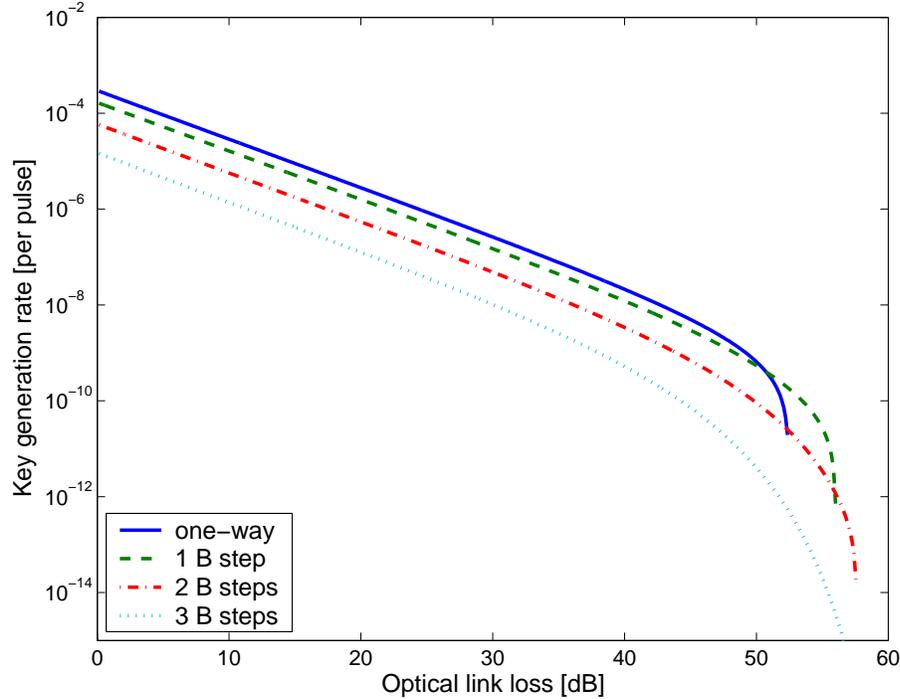}} \caption{Plot of the key generation rate in terms of the optical loss. We take a realistic
$\mu=2\lambda=0.053$, and consider a fluctuation with a data size (the number of the pumping pulses) of $N=1.5\times10^{11}$ and a confident interval of $1-P_{\epsilon} \ge 1-e^{-50}$.}
\label{Fig:Fluctuation}
\end{figure}

We have a couple remarks about Figure~\ref{Fig:Fluctuation}.
\begin{enumerate}
\item
In Figure~\ref{Fig:Fluctuation}, if we use the key rate of $10^{-10}$ as the cut-off point\footnote{Then the final key length is 15 bits. One should also consider the cost in
the authentication procedure. Thus this is a reasonable cut off point.}, the
entanglement PDC QKD with one B step can tolerate up to a 53 dB transmission loss.

\item
We have tried simulations with various $\mu$s. We find that the key rate is stable
with moderate changes of $\mu$. With the above fluctuation analysis, if we numerically
optimize $\mu$ for each curve, the maximal tolerable channel loss (with cut off key
rate of $10^{-10}$) is only 1 dB larger than the one given by $\mu=0.053$. Thus, one
cannot significantly improve the maximal tolerable channel loss by just using a
better PDC source in the 144 km PDC experiment setup \cite{PDC144_07}.


\end{enumerate}

\section{Conclusion}
We proposed a model and post-processing protocol for the entanglement PDC QKD. We find
that the post-processing is simple by applying Koashi-Preskill's security proof due
to the fact that the entanglement PDC QKD is a basis independent QKD. Specifically,
only directly measured data (the overall gain and the overall QBER) are needed to
perform the post-processing. By simulating a recent experiment, we compare three QKD
schemes: coherent state QKD+aysmptotic decoy, triggering PDC+asymptotic decoy, and
entanglement PDC QKD (source in the middle and on Alice's side). We find that a) the
entanglement PDC (with source in the middle) can tolerate the highest channel loss;
b) the coherent state QKD with decoy states can achieve the highest key rate in the
medium- and low-loss regions; c) asymptotically, with a realistic PDC experiment
setup, the entanglement PDC QKD can tolerate up to a 70 dB channel loss by applying
post-processing schemes with two-way classical communication; d) the PDC setup can
tolerate up to a 53 dB channel loss when statistical fluctuations are taken into
account.

\chapter{Quantum cryptanalysis} \label{Chpt:Attack}
In this chapter, we will discuss existing security loopholes in current QKD setups. We propose a technologically feasible attack and present possible solutions. Note that although the attack is proposed for the BB84 coherent state QKD implementation, the attack works for many other protocols as well. 

The theoretical work of the time-shift attack is published in Ref.~\cite{QFLM_TimeShift_07}. The security proof of efficiency mismatch is presented in Ref.~\cite{Mismatch_security_08}. Aside from the decoy state method, we also studied other methods to improve the QKD performance, such as dual detector scheme \cite{QZMLQ_dual_07,QZMLQ_eff_07}. Note that I am not the main contributor of these projects. I joined in discussions and helped work out the details.

\section{Side information}
In Chapter \ref{Chpt:Security}, we discussed various security analyses of QKD. In many cases, we assumed that Eve cannot learn about bit values or basis information directly from Alice and Bob's systems, e.g., by breaking into Alice or Bob's box. As we pointed out in Section \ref{Sc:Squash}, in the security proofs, many rely on the assumption of the squash model. In reality, the bit value or basis information might be revealed to Eve through some side channels. For example, two detectors used in QKD systems may have different properties, which might reveal to Eve partial information about the bit values.

\subsection{Detector inefficiency loophole}
Before examining the details of possible side information channels in current QKD setups, let us take a look at a fundamental reason for existence of these loopholes.

An important piece of evidence that indicates the validity of quantum mechanics is shown by the violation of the Bell inequality \cite{Bell_Ineq_64} and its descendant experiment verifications (see for example, Ref.~\cite{Bellin_exp_81}). The experiments show that the concept of traditional local realism is inconsistent with quantum mechanics and then, with the real world. However, this verification has not been completely conclusive, since there exists certain loopholes in these experiments. See for example, \cite{Pearle_Bell_70,CH_Bell_74,GM_Bell_87}.

Since entanglement is the precondition of QKD security \cite{CLL_Precondition_04} and the concept of entanglement is closely related to Bell's inequality\footnote{Although entanglement does not promise violation of the Bell's inequality.}, a natural question is ``Does this detector inefficiency loophole affect the security of QKD?" As we will show shortly, the answer is \emph{yes}.

\subsection{Timing information}
In many QKD systems, detectors are operated in a gated mode in order to reduce the dark count rate. In general, the width of SPD's open window (a few ns) is often substantially larger than the laser pulse duration (a few hundreds ps). Here, we treat the signal pulse as a delta function in time-domain.

\begin{figure}[hbt]
\centering \resizebox{12cm}{!}{\includegraphics{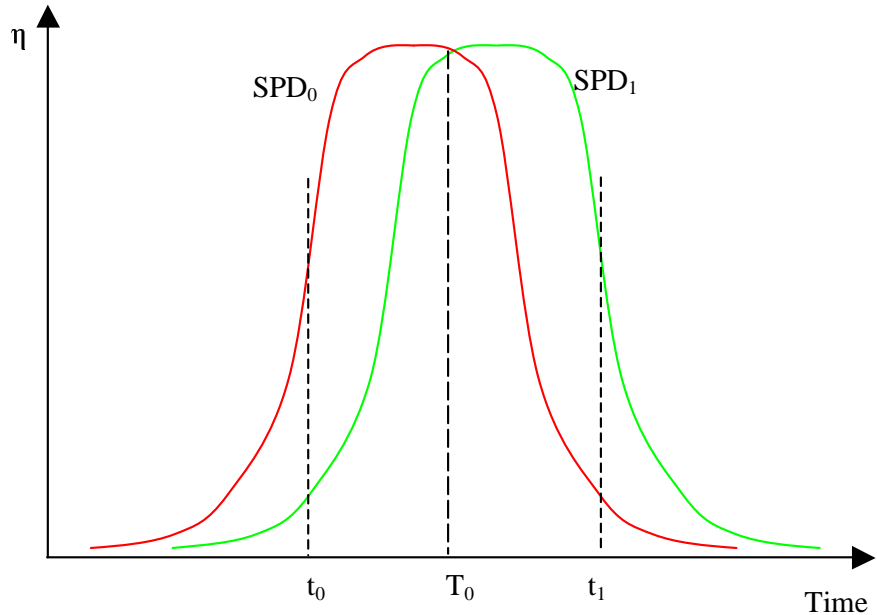}}
\caption{The time-dependence efficiencies of single photon detectors (SPDs).} \label{Fig:SPD-a}
\end{figure}

Typically, Bob uses two separate single photon detectors, which are labeled as SPD0 and SPD1, to detect bit ``0'' and bit ``1'', respectively.  In real life, due to device imperfections, the time-dependent efficiencies of the two detectors are not identical in general as shown in Figure~\ref{Fig:SPD-a}.

Ideally, Alice and Bob can synchronize the laser pulse with the center of the time window ($T_0$ in Figure~\ref{Fig:SPD-a}). This ensures that a small detector efficiency mismatch will not affect the normal operation of the QKD system. In reality, the timing may be shifted by a small amount due to fluctuations or device imperfections\footnote{Shortly, we will see that Eve may shift the pulse large for her attack.}. Thus, the pulse timing contains information about the detector efficiencies, which may reveal the detection bit values.

Note that other freedoms of the signal may also introduce similar problems. For example, two detectors may respond differently in the frequency domain \cite{QFZMTCL_attack_07}. In the following discussions, we will focus on the efficiency mismatch due to signal timing.

\section{Time-shift attack} \label{Sc:Timeshift}
Recently, an eavesdropping attack that exploits this efficiency mismatch of detectors in the QKD system has been proposed \cite{MAS_Eff_06}. In this attack, Eve intercepts and performs a complete von Neumann measurement on each quantum state sent out by Alice. She then generates a new time-shifted signal based on her measurement result and sends it to Bob.

Note that to implement this attack in Ref.~\cite{MAS_Eff_06}, Eve will need a complicated detection (similar to Bob's system) and resend (similar to Alice's system) system. If we assume that Eve builds her ``practical'' eavesdropping device based on today's technology, she will also experience the problem of low detection efficiency and will introduce additional errors due to imperfections in her setup.

Based on this work, we propose a simple practical attack: time-shift attack \cite{QFLM_TimeShift_07}. In our attack, Eve does not measure the quantum state that is sent to Alice. Instead, Eve simply shifts the arrival time of either the signal pulse or the synchronization (reference) pulse or both between Alice and Bob. Consequently, Eve has control of the arriving time of the pulse. For example, she shifts the pulse to $t_0$ in Figure~\ref{Fig:SPD-a} and then Bob claims a detection event of that pulse. Now, Eve knows with a high probability that SPD0 clicks. Hence, she can guess Bob's measurement result 0. In an extreme case where there is a complete detector efficiency mismatch\footnote{That is to say, there is a time window where SPD0 (or SPD1) is active while SPD1 (or SPD0) is completely inactive.}, Eve can acquire full information on the final key without introducing any error. In other words, a na\"{i}ve application of standard security proofs, for instance, the GLLP \cite{GLLP_04} security analysis, without taking into account the detector efficiency mismatch is invalid.

\begin{figure}[hbt]
\centering \resizebox{12cm}{!}{\includegraphics{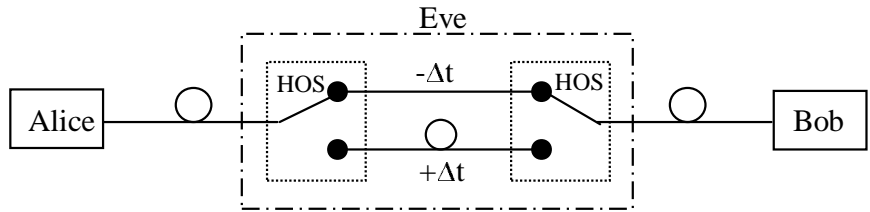}}
\caption{A schematic diagram of Eve's attack. HOS: high-speed optical switch.} \label{Fig:attack1}
\end{figure}

Figure~\ref{Fig:attack1} shows a schematic diagram for the experimental realization of the time-shift attack. Instead of measuring Alice's quantum state, Eve just randomly shifts the time of Alice's quantum state to make sure that it arrives at Bob's detector at either time $t_0$ or $t_1$.  When Eve chooses time $t_0$ and Bob detects a signal, with the probability of $\eta_0/(\eta_0+\eta_1)$, the bit value will be ``0''. Here, we assume that the detector efficiencies of SPD0 and SPD1 are $\eta_0$ and $\eta_1$ at time $t_0$ and Alice chooses bit ``0'' and ``1'' with an equal prior probability. 
Because the probability that Eve incorrectly guesses Bob's bit value is $\eta_1/(\eta_0+\eta_1)$, therefore, Eve's knowledge about the final key is given by:
\begin{equation}\label{Mismatch:EveInfo}
\begin{aligned}
I(B:E) = 1 - H_2(\frac{\eta_1}{\eta_0+\eta_1}).
\end{aligned}
\end{equation}
Note in this attack, Eve does not measure Alice's state. Therefore, Eve will not introduce extra errors. Due to the symmetry, the same analysis can also applied to the case when Eve chooses $t_1$.

In comparison with the attack in Ref.~\cite{MAS_Eff_06}, our attack is simpler and can be easily realized with today's technology: Eve can use high speed optical switches to re-route Alice's signal through either a long or short optical path to achieve the desired time shift. Another advantage of our attack is that Eve will never introduce errors. Therefore, it is difficult for Alice and Bob to detect Eve's presence.

For details of the time-shift attack, one can refer to Ref.~\cite{QFLM_TimeShift_07}. Note that our time-shift attack was experimentally realized in our lab \cite{Timeshift_Exp_07}.


%
%
%

\section{Security against time-shift attack}
Now that we know about the time-shift attack, we can provide a secure QKD against the attack. There are two approaches: hardware based and software based. In the hardware based approach, we perform some counter measurements or improve the system setups. In the software based approach, we provide a security analysis with detector efficiency mismatch.

\subsection{A simple solution}
To counter Eve's attack, Alice and Bob could develop various countermeasures, such as those discussed in Ref.~\cite{MAS_Eff_06}. Note that a recently proposed single SPD QKD system is also immune to this attack \cite{USpa_Mismatch_05}. In a phase encoding BB84 version of this design, instead of randomly selecting from a set of two values, Bob's phase modulation is randomly selected from a set of four values, which is identical to the set for Alice's phase modulation. In this case, Bob not only randomly chooses his measuring basis for each incoming pulse, he also randomly determines which SPD is used for detecting bit ``0'' or bit ``1''. Bob broadcasts his basis choice, but keeps his choice of detector (for the bit ``0'' or ``1'') secretly. In such a set-up, even if Eve has information about which detector clicks, Eve still cannot work out Bob's bit value because she does not know which detector corresponds to the bit ``0". Bob's random choice of detectors to detect the bit ``0'' or ``1'' will even out the efficiency mismatch.

\subsection{Security proof for a QKD system with detector efficiency mismatch}
Here, we will only discuss the security proof for a simple scenario: single photon source, noiseless channel and the efficiencies of two detectors, which are $\eta_0$ and $\eta_1$, to detect the bit ``0" and ``1"\footnote{In real time-shift attack, Eve might shift the pulse in various positions. Here, we only consider one point that will cause a detector efficiency mismatch. In general, $\eta_0$ and $\eta_1$ can be characterized by a tensor in the auxiliary dimension (for instance, time domain).}. For a full discussion of the security proof for a QKD system with efficiency mismatch detectors, one can refer to Ref.~\cite{Mismatch_security_08}.

In this simple QKD picture, Eve does not introduce any bit or phase errors, but only intervenes in the auxiliary dimension to gain side information. As discussed in Section \ref{Sc:SingleEDP}, the state shared by Alice and Bob after transmission (Eve's intervention) and basis reconciliation is
\begin{equation}\label{Mismatch:Source}
\begin{aligned}
(\ket{00}+\ket{11})_{AB} \mapsto (\sqrt\eta_0\ket{00}+\sqrt\eta_1\ket{11})_{AB}
\end{aligned}
\end{equation}
Eve does not introduce any bit errors and she simply attaches an extra system $T$, by shifting the timing of the signals that represents her intervention in the auxiliary dimension.

With a hashing based EDP \cite{BDSW_96}, the amount of EPR pairs that Alice and Bob can distill from the final state is $H_2(\eta_0/(\eta_0+\eta_1))$, which is consistent with the result of Eq.~\eqref{Mismatch:EveInfo}. Note that when $\eta_0\neq\eta_1$, the key rate is less than 1 in comparison to the perfect case of $R=1$.

\section{Discussion}
From this cryptanalysis exercise, we learn that a security proof is only as good as its underlying assumptions. Once a security loophole has been discovered, it is often not very difficult to develop countermeasures that will plug the loophole and regain unconditional proofs of security of the QKD system. One example is the time-shift attack that we mentioned above. However, the difficult part is how to identify such security loopholes in the first place. A QKD system is a complicated system with many intrinsic imperfections. It is, thus, very important to conduct extensive research on such imperfections carefully to determine if they are innocent or fatal for security. We need more quantum hackers in the field. The investigation of loopholes and countermeasures in practical QKD systems plays a complementary role to security proofs.

Given that a practical QKD system will always have imperfections, one might wonder if QKD systems offer any real advantages over conventional systems. Our answer is three-fold. First of all, implementation loopholes are a fact of life. Even conventional security systems, such as smart cards, suffer implementation loopholes. For instance, Eve may attempt to read off a private key from a smart card by using various techniques (including X-ray) to reverse-engineer the circuit embedded in a smart card. Secondly, QKD can be used in concatenation with a conventional system to ensure security. By defending in depth, QKD can only increase security, not reduce it. Thirdly, QKD has an important advantage of being future-proof: The signals are quantum. Once the transmission is done, there is no transcript for the transmission. For an eavesdropper to launch a quantum attack, she has to possess much of the quantum technology during the quantum transmission. In contrast, in a standard Diffie-Hellman public-key key exchange scheme, Eve has a complete transcript of the transmission and can save such a transcript for decades to wait for unexpected future advances in hardware and algorithms. Given that public key crypto-systems are an unexpected discovery made only three decades ago, our view is that it will be complacent to believe that our standard public key crypto-systems will be safe forever. Therefore, it pays to reduce one's risk by defending in depth with a QKD system in concatenation with a conventional cryptosystem.

\chapter{Conclusions and outlook} \label{Chpt:Conclusion}
In this chapter, I will conclude my thesis by summarizing the results of my Ph.D.~study and stating some interesting topics for future research.

\section{Decoy state QKD}
The major topic in my Ph.D. study is decoy state quantum key distribution (QKD). The main results are presented in Chapters \ref{Chpt:Decoy}, \ref{Chpt:Practical}, \ref{Chpt:TwoWay} and \ref{Chpt:Trig}.

Recall that the motivation of this thesis is to bridge the gap between theory and practice of QKD. One of the major problems in a practical QKD system is that a single photon source is difficult to obtain with current technology. Now, with the decoy state method, the key rate is linearly dependent on the channel transmission. Note that this is the highest order that the key rate can reach even with a perfect single photon source. Hence, with decoy states, one can treat weak coherent state sources and triggering parametric down-conversion (PDC) sources as good single photon sources for QKD setups. 

For practical implementations, we showed that with only one or two decoy states, one can achieve most of the benefits of the decoy state method. Further improvement for the decoy state QKD was studied by considering two-way classical communication in the post-processing step. With our two-way classical communication based schemes, one can obtain a performance that is close to the theoretical limit. We also investigated the decoy state method for other photon sources, triggering PDC source. With similar results concluded, we expected the decoy state QKD to become a standard technique not only in the coherent state QKD, but also in QKD with triggering PDC sources.

All the decoy state QKD experiment demonstrations, including our first realization, showed that the decoy state idea is easy to implement in real system setups. Therefore, we conclude that the practical quantum cryptography is close to real-life applications.

\section{Other topics}
As an extension of the decoy state QKD work, we searched for other techniques to improve the QKD performance of practical systems. We proposed a dual detector scheme to improve the case when fast and noisy detectors are in use.

We also investigated other QKD protocols, such as the entanglement based QKD protocols. By simulating a recent experiment, we showed that a) with an entangled PDC source in the middle, the QKD setup can tolerate highest channel loss comparing to decoy state QKD protocols; b) the coherent state QKD with decoy states can achieve the highest key rate in the medium- and low-loss regions.

Security is the most important issue in QKD. We studied various eavesdropping attack schemes in quantum cryptography. We proposed a technologically feasible attack scheme and presented possible solutions. Note that although the attack is proposed for the BB84 coherent state QKD implementation, the attack works for many other protocols as well. We also studied the countermeasures against this attack. We provided a security proof for a QKD system with detector efficiency mismatch.

\section{Future work outlook}
In the future, one interesting topic is a natural extension of my previous work: enhancing the performance of practical QKD systems. Further improvements, both in key rate and secure transmission distance, are required for some applications. Another crucial point is that, in real life, one needs to consider some extra disturbances (e.g., quantum signals may share the channel with regular classical signals). The final goal is to achieve a customer friendly QKD system that can be easily integrated with the Internet.

To achieve an intercontinental transmission distance, ground-satellite QKD is a promising proposal. One interesting project is to test the feasibility of ground-satellite QKD. In Chapter \ref{Chpt:Ent}, we have preliminarily studied the feasibility of ground-satellite QKD with the current entangled photon source. Previously, we used a beam splitter as a channel model for ground-satellite QKD. A study of the disturbance of atmosphere is needed to develop a more realistic model for the ground-satellite channel. By modeling and simulating, one can investigate the requirement for QKD components. For example, what efficiency and noise level of single photon detectors are required and how large the telescope is needed. Meanwhile, it is interesting to explore good QKD schemes for ground-satellite QKD.

To achieve a higher QKD key rate, one can consider other QKD protocols. Continuous variable QKD is proposed to achieve a higher key rate in the short and medium transmission distance. One open question is the security of continuous variable QKD. This is an appealing topic in the field. Modeling and simulations for continuous variable QKD are also interesting.

Statistical fluctuations need to be considered in QKD with a finite key length. There is some work on this topic recently (e.g., by Renner \cite{Renner_Thesis_05}). One interesting topic is to apply Koashi's complementary idea \cite{Koashi_Compl_07} to finite key QKD and compare it with prior results.

It has already been known that one can realize quantum gates by quantum teleportation \cite{GottesmanChuang_99}. There are some proposals for the experimental quantum computation with linear optics \cite{KLM_01}. However, the scalability is a huge challenge. As yet, no one knows how to build a large scale quantum computer. A long-term challenge in the field is to find a practical proposal for a quantum factoring machine with current technology. Here a interesting topic is that whether those techniques developed in QKD could be useful to quantum computing. For instance, can the restrictions in single photon source be loosened by applying decoy idea? 


\begin{appendix}

\chapter{Abbreviations and mathematical derivations} \label{AppCh:Abbmath}

\section{Abbreviations} \label{App:Abb}
The following abbreviations are used in this thesis.
\begin{itemize}
\item
QKD: quantum key distribution

\item
BB84: the QKD protocol presented by Bennett and Brassard in 1984 \cite{BB_84}

\item
EPR pair: a maximally entangled photon pair that originated from the Einstein-Podolsky-Rosen paradox \cite{EPR_35}

\item
EDP: entanglement distillation protocol

\item
LOCC: local operations and classical communication; 1-LOCC: local operations and one-way classical communication; 2-LOCC: local operations and two-way classical communication

\item
PDC: parametric down-conversion

\item
GLLP: the security proof of QKD with imperfect devices proposed by Gottesman, Lo, L\"utkenhaus, and Preskill \cite{GLLP_04}
\end{itemize}

\section{Key rate of the recurrence scheme with an ideal single photon source} \label{App:review-recurrence}
In this section, we will review the recurrence EDP and develop the key generation rate formula given by:
\begin{equation}
R=q \cdot r,
\end{equation}
where $q$ is the basis reconciliation factor and $r$ is the residue of post-processing which we will find in the sequel. In the following, we use the same notation as in Section \ref{Sc:SingleEDP} and consider a Bell diagonal state $(q_{00}, q_{10}, q_{11}, q_{01})$.

\subsection{Parity check}
As the first step of recurrence, Alice and Bob check the parity of two pairs (labeled by control qubit $C$ and target qubit $T$). They will get an even parity if the two pairs are in one of the following states:
$$
0000,0001,0100,0101,1010,1011,1110,1111,
$$
and will get odd parity if they are in one of the following states:
$$
0010,0011,0110,0111,1000,1001,1100,1101,
$$
where the first two bits represent the control qubit, and the last
two bits represent the target qubit. That is, $ij$ represents the
Bell state $|\psi_{ij}\rangle$ as given in
Eq.~\eqref{EDP:BellStates} with $i,j=0,1$. For example, $1110$ means
that there is a bit error and a phase error in the control qubit
($|\psi_{11}\rangle$), and a bit error and no phase error in the
target qubit ($|\psi_{10}\rangle$). Thus, the probability to get an
even parity is given by:
\begin{equation} \label{Rec:Survival}
\begin{aligned}
p_S &= (q_{00}^C+q_{01}^C)(q_{00}^T+q_{01}^T) + (q_{10}^C+q_{11}^C)(q_{10}^T+q_{11}^T)\\
    &= (1-\delta_b^C)(1-\delta_b^T) + \delta_b^C\delta_b^T,
\end{aligned}
\end{equation}
where $\delta_b^C=q_{10}^C+q_{11}^C$ and
$\delta_b^T=q_{10}^T+q_{11}^T$ are the bit error rates of the input
control and target qubits, respectively. During the parity check,
the number of pure EPR pairs (or secret bits) that Alice and Bob
need to sacrifice is given by:
\begin{equation} \label{Rec:ParityCh}
\begin{aligned}
\frac12H_2(p_S),
\end{aligned}
\end{equation}
where the factor 1/2 is for the reason that Alice and Bob compute the parity of two-qubit pairs at one time.

After the parity check, the qubits are divided into two groups,
qubits with even parity and odd parity. In the following, we will
discuss the error correction and privacy amplification on these two
groups separately. The recurrence protocol appearing in Ref.~\cite{VV_Recurrence_05}
only performs error correction on qubits with even parity.

\subsection{Error correction}
For even parity qubits, we can see that the bit error syndrome of
control qubits will be the same as that of target qubits. Thus,
Alice and Bob only need to do error correction on the control (or
target) qubits. Similar to Eq.~\eqref{Twoway:AfterBstepErr}, the bit error rate of control qubits after recurrence is given by:
\begin{equation} \label{Rec:ErrControl}
\begin{aligned}
\tilde\delta_b^{C} =
\frac{(q_{10}^C+q_{11}^C)(q_{10}^T+q_{11}^T)}{p_S} =
\frac{\delta_b^C\delta_b^T}{p_S}
\end{aligned}
\end{equation}
where $p_S$ is the probability of even parity in the recurrence
given by Eq.~\eqref{Rec:Survival}. Therefore, Alice and Bob need to
sacrifice a fraction:
\begin{equation} \label{Rec:OverallErrC}
\begin{aligned}
\frac12p_SH_2(\tilde\delta_b^{C}) =
\frac12p_SH_2(\frac{\delta_b^C\delta_b^T}{p_S})
\end{aligned}
\end{equation}
to do the overall error correction. The factor 1/2 is due to the fact that control qubits have the same error syndrome as target
qubits.

Therefore, the residue of data post-processing can be expressed as:
\begin{equation} \label{Rec:Residue0}
\begin{aligned}
r = -\frac12H_2(p_S)
-\frac12p_SH_2(\frac{\delta_b^C\delta_b^T}{p_S}) + K
\end{aligned}
\end{equation}
where $p_S$ is given in Eq.~\eqref{Rec:Survival}, $\delta_b^C$ and $\delta_b^T$ are the QBER of control and target qubits respectively, and K is the residue of privacy amplification, which we will focus on in the following discussion.

\subsection{Privacy amplification}
Alice and Bob perform privacy amplification to the qubits with even and odd parities separately.

\textbf{Even parity:} Now, Alice and Bob already know the bit error
syndrome. The control and target qubits have the same bit error
syndromes, but may have different phase error syndromes. Thus, Alice
and Bob can divide the even parity qubits into four groups: control
qubits with bit error syndrome 0 and 1,  and target qubits with bit
error syndrome 0 and 1, and treat these groups separately in the
privacy amplification step. The probability of each group (summing
together the even parity probabilities given in
Eq.~\eqref{Rec:Survival}) is given by:
$$
\frac{(q_{00}^C+q_{01}^C)(q_{00}^T+q_{01}^T)}{2},
\frac{(q_{10}^C+q_{11}^C)(q_{10}^T+q_{11}^T)}{2},
\frac{(q_{00}^C+q_{01}^C)(q_{00}^T+q_{01}^T)}{2},
\frac{(q_{10}^C+q_{11}^C)(q_{10}^T+q_{11}^T)}{2}
$$
with phase error rate:
$$
\frac{q_{01}^C}{q_{00}^C+q_{01}^C},
\frac{q_{11}^C}{q_{10}^C+q_{11}^C},
\frac{q_{01}^T}{q_{00}^T+q_{01}^T},
\frac{q_{11}^T}{q_{10}^T+q_{11}^T}.
$$
Since the error syndrome of each group of qubits is known to Alice and Bob,
privacy amplification can be applied to the different
groups separately.
Then, Alice
and Bob should sacrifice a fraction:
\begin{equation} \label{Rec:PriAmpPri}
\begin{aligned}
&\frac{(q_{00}^C+q_{01}^C)(q_{00}^T+q_{01}^T)}{2}H_2(\frac{q_{01}^C}{q_{00}^C+q_{01}^C})
+
\frac{(q_{10}^C+q_{11}^C)(q_{10}^T+q_{11}^T)}{2}H_2(\frac{q_{11}^C}{q_{10}^C+q_{11}^C}) + \\
&\frac{(q_{00}^C+q_{01}^C)(q_{00}^T+q_{01}^T)}{2}H_2(\frac{q_{01}^T}{q_{00}^T+q_{01}^T})
+
\frac{(q_{10}^C+q_{11}^C)(q_{10}^T+q_{11}^T)}{2}H_2(\frac{q_{11}^T}{q_{10}^T+q_{11}^T}) \\
\end{aligned}
\end{equation}
to do the privacy amplification. Given the bit and phase error rates
of input control and target qubits $\delta_p^C=q_{11}^C+q_{01}^C$
and $\delta_p^T=q_{11}^T+q_{01}^T$, Eq.~\eqref{Rec:PriAmpPri} can be
written as:
\begin{equation} \label{Rec:PriAmp}
\begin{aligned}
\frac12(1-\delta_b^C)(1-\delta_b^T)[H_2(\frac{\delta_p^C-q_{11}^C}{1-\delta_b^C})+H_2(\frac{\delta_p^T-q_{11}^T}{1-\delta_b^T})] + \frac12\delta_b^C\delta_b^T[H_2(\frac{q_{11}^C}{\delta_b^C})+H_2(\frac{q_{11}^T}{\delta_b^T})]. \\
\end{aligned}
\end{equation}

Thus, the privacy amplification residue of even parity qubits is
given by:
\begin{equation} \label{Rec:EvenRes}
\begin{aligned}
K_{even} =  p_S
-\frac12(1-\delta_b^C)(1-\delta_b^T)[H_2(\frac{\delta_p^C-q_{11}^C}{1-\delta_b^C})+H_2(\frac{\delta_p^T-q_{11}^T}{1-\delta_b^T})]
- \frac12\delta_b^C\delta_b^T[H_2(\frac{q_{11}^C}{\delta_b^C})+H_2(\frac{q_{11}^T}{\delta_b^T})]. \\
\end{aligned}
\end{equation}

\textbf{Odd parity:}
It turns out that pairs with odd parity during the recurrence can
also contribute to the final key \cite{VV_Recurrence_05}. Instead of including
them in the error correction, Alice and Bob measure one of the two
qubits and hence, they know the bit error syndrome of the remaining
qubit. They can then proceed with privacy amplification on these
qubits. 

Suppose Alice and Bob always choose to measure the target qubits and
obtain the error syndrome of the control qubits. Similar to the even
parity case, now, Alice and Bob can divide the control qubits with
odd parity into two parts in accordance to the bit error syndrome. The
probability of each part is given by:
$$
\frac{(q_{00}^C+q_{01}^C)(q_{10}^T+q_{11}^T)}{2},
\frac{(q_{10}^C+q_{11}^C)(q_{00}^T+q_{01}^T)}{2},
$$
with a phase error rate:
$$
\frac{q_{01}^C}{q_{00}^C+q_{01}^C},
\frac{q_{11}^C}{q_{10}^C+q_{11}^C}.
$$

With the same argument as Eq.~\eqref{Rec:PriAmpPri}, the number of
qubits that need to be sacrificed to privacy amplification is given by:
\begin{equation} \label{Rec:PriAmpPriOdd}
\begin{aligned}
&\frac{(q_{00}^C+q_{01}^C)(q_{10}^T+q_{11}^T)}{2}H_2(\frac{q_{01}^C}{q_{00}^C+q_{01}^C})
+
\frac{(q_{10}^C+q_{11}^C)(q_{00}^T+q_{01}^T)}{2}H_2(\frac{q_{11}^C}{q_{10}^C+q_{11}^C}) \\
&=\frac12[(1-\delta_b^C){\delta_b^T}H_2(\frac{\delta_p^C-q_{11}^C}{1-\delta_b^C})
+ \delta_b^C(1-\delta_b^T)H_2(\frac{q_{11}^C}{\delta_b^C})]
\\
\end{aligned}
\end{equation}
Hence, the privacy amplification residue of odd parity qubits is given
by:
\begin{equation} \label{Rec:OddRes}
\begin{aligned}
K_{odd}=\frac12(1-\delta_b^C){\delta_b^T}[1-H_2(\frac{\delta_p^C-q_{11}^C}{1-\delta_b^C})]
+
\frac12\delta_b^C(1-\delta_b^T)[1-H_2(\frac{q_{11}^C}{\delta_b^C})]
\end{aligned}
\end{equation}

Therefore, the privacy amplification residue, $K$ in
Eq.~\eqref{Rec:Residue0}, by adding Eqs.~\eqref{Rec:EvenRes} and
\eqref{Rec:OddRes} and substituting Eq.~\eqref{Rec:Survival}, is
given by:
\begin{equation} \label{Rec:PriRes2}
\begin{aligned}
K =& K_{even}+K_{odd} \\
=&1-\frac12(1-\delta_b^C){\delta_b^T}-\frac12\delta_b^C(1-\delta_b^T)
-\frac12(1-\delta_b^C)H_2(\frac{\delta_p^C-q_{11}^C}{1-\delta_b^C})
-\frac12\delta_b^CH_2(\frac{q_{11}^C}{\delta_b^C})
\\
&-\frac12(1-\delta_b^C)(1-\delta_b^T)H_2(\frac{\delta_p^T-q_{11}^T}{1-\delta_b^T})
-\frac12\delta_b^C\delta_b^TH_2(\frac{q_{11}^T}{\delta_b^T}).
\end{aligned}
\end{equation}
Note that there are two free parameters $q_{11}^C$ and $q_{11}^T$ in
Eq.~\eqref{Rec:PriRes2}, which should be minimized over to
lower-bound the key rate.

\section{Security against basis dependent source} \label{Koashi}
Here, we derive Eq.~\eqref{Bounds:bp} in Section \ref{Boundary}. Rewriting Eq.~(9) of \cite{Koashi_05} gives:
\begin{equation} \label{Koashi:eq9}
\sqrt{F}\le\sqrt{(1-\delta_{bx})(1-\delta_{pz})}+\sqrt{\delta_{bx}
\delta_{pz}},
\end{equation}
where $F$ is the fidelity between the two states with two bases ($X$
and $Z$) sent by Alice, $\delta_{bx}$ is the QBER of $X$-basis
states from error testing, and $\delta_{pz}$ is the phase error rate
of the $Z$-basis states\footnote{Note that we have used different notations from those in Ref.~\cite{Koashi_05}.
By letting
$\delta_1=\delta_{bx}$ and $\delta_{ph}=\delta_{pz}$, and
substituting Eq.~(3) of \cite{Koashi_05}, we can recover Eq.~(9) of
\cite{Koashi_05} from Eq.~\eqref{Koashi:eq9}.}.
Similarly, we
have another inequality between the QBER of $Z$-basis states
$\delta_{bz}$, and the phase error rate of $X$-basis states
$\delta_{px}$:
\begin{equation} \label{Koashi:eq9other}
\sqrt{F}\le\sqrt{(1-\delta_{bz})(1-\delta_{px})}+\sqrt{\delta_{bz}
\delta_{px}}.
\end{equation}


Adding Eqs.~\eqref{Koashi:eq9} and \eqref{Koashi:eq9other} gives:
\begin{equation} \label{Koashi:bp}
\begin{aligned}
\sqrt{F} &\le
\frac12\left(\sqrt{(1-\delta_{bx})(1-\delta_{pz})}+\sqrt{\delta_{bx}
\delta_{pz}}+\sqrt{(1-\delta_{bz})(1-\delta_{px})}+\sqrt{\delta_{bz}
\delta_{px}}\right) \\
&\le
\sqrt{(1-(\delta_{bx}+\delta_{bz})/2)(1-(\delta_{pz}+\delta_{px})/2)}+\sqrt{(\delta_{bx}+\delta_{bz})/2
(\delta_{pz}+\delta_{px})/2} \\
&= \sqrt{(1-\delta_b)(1-\delta_p)}+\sqrt{\delta_b \delta_p},
\end{aligned}
\end{equation}
where the second inequality is due to the concavity of the function
$\sqrt{(1-x)(1-y)}+\sqrt{xy}$ in $[0,1]\times[0,1]$ and
we have used the definitions
$\delta_b\equiv(\delta_{bx}+\delta_{bz})/2$ and $\delta_p\equiv(\delta_{pz}+\delta_{px})/2$. Here, we assume the number of received qubits with Z basis and X basis is the same.

\section{Residue for the Decoy+GLLP+Recurrence scheme} \label{App:residue}
We calculate the residues, $K_i$, in Eq.~\eqref{Rec:Residue1} for
the five cases: $V \bigoplus S$, $S \bigoplus V$, $S\bigoplus S$,
$S\bigoplus M$, $M\bigoplus S$. Here, we apply each case, with
parameters shown in Table \ref{Rec:Tab:Input} into
Eq.~\eqref{Rec:PriRes2} to calculate each $K_i$.

\textbf{$V \bigoplus S$:}  the probability of this case is
$\Omega_{VS}=\Omega_V\Omega$.
\begin{equation} \label{Rec:KVS}
\begin{aligned}
K_{VS} &= 1-\frac14-\frac14H_2(1-2q_{11}^V)-\frac14H_2(2q_{11}^V)
-\frac14(1-e_1)H_2\left(\frac{e_1-a}{1-e_1}\right) -
\frac14e_1H_2\left(\frac{a}{e_1}\right) \\
&\ge \frac14 -\frac14(1-e_1)H_2\left(\frac{e_1-a}{1-e_1}\right)
-\frac14e_1H_2\left(\frac{a}{e_1}\right)
\end{aligned}
\end{equation}
with equality when $q_{11}^V=1/4$. This is due to the concavity of
function $H_2(\cdot)$.

\textbf{$S \bigoplus V$:} the probability of this case is
$\Omega_{VS}=\Omega_V\Omega$.
\begin{equation} \label{Rec:KSV}
\begin{aligned}
K_{SV} &\ge
1-\frac14-\frac12(1-e_1)H_2\left(\frac{e_1-a}{1-e_1}\right)-\frac12e_1H_2\left(\frac{a}{e_1}\right)
-\frac14(1-e_1)H_2\left(1-2q_{11}^V\right) -
\frac14e_1H_2\left(2q_{11}^V\right) \\
&\ge
\frac12-\frac12(1-e_1)H_2\left(\frac{e_1-a}{1-e_1}\right)-\frac12e_1H_2\left(\frac{a}{e_1}\right)
\end{aligned}
\end{equation}
with equality when $q_{11}^V=1/4$.

\textbf{$S \bigoplus S$:} the probability of this case is
$\Omega_{VV}=\Omega^2$.
\begin{equation} \label{Rec:KSS}
\begin{aligned}
K_{SS} = 1 &- e_1(1-e_1) -
\frac12(1-e_1)H_2\left(\frac{e_1-a}{1-e_1}\right) -
\frac12e_1H_2\left(\frac{a}{e_1}\right) \\
&-\frac12(1-e_1)^2H_2\left(\frac{e_1-a}{1-e_1}\right) -
\frac12e_1^2H_2\left(\frac{a}{e_1}\right). \\
\end{aligned}
\end{equation}

\textbf{$S \bigoplus M$:} the probability of this case is
$\Omega_{SM}=\Omega\Omega_M$.
\begin{equation} \label{Rec:KSM}
\begin{aligned}
K_{SM} &=
1-\frac12e_1(1-e_M)-\frac12e_M(1-e_1) -\frac12(1-e_1)H_2\left(\frac{e_1-a}{1-e_1}\right)-\frac12e_1H_2\left(\frac{a}{e_1}\right)\\
& -\frac12(1-e_1)(1-e_M)H_2\left(\frac{1-2q_{11}^M}{2-2e_M}\right)-\frac12e_1e_MH_2\left(\frac{q_{11}^M}{e_M}\right) \\
&\ge\frac12-\frac12(1-e_1)H_2\left(\frac{e_1-a}{1-e_1}\right)-\frac12e_1H_2\left(\frac{a}{e_1}\right),
\end{aligned}
\end{equation}
with equality when $q_{11}^M=e_M/2$.

\textbf{$M \bigoplus S$:} the probability of this case is
$\Omega_{MS}=\Omega_M\Omega$.
\begin{equation} \label{Rec:KMS}
\begin{aligned}
K_{MS} &=
1-\frac12e_M(1-e_1)-\frac12e_1(1-e_M) -\frac12(1-e_M)H_2\left(\frac{1-2q_{11}^M}{2-2e_M}\right)-\frac12e_MH_2\left(\frac{q_{11}^M}{e_M}\right)\\
& -\frac12(1-e_1)(1-e_M)H_2\left(\frac{e_1-a}{1-e_1}\right)-\frac12e_1e_MH_2\left(\frac{a}{e_1}\right) \\
&\ge\frac12-\frac12e_M(1-e_1)-\frac12e_1(1-e_M)\\
& -\frac12(1-e_1)(1-e_M)H_2\left(\frac{e_1-a}{1-e_1}\right)-\frac12e_1e_MH_2\left(\frac{a}{e_1}\right), \\
\end{aligned}
\end{equation}
with equality when $q_{11}^M=e_M/2$.

Therefore, the data post-processing residue of the Decoy+GLLP+Recurrence scheme will be
given by substituting Eqs.~\eqref{Rec:KVS}, \eqref{Rec:KSV},
\eqref{Rec:KSS}, \eqref{Rec:KSM} and \eqref{Rec:KMS} into
Eq.~\eqref{Rec:Residue1}:
\begin{equation} \label{App:Residue2}
\begin{aligned}
r =&  -\frac12f(p_S)H_2(p_S)-\frac12p_Sf(\frac{\delta^2}{p_S})H_2(\frac{\delta^2}{p_S})+K_{VS}+K_{SV}+K_{SS}+K_{SM}+K_{MS} \\
\ge&-\frac12f(p_S)H_2(p_S)-\frac12p_Sf(\frac{\delta^2}{p_S})H_2(\frac{\delta^2}{p_S})\\
& +\Omega_V\Omega\left[\frac14
-\frac14(1-e_1)H_2\left(\frac{e_1-a}{1-e_1}\right)
-\frac14e_1H_2\left(\frac{a}{e_1}\right)\right]
\\
& +
\Omega_V\Omega\left[\frac12-\frac12(1-e_1)H_2\left(\frac{e_1-a}{1-e_1}\right)-\frac12e_1H_2\left(\frac{a}{e_1}\right)\right]
\\
&+ \Omega^2 [1 - e_1(1-e_1) -
\frac12(1-e_1)H_2\left(\frac{e_1-a}{1-e_1}\right)-\frac12e_1H_2\left(\frac{a}{e_1}\right) \\
&-\frac12(1-e_1)^2H_2\left(\frac{e_1-a}{1-e_1}\right)-\frac12e_1^2H_2\left(\frac{a}{e_1}\right)]
\\
& +
\Omega\Omega_M[\frac12-\frac12(1-e_1)H_2\left(\frac{e_1-a}{1-e_1}\right)-\frac12e_1H_2\left(\frac{a}{e_1}\right)]
\\
& + \Omega\Omega_M[\frac12-\frac12e_M(1-e_1)-\frac12e_1(1-e_M)\\
&-\frac12(1-e_1)(1-e_M)H_2\left(\frac{e_1-a}{1-e_1}\right)-\frac12e_1e_MH_2\left(\frac{a}{e_1}\right)]\\
\end{aligned}
\end{equation}
with equality when $q_{11}^V=1/4$ and $q_{11}^M=e_M/2$. In order to
simplify this formula, we define some variables:
\begin{equation} \label{Rec:Const}
\begin{aligned}
B &=
\frac12f(p_S)H_2(p_S)+\frac12p_Sf(\frac{\delta^2}{p_S})H_2(\frac{\delta^2}{p_S})
\\
C &= \frac34\Omega_V\Omega + \Omega^2 (1-e_1+e_1^2) +
\frac12\Omega\Omega_M(2-e_1-e_M+2e_1e_M)
\\
D_1 &=
\frac34\Omega_V\Omega+\frac12\Omega^2(2-e_1)+\frac12\Omega\Omega_M(2-e_M)
\\
D_2 &=
\frac34\Omega_V\Omega+\frac12\Omega^2(1+e_1)+\frac12\Omega\Omega_M(e_M+1)
\end{aligned}
\end{equation}
Thus, Eq.~\eqref{Rec:Residue2} can be expressed as:
\begin{equation} \label{App:Residue3}
\begin{aligned}
r =& -B+K_{VS}+K_{SV}+K_{SS}+K_{SM}+K_{MS} \\
\ge& -B+C-F_a
\\
\end{aligned}
\end{equation}
where
\begin{equation} \label{Rec:Fab}
\begin{aligned}
F_a&=D_1(1-e_1)H_2(\frac{e_1-a}{1-e_1})+D_2e_1H_2(\frac{a}{e_1}) \\
\end{aligned}
\end{equation}
with equality when $q_{11}^V=1/4$ and $q_{11}^M=e_M/2$.

To obtain the lower bound $r$ in Eq.~\eqref{App:Residue3}, we need to find the
maximum value of $F_a$ over the free variable $a$.
We are interested in the range of $a\in[0,e_1]$ with $e_1\le 1/2$.
Note that $F_a$ is a concave function of $a$ in the valid range,
since a sum of two concave functions is also a concave function, and
reflecting and shifting a concave function is also a concave
function. Thus, we can take the derivative of $F_a$ with respect to
$a$ and set it to zero to find the maximum of $F_a$. Differentiating
$F_a$ with respect to $a$ gives:
\begin{eqnarray*}
\frac{d F_a}{d a} &=& D_1 \left[ \log_2 \left( \frac{e_1-a}{1-e_1}
\right) - \log_2 \left( 1-\frac{e_1-a}{1-e_1} \right) \right] + D_2
\left[ \log_2 \left( 1-\frac{a}{e_1} \right) - \log_2 \left(
\frac{a}{e_1} \right) \right]
\end{eqnarray*}
Setting $2^{\frac{d F_a}{d a}} = 1$ gives
\begin{eqnarray*}
\left(\frac{1-e_1}{e_1-a}-1\right)^{-D_1}
\left(\frac{e_1}{a}-1\right)^{D_2} &=& 1.
\end{eqnarray*}
Denoting the left-hand side to be $f(a)$, $f(a)$ is a decreasing
function of $a$ since $\frac{d F_a}{d a}$ is a decreasing function
of $a$. Therefore, we can use the bisection method to find $a$ such
that $f(a)=1$. The initial range for the bisection method is
$[0,e_1]$.


\section{QBER for entanglement PDC QKD} \label{QBER}
Here, we will study the quantum bit error rate (QBER) of the entanglement PDC QKD. Our objective is
to derive the QBER formula given in Eq.~\eqref{Model:Enta:QBER} used in the simulation. The QBER has
three main contributions:
\begin{enumerate}
\item background counts, which are random noises $e_0=1/2$;

\item intrinsic detector errors, $e_{d}$, which is the probability that a photon hits the
erroneous detector. $e_{d}$ characterizes the alignment and stability of the optical
system between the detection systems of Alice and Bob;

\item errors introduced by multi-photon-pair states: a) Alice and Bob may detect different photon
pairs; b) double clicks.  Due to the strong pulsing attack \cite{Lutkenhaus_99DoubleClick}, we
assume that Alice and Bob will assign a random bit when they get a double click. In either case,
the error rate will be $e_0=1/2$.
\end{enumerate}

Let us start with the single-photon-pair case, a Bell state given in Eq.~\eqref{Model:EPR}. The
error rate of single-photon-pair $e_1$ has two sources: background counts and intrinsic detector
errors:
\begin{equation}\label{Model:e1}
\begin{aligned}
e_1 = e_0-\frac{(e_0-e_{d})\eta_{A}\eta_{B}}{Y_1}
\end{aligned}
\end{equation}
If we neglect the case where both background and true signal cause clicks, then $e_1$ can be written as:
\begin{equation}\label{Model:e1app}
\begin{aligned}
e_1 \approx \frac{e_0(Y_{0A}Y_{0B}+Y_{0A}\eta_{B}+\eta_{A}Y_{0B})+e_{d}\eta_{A}\eta_{B}}{Y_1}.
\end{aligned}
\end{equation}
where $e_0=1/2$ is the error rate of background counts. The first term of the numerator is the
background contribution and the second term comes from the errors of true signals.

In the following, we will discuss the errors introduced by multi-photon pair states, $e_n$ with
$n\ge2$. Here, we assume that Alice and Bob use threshold detectors.  One can imagine the detection of an $n$-photon-pair state
as follows.
\begin{enumerate}

\item Alice and Bob
project the $n$-photon-pair state, Eq.~\eqref{Model:PDCn}, into $Z^{\otimes n}$ basis.

\item Afterwards, they detect each photon with certain probabilities
($\eta_A$ for Alice and $\eta_B$ for Bob).

\item If either Alice or Bob detects vacuum, then we regard it as a \emph{loss}.
If Alice and Bob both detect non-vacuum only in one polarization ($\leftrightarrow$ \emph{or}
$\updownarrow$), we regard it as a \emph{single click} event. Otherwise, we regard it as a
\emph{double click} event.
\end{enumerate}

The state of a $2$-photon-pair state, according to Eq.~\eqref{Model:PDCn}, can be written as:
\begin{equation}\label{Model:PDC2pair}
\begin{aligned}
|\Phi_2\rangle&=\frac{1}{\sqrt{3}}(|2,0\rangle_a|0,2\rangle_b-|1,1\rangle_a|1,1\rangle_b+|0,2\rangle_a|2,0\rangle_b \\
&= \frac{1}{\sqrt{3}}[|\leftrightarrow\leftrightarrow\rangle_a|\updownarrow\updownarrow\rangle_b
-\frac12(|\leftrightarrow\updownarrow\rangle+|\updownarrow\leftrightarrow\rangle)_a\otimes(|\updownarrow\leftrightarrow\rangle+|\leftrightarrow\updownarrow\rangle)_b
+|\updownarrow\updownarrow\rangle_a|\leftrightarrow\leftrightarrow\rangle_b]. \\
\end{aligned}
\end{equation}
As discussed above, Alice and Bob project the state into $Z \otimes Z$ basis.
If they end up with the first or the third state in the bracket of Eq.~\eqref{Model:PDC2pair}, they
will get perfect anti-correlation, which will not contribute to errors. If they get the second state in
the bracket of Eq.~\eqref{Model:PDC2pair}, their results are totally independent, which will cause
an error with a probability $e_0=1/2$. Thus, the error probability introduced by a $2$-photon-pair
state is 1/6. Here, we have only considered the errors introduced by multi photon states, which is item 3 discussed in the beginning of this Appendix.
We should also take into account the effects of background counts and intrinsic detector errors.
With these modifications, the error rate of $2$-photon-pair state is given by:
\begin{equation}\label{Model:e2}
\begin{aligned}
e_2 = e_0-\frac{2(e_0-e_{d})[1-(1-\eta_A)^2][1-(1-\eta_B)^2]}{3Y_2}
\end{aligned}
\end{equation}
where $Y_2$ is given in Eq.~\eqref{Model:Yn}. Eq.~\eqref{Model:e2} can be understood as follows.
Only when Alice and Bob project Eq.~\eqref{Model:PDC2pair} into
$|\leftrightarrow\leftrightarrow\rangle_a|\updownarrow\updownarrow\rangle_b$ or
$|\updownarrow\updownarrow\rangle_a|\leftrightarrow\leftrightarrow\rangle_b$ and no background
count occurs, they have a probability of $e_d$ to get the wrong answer. Given a coincident
detection, the conditional probability for this case is
${2[1-(1-\eta_A)^2][1-(1-\eta_B)^2]}/{3Y_2}$. All other cases, a background count, a double click
and measuring different photon pairs,
will contribute to an error probability $e_0=1/2$.


Next, let us study the errors coming from the state $|n-m,m\rangle_a|m,n-m\rangle_b$. When Alice
detects at least one of $n-m$ $|\updownarrow\rangle$ photons, but none of $m$
$|\leftrightarrow\rangle$ photons, and Bob detects at least one of $n-m$ $|\leftrightarrow\rangle$
photons, but none of $m$ $|\updownarrow\rangle$ photons, or both Alice and Bob have bit flips of
this case, they will end up with an error probability of $e_d$. Given a coincident detection, the
conditional probability for these two cases is:
\begin{displaymath}
\begin{aligned}
\frac{1}{Y_n}& \{[1-(1-\eta_A)^{n-m}](1-\eta_A)^{m}[1-(1-\eta_B)^{n-m}](1-\eta_B)^{m}\\
&+[1-(1-\eta_A)^m](1-\eta_A)^{n-m}[1-(1-\eta_B)^m](1-\eta_B)^{n-m}\}. \\
\end{aligned}
\end{displaymath}
When Alice detects at least one of $n-m$ $|\updownarrow\rangle$ polarizations, but none of $m$
$|\leftrightarrow\rangle$ polarizations, and Bob detects at least one of $m$ $|\updownarrow\rangle$ polarizations, but none of $n-m$ $|\leftrightarrow\rangle$ polarizations, or both Alice and Bob have
bit flips of this case, they will end up with an error probability of $1-e_d$. Given a coincident
detection, the conditional probability for these two cases is:
\begin{displaymath}
\begin{aligned}
\frac{1}{Y_n}&\{[1-(1-\eta_A)^m](1-\eta_A)^{n-m}[1-(1-\eta_B)^{n-m}](1-\eta_B)^{m}  \\
&+[1-(1-\eta_A)^{n-m}](1-\eta_A)^{m}[1-(1-\eta_B)^{m}](1-\eta_B)^{n-m}\}. \\
\end{aligned}
\end{displaymath}
For all other cases, the error probability is $e_0$. Thus, the error probability for the state
$|n-m,m\rangle_a|m,n-m\rangle_b$ is:
\begin{equation}\label{Model:errn}
\begin{aligned}
e_{nm}
=& e_0-\frac{e_0-e_{d}}{Y_n} \{(1-\eta_A)^{n-m}(1-\eta_B)^{n-m} [1-(1-\eta_A)^m][1-(1-\eta_B)^m] \\
&+(1-\eta_A)^{m}(1-\eta_B)^{m}[1-(1-\eta_A)^{n-m}][1-(1-\eta_B)^{n-m}] \\
&-(1-\eta_A)^{n-m}(1-\eta_B)^{m} [1-(1-\eta_A)^m][1-(1-\eta_B)^{n-m}] \\
&-(1-\eta_A)^{m}(1-\eta_B)^{n-m}[1-(1-\eta_A)^{n-m}][1-(1-\eta_B)^{m}]\} \\
=& e_0-\frac{e_0-e_{d}}{Y_n} [(1-\eta_A)^{n-m}-(1-\eta_A)^{m}][(1-\eta_B)^{n-m}-(1-\eta_B)^m] \\
\end{aligned}
\end{equation}

In general, for an $n$-photon-pair state described by Eq.~\eqref{Model:PDCn}, the error rate is
given by:
\begin{equation}\label{Model:en}
\begin{aligned}
e_n &= \frac{1}{n+1}\sum_{m=0}^{n} e_{nm} \\
     &= \frac{1}{n+1}\sum_{m=0}^{n} e_0-\frac{e_0-e_{d}}{Y_n} [(1-\eta_A)^{n-m}-(1-\eta_A)^{m}][(1-\eta_B)^{n-m}-(1-\eta_B)^m] \\
     &= e_0-\frac{e_0-e_{d}}{(n+1)Y_n}\sum_{m=0}^{n}[(1-\eta_A)^{n-m}-(1-\eta_A)^{m}][(1-\eta_B)^{n-m}-(1-\eta_B)^m] \\
     &= e_0-\frac{2(e_0-e_{d})}{(n+1)Y_n} [\frac{1-(1-\eta_A)^{n+1}(1-\eta_B)^{n+1}}{1-(1-\eta_A)(1-\eta_B)}-\frac{(1-\eta_A)^{n+1}-(1-\eta_B)^{n+1}}{\eta_B-\eta_A}] \\
\end{aligned}
\end{equation}

The overall QBER is given by:
\begin{equation}\label{Model:QBERApp}
\begin{aligned}
E_{\lambda}Q_{\lambda} =& \sum_{n=0}^{\infty} e_nY_nP(n) \\
               =& e_0Q_{\lambda}-\sum_{n=0}^{\infty} \frac{2(e_0-e_{d})\lambda^n}{(1+\lambda)^{n+2}}
               [\frac{1-(1-\eta_A)^{n+1}(1-\eta_B)^{n+1}}{1-(1-\eta_A)(1-\eta_B)}-\frac{(1-\eta_A)^{n+1}-(1-\eta_B)^{n+1}}{\eta_B-\eta_A}] \\
               =&e_0Q_{\lambda}-\frac{2(e_0-e_{d})\eta_A\eta_B\lambda(1+\lambda)}{(1+\eta_A\lambda)(1+\eta_B\lambda)(1+\eta_A\lambda+\eta_B\lambda-\eta_A\eta_B\lambda)} \\
\end{aligned}
\end{equation}
where $Q_{\lambda}$ is the gain given in Eq.~\eqref{Model:Enta:Gain}.

\chapter{Optimal $\mu$} \label{Ap:Optimalmu}
In this appendix, we will discuss the optimal expected photon number $\mu$ for various protocols.


\section{Coherent state QKD} \label{ApSc:muCoherent}
Here, we will discuss the optimal choice of the expected photon number $\mu$ of the coherent state QKD with and without decoy states.

Let us start with a generic discussion. On the one hand, we need to maximize the probability of a single photon detection, which is the only source of the final secure key (for BB84). To achieve this point, we should maximize the single photon sources. Considering a weak coherent state photon sources in accordance to the Poisson distribution of the photon number as shown in Eq.~\eqref{Model:Poisson}, the single photon source reaches its maximum when $\mu=1$. On the other hand, we have to control the probability of the multi photon detection to ensure the security of the system. Thus, we should keep the untagged states (single photon states) ratio large, which requires $\mu$ to be not too large.
Therefore, intuitively we have:
$$
\mu \in (0,1].
$$

\subsection{Without decoy states} \label{ApSub:nondecoymu}
Here, we will consider the case of the coherent state QKD without decoy states, following the discussion in Ref.~\cite{IndividualAttack_00}. Assume that Alice and Bob apply the GLLP security analysis as discussed in Section \ref{Sc:GLLP}. We desire to get an optimal value of $\mu$ that maximizes the key generation rate $R$ in Eq.~\eqref{Post:KeyRate} with other parameters fixed. The key parameters here are the overall transmittance $\eta$ given in Eq.~\eqref{Model:Eta}, background rate $Y_0$, and the intrinsic detection error rate $e_{d}$.

Let us make an approximation first: if the background contribution is negligible, that is, $Y_0\ll \eta$, then from Eqs.~\eqref{Model:WithoutEve}:
\begin{equation}\label{Optimalmu:ApproxNondecoy}
\begin{aligned}
Q_\mu &\cong 1-e^{-\eta\mu}\\
E_\mu &\cong e_{d}\\
\end{aligned}
\end{equation}
Then according to Eq.~\eqref{Decoy:nondecoyQ1e1}, the estimation of $Q_1$ and $e_1$ is:
\begin{equation}\label{Optimalmu:ApproxQ1e1}
\begin{aligned}
Q_1 &\ge Q_\mu-\sum_{i=2}^{\infty}\frac{\mu^{i}}{i!}e^{ - \mu} \\
&\cong (1+\mu)e^{-\mu}-e^{-\eta\mu} \\
e_1 &\le \frac{e_d(1-e^{-\eta\mu})}{(1+\mu)e^{-\mu}-e^{-\eta\mu}} \\
\end{aligned}
\end{equation}
Then we can substitute these approximations into the key rate formula Eq.~\eqref{Post:KeyRate} and take the derivative of $\mu$ to get the optimal $\mu$.

$$
\begin{aligned}
R & \le \frac12(Q_{\mu}-p_M)\\
& = \frac12[(1+\mu)\exp(-\mu)-\exp(-\eta\mu)]
\end{aligned}
$$
with the pessimistic assumption Eq.~\eqref{Decoy:nondecoyQ1e1}. This
expression is optimized if we choose $\mu=\mu_{Optimal}$, which
fulfills:
$$
-\mu\exp(-\mu)+\eta\exp(-\eta\mu)=0.
$$
Since for a realistic setup, we expect that $\eta\mu \ll 1$, we find:
\begin{equation} \label{AppendixB:PriorMu}
\eta_{Optimal}\approx\eta.
\end{equation}

\begin{figure}[hbt]
\centering
\resizebox{12cm}{!}{\includegraphics{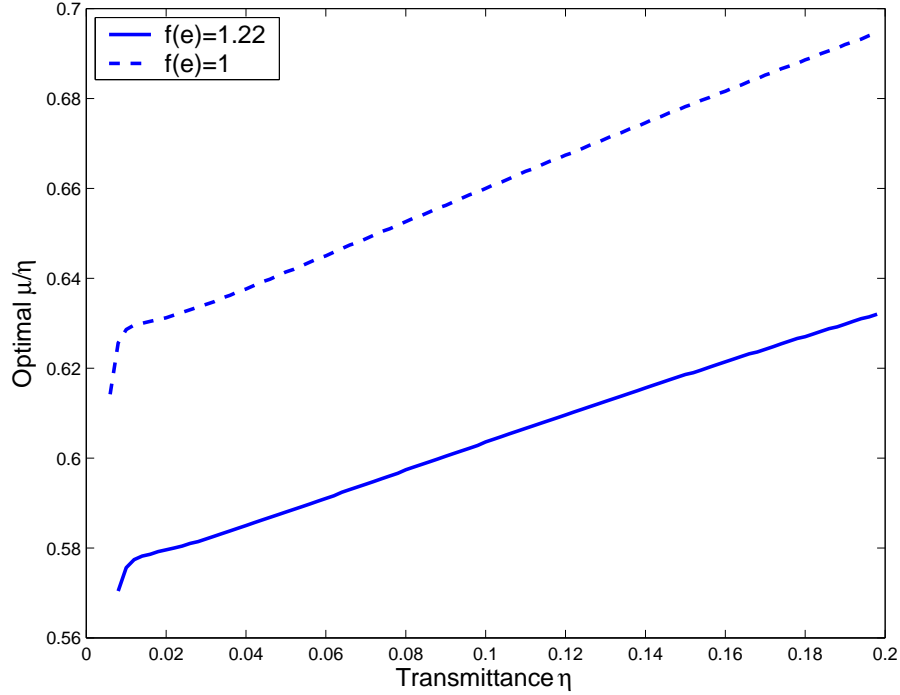}}
\caption{Plot of the optimal expected photon number $\mu$ as a function of transmittance $\eta$ for the coherent state QKD+non decoy. The parameters used in the simulation are listed in Table~\ref{Tab:GYSpara}. Here, we numerically calculate the optimal $\mu$ that maximizes the key generation rate by Eqs.~\eqref{Post:KeyRate} and \eqref{Decoy:nondecoyQ1e1}. In the regime around $\eta\approx0$, the key rate is 0. Thus, there is no point to talk about optimal $\mu$ in that regime.}
\label{AppendixB:fig:Prior}
\end{figure}

We use the numerical analysis to verify Eq.~\eqref{AppendixB:PriorMu}. When we keep all parameters fixed and vary the expected photon number $\mu$ of the signal, we can determine the $\mu_{Optimal}$ to maximize the key generation rate by Eq.~\eqref{Post:KeyRate}. If we fix the background rate $Y_0$ and the probability of erroneous detection $e_{d}$, and vary the transmittance $\eta$, we can draw the relationship between the optimal $\mu_{Optimal}$ and $\eta$. The result is shown in Figure~\ref{AppendixB:fig:Prior}, from which we can see that Eq.~\eqref{AppendixB:PriorMu} is a good approximation.

\subsection{With decoy state} \label{ApSub:decoymu}
In principle, Alice and Bob can estimate $Q_1$ and $e_1$ accurately with the decoy state. Hence, $\mu_{Optimal}$ should maximize the untagged states ratio $\Omega=Q_1/Q_\mu$. Thus, we can expect that $\mu_{Optimal}$ should be greater than \eqref{AppendixB:PriorMu}.

\begin{figure}[hbt]
\centering \resizebox{12cm}{!}{\includegraphics{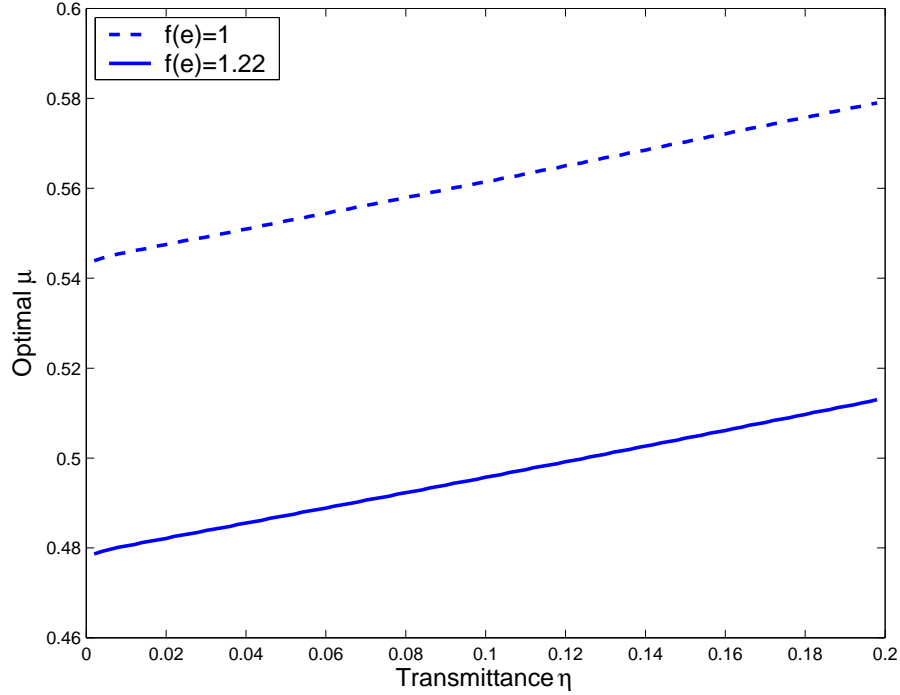}}
\caption{Plot of the optimal expected photon number $\mu$ as a function of transmittance $\eta$ for the coherent state QKD+infinite decoy. The parameters used in the simulation are listed in Table~\ref{Tab:GYSpara}.}
\label{AppendixB:fig:DecoyGen}
\end{figure}

Let us start with a numerical analysis on Eq.~\eqref{Post:KeyRate} directly. For each distance, we determine the optimal $\mu$ that maximizes the key generation rate. The result is shown in Figure~\ref{AppendixB:fig:DecoyGen}. We can see that the optimal $\mu$ for GYS is around $0.48$ when $f(\delta)=1.22$.

Now, we would like to do an analytical discussion under some approximations. We take the approximations $Y_0\ll\eta\ll1$. Then Eqs.~\eqref{Model:Qi}, \eqref{Model:ei}, \eqref{Model:Gain} and \eqref{Model:QBER} are reduced to:
$$
\begin{aligned}
Q_1   &\cong \eta\mu e^{-\mu}\\
e_1   &\cong e_{d}\\
Q_\mu &\cong \eta\mu\\
E_\mu &\cong e_{d}\\
\end{aligned}
$$
Substituting these formulas into Eq. \eqref{Post:KeyRate}, the key generation rate is given by:
$$
R\approx\frac12\{-\eta\mu f(e_{d})H_2(e_{d})+\eta\mu
e^{-\mu}[1-H_2(e_{d})]\}
$$
The expression is optimized if we choose $\mu=\mu_{Optimal}$ which fulfills:
\begin{equation} \label{Optmu:Decoymu}
(1-\mu)\exp(-\mu)=\frac{f(e_{d})H_2(e_{d})}{1-H_2(e_{d})}.
\end{equation}
Then we can solve this equation and obtain, by using $f(\delta)=1.22$:
$$
\begin{aligned}
\mu_{Optimal}^{GYS} &\approx 0.48
\end{aligned}
$$
where for the GYS experiment, $e_{d}=3.3\%$, as listed in Table~\ref{Tab:GYSpara}. In comparison of this result to Figure~\ref{AppendixB:fig:DecoyGen}, we can see that Eq.~\eqref{Optmu:Decoymu} is a good approximation.

\section{Triggering PDC QKD} \label{ApSc:muTrig}
Here, instead of numerically optimizing $\mu$ as implemented for Figure~\eqref{Fig:Toytr}, we qualitatively investigate the optimal $\mu$ for the triggering PDC QKD with and without decoy states. We are interested in the case where Alice uses a threshold detector.

\subsection{Without decoy states}
Let us begin with the optimal $\mu$ of the case without decoy states. Here, we will apply the GLLP
\cite{GLLP_04} security analysis. As shown in Ref.~\cite{Low_06}, GLLP and L\"utkenhaus's
\cite{IndividualAttack_00} security analyses achieve similar performances for the coherent state
QKD. Intuitively, we should get a similar optimal $\mu$ as given in
Ref.~\cite{IndividualAttack_00}, $\mu\approx\eta/2$.

\begin{figure}[hbt]
\centering \resizebox{12cm}{!}{\includegraphics{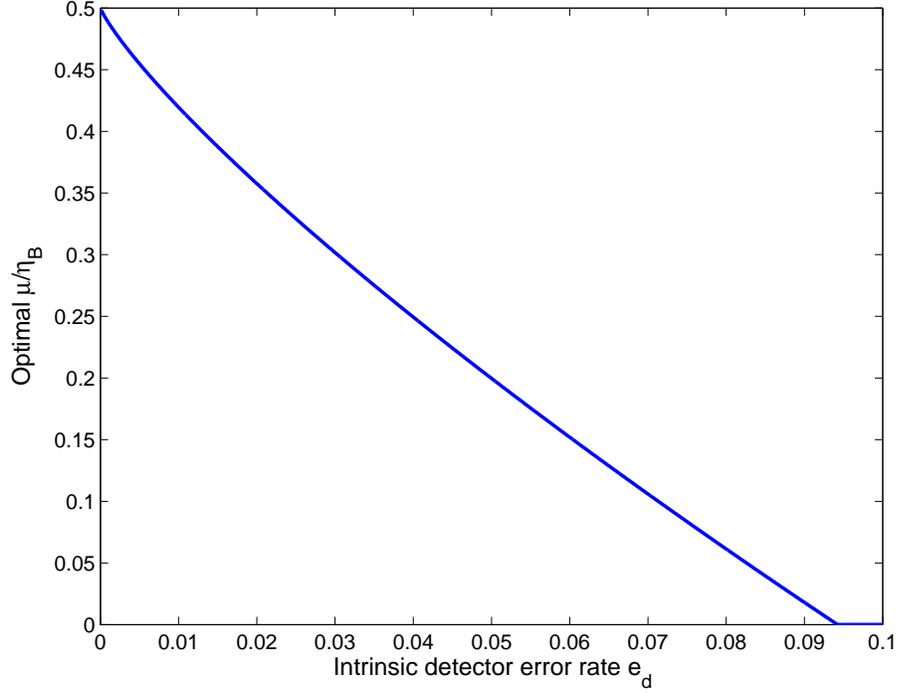}} \caption{Plot of the optimal $\mu$
in terms of $e_d$ for triggering PDC+non-decoy. Here, we use $f(e_d)=1.22$.} \label{Fig:optmunon}
\end{figure}

From Eq.~\eqref{Model:QEThreshold}, we can see that the gain $Q_{\mu,j}$ ($j=0,1$) is in the order of $\mu\eta$. To keep $Q_{1,0}$ or $Q_{1,1}$ in Eq.~\eqref{Post:Q1e1nondecoy} positive, $\mu$ should be in the order of $\eta$. By assuming $\mu$, $\eta$ and $Y_{0B}$ are small, we can simplify Eq.~\eqref{Model:QEThreshold}:
\begin{equation}\label{Simulation:nondecoyApp}
\begin{aligned}
Q_{\mu,0}+Q_{\mu,1} &\approx \eta\mu \\
E_{\mu,0}\approx E_{\mu,0} &\approx e_d \\
Q^L_{1,0}+Q^L_{1,1} &\approx \eta\mu - \mu^2 \\
e^U_1 &\approx \frac{\eta e_d}{\eta-\mu} \\
\end{aligned}
\end{equation}
where $Q^L_{1,0}+Q^L_{1,1}$ is the lower bound of $Q_{1,0}+Q_{1,1}$ and $e^U_1$ is the upper bound of $e_1$ from Eq.~\eqref{Post:Q1e1nondecoy}. Since the error rates from triggered ($j=1$) and non-triggered ($j=0$) detection events are the same, the key generation rate given by Eq.~\eqref{Post:KeyTrig} can be simplified to:
\begin{equation} \label{Simulation:KeySmartApp}
\begin{aligned}
R &\ge q \{-f(E_{\mu})Q_\mu H_2(E_{\mu})+Q_1[1-H_2(e_1)]+Q_0\} \\
  &\approx q \{-f(e_d)\eta\mu H_2(e_d)+(\eta\mu - \mu^2)[1-H_2(\frac{\eta e_d}{\eta-\mu})]\}
\end{aligned}
\end{equation}
By taking the derivative of $R$, the optimal $\mu\equiv x\eta$ satisfies:
\begin{equation} \label{Simulation:Keynondiffmu} 
\begin{aligned}
&-f(e_{d}) H_2(e_{d})+1-2x+e_d\log_2\frac{e_d}{1-x}+(1-e_d-2x)\log_2(1-\frac{e_d}{1-x})=0. \\
\end{aligned}
\end{equation}
Here if set $e_{d}=0$, then we get $x=1/2$, which is compatible with L\"ukenthaus' result
\cite{IndividualAttack_00}. Note that $x=1/2$ essentially maximizes the probability of the single photon source $Q^L_{1,0}+Q^L_{1,1}$ in Eq.~\eqref{Simulation:nondecoyApp}. More precisely, we can solve Eq.~\eqref{Simulation:Keynondiffmu} numerically, see Figure~\ref{Fig:optmunon}.

From Figure~\ref{Fig:optmunon}, we can see that the optimal $\mu$ for triggering PDC+non-decoy is
$\mu=O(\eta)$, which will lead the final key generation rate $R=O(\eta^2)$.

\subsection{With decoy states}
With decoy states, Alice and Bob can estimate $Q_1$ and $e_1$ better. Here, we consider the
infinite decoy state case with threshold detectors. Under the assumption that $\eta$ and $Y_{0B}$ are small, we can simplify Eqs.~\eqref{Model:QEThreshold} and \eqref{Model:qe1Threshold}:
\begin{equation}\label{Simulation:AsymdecoyApp}
\begin{aligned}
Q_{\mu,0}+Q_{\mu,1} &\approx \eta\mu \\
E_{\mu,0} \approx E_{\mu,0} &\approx e_d \\
Q_{1,0}+Q_{1,1} &\approx \frac{\eta\mu}{(1+\mu)^2} \\
e_1 &\approx e_d \\
\end{aligned}
\end{equation}

\begin{figure}[hbt]
\centering \resizebox{12cm}{!}{\includegraphics{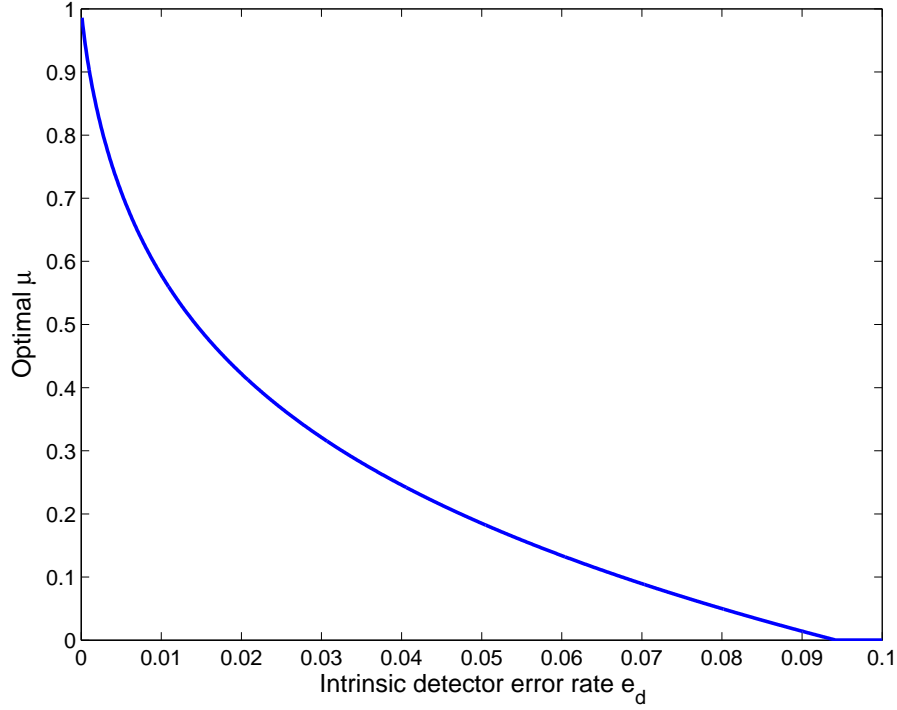}} \caption{Plot of the optimal $\mu$
in terms of $e_d$ for the triggering PDC+infinite decoy. Here, we use $f(e_d)=1.22$.} \label{Fig:optmudec}
\end{figure}

With these approximations, the key generation rate given in Eq.~\eqref{Post:KeyTrig} can be
simplified to:
\begin{equation} \label{Simulation:KeyTrigapp}
R \approx q \{-f(e_{d})\eta\mu H_2(e_{d})+ \frac{\eta\mu}{(1+\mu)^2}[1-H_2(e_{d})]\}.
\end{equation}
The optimal $\mu$ satisfies:
\begin{equation} \label{Simulation:Keydecdiffmu}
\frac{1-\mu}{(1+\mu)^3}=\frac{f(e_{d})H_2(e_{d})}{1-H_2(e_{d})}
\end{equation}
Here, if set $e_{d}=0$, then we get $\mu=1$ with which the probability to getting a single photon state is maximized. The numerical result of Eq.~\eqref{Simulation:Keydecdiffmu} is shown in
Figure~\ref{Fig:optmudec}.


From Figure~\ref{Fig:optmudec}, which is similar to the case coherent state QKD with decoy states
\cite{Practical_05}, one can see that the optimal $\mu$ is independent of channel loss $\eta$ for the infinite decoy state case with threshold detectors, i.e., $\mu=O(1)$, which will lead the final key generation rate $R=O(\eta)$.

\subsection{Numerical checking}
\begin{figure}[hbt]
\centering \resizebox{12cm}{!}{\includegraphics{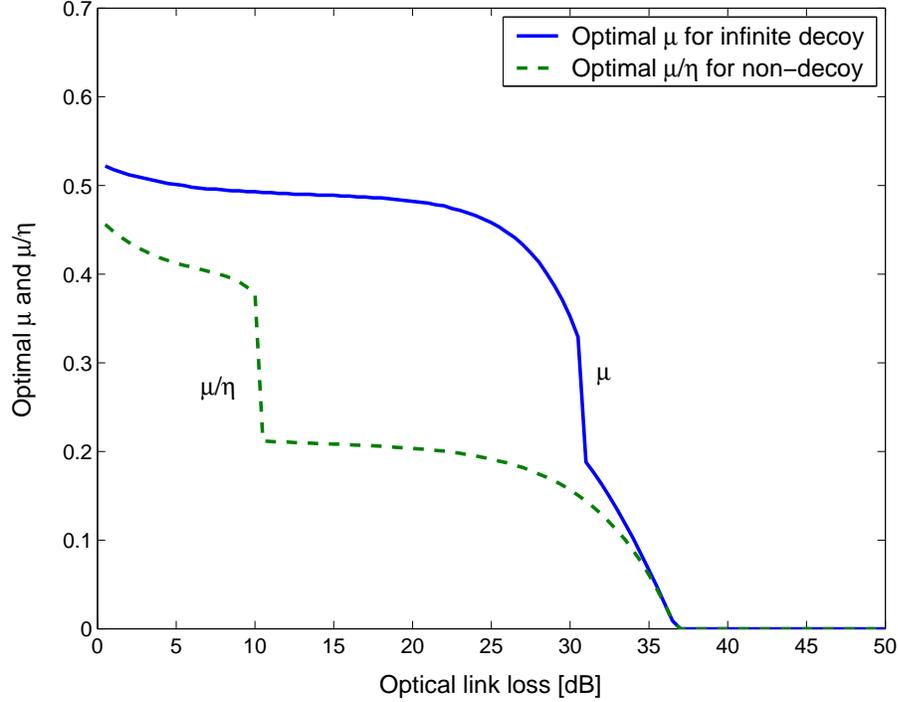}} \caption{Plot of the optimal $\mu$
in terms of optical loss for triggering PDC+non-decoy and triggering PDC+infinite-decoy. Here, we use $q=1/2$ and $f(E_\mu)=1.22$. Simulation parameters are listed in Table \ref{Tab:PDC144}.} \label{Fig:munumer}
\end{figure}

Now we would like to numerically compare the optimal $\mu$ with and without decoy states by simulating a recent PDC experiment \cite{PDC144_07}, with parameters listed in Table \ref{Tab:PDC144}. In the simulation, we numerically optimize $\mu$ for the key rate given by Eq.~\eqref{Post:Key01} for the non-decoy and infinite decoy methods. For this particular setup, the optimal $\mu$ is shown in Figure \ref{Fig:munumer}.

From the figure, we can see that the optimal $\mu$ for the non-decoy case is in the order of $\eta$, while the optimal $\mu$ for the infinite-decoy case is in the order of 1. This is consistent with the results of the analysis in the two previous subsections.

\section{Entanglement PDC QKD} \label{ApSc:muEntangle}
The optimal $\mu$ for the coherent state QKD has already been discussed
\cite{IndividualAttack_00,Practical_05}. Here, we need to determine the optimal $\mu$ for the
entanglement PDC QKD. In the following calculation, we will focus on optimizing the parameter
$\lambda$ ($=\mu/2$) for the key generation rate given in Eq.~\eqref{Post:KeyEn}.

By assuming $\eta_B$ to be small and neglecting $Y_0$, we can simplify Eq.~\eqref{Model:Enta:Gain}:
\begin{equation}\label{Simulation:Gainapp}
\begin{aligned}
Q_{\lambda} 
        &\approx 2\eta_B\lambda[1-\frac{1-\eta_A}{(1+\eta_A\lambda)^3}]. \\
\end{aligned}
\end{equation}

The overall QBER given in Eq.~\eqref{Model:Enta:QBER} can be simplified to:
\begin{equation}\label{Simulation:QBERapp}
\begin{aligned}
E_{\lambda} 
        &\approx\frac12-\frac{(1-2e_{d})(1+\lambda)(1+\eta_A\lambda)}{2(1+3\lambda+3\eta_A\lambda^2+\eta_A^2\lambda^3)}. \\
\end{aligned}
\end{equation}

In order to maximize the key generation rate given by Eq.~\eqref{Post:KeyEn}, the optimal
$\lambda$ satisfies:
\begin{equation} \label{Simulation:KeyEndiffmu}
\frac{\partial Q_\lambda}{\partial \lambda}[1-(1+f(E_\lambda))H_2(E_\lambda)]
-Q_\lambda[1+f(E_\lambda)]\frac{\partial E_\lambda}{\partial
\lambda}\log_2\frac{1-E_\lambda}{E_\lambda}=0.
\end{equation}
Here, we treat $f(E_\lambda)$ as a constant. In the following, we will consider two extremes:
$\eta_A\approx1$ and $\eta_A\ll1$.

When $\eta_A\approx1$, the overall gain and QBER are given by:
\begin{equation}\label{Simulation:GainappA1}
\begin{aligned}
Q_{\lambda} 
        &\approx 2\eta_B\lambda \\
E_{\lambda} 
        &\approx\frac{2e_d+\lambda}{2+2\lambda}. \\
\end{aligned}
\end{equation}
Thus, Eq.~\eqref{Simulation:KeyEndiffmu} can be simplified to:
\begin{equation} \label{Simulation:KeyEndiffmuA1}
1-[1+f(E_\lambda)]H_2(E_\lambda)-\lambda[1+f(E_\lambda)]\frac{1-2e_d}{2(1+\lambda)^2}
\log_2\frac{1-E_\lambda}{E_\lambda}=0.
\end{equation}

\begin{figure}[hbt]
\centering \resizebox{12cm}{!}{\includegraphics{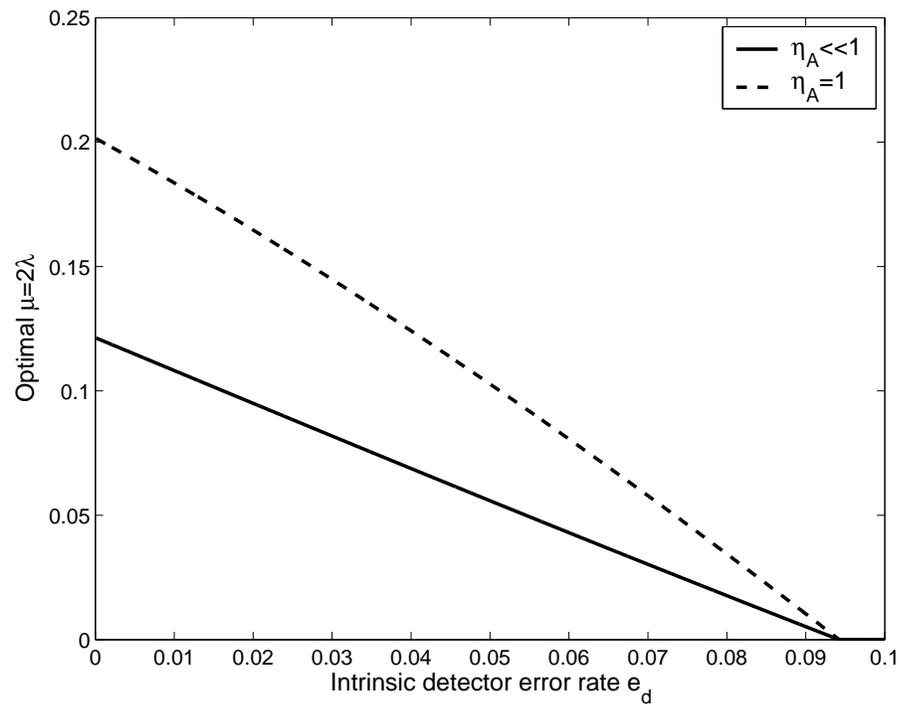}} \caption{Plot of the optimal $\mu$ in
terms of $e_d$ for the entanglement PDC QKD. $f(e_d)=1.22$.} \label{Fig:optmuen}
\end{figure}

When $\eta_A\ll1$,
\begin{equation}\label{Simulation:GainappA0}
\begin{aligned}
Q_{\lambda} 
        &\approx 2\eta_A\eta_B\lambda(1+3\lambda) \\
E_{\lambda} 
        &\approx \frac{e_d+\lambda+e_d\lambda}{1+3\lambda}.
\end{aligned}
\end{equation}
Thus, Eq.~\eqref{Simulation:KeyEndiffmu} can be simplified to:
\begin{equation} \label{Simulation:KeyEndiffmuA0}
(1+6\lambda)\{1-[1+f(E_\lambda)]H_2(E_\lambda)\}-\lambda[1+f(E_\lambda)]\frac{1-2e_d}{1+3\lambda}
\log_2\frac{1-E_\lambda}{E_\lambda}=0.
\end{equation}
The solutions to Eqs.~\eqref{Simulation:KeyEndiffmuA1} and \eqref{Simulation:KeyEndiffmuA0} are
shown in Figure~\ref{Fig:optmuen}.

From Figure~\ref{Fig:optmuen}, we can see that the optimal
$\mu=2\lambda$ for the entanglement PDC is in the order of 1,
$\mu=2\lambda=O(1)$, which will lead the final key generation rate
to be $R=O(\eta_A\eta_B)$.

\end{appendix}

\addcontentsline{toc}{chapter}{Bibliography}
\bibliographystyle{abbrv}
\bibliography{Bibli}

%
%


\end{document}